\documentclass[twocolumn,showpacs,preprintnumbers,amsmath,amssymb,floatfix,prd]{revtex4}
\usepackage{epsf}
\usepackage{graphicx}
\usepackage{dcolumn}
\usepackage{bm}

\newcommand{\be}{\begin{equation}}
\newcommand{\ee}{\end{equation}}
\newcommand{\ba}{\begin{eqnarray}}
\newcommand{\ea}{\end{eqnarray}}
\newcommand{\bi}{\begin{itemize}}
\newcommand{\ei}{\end{itemize}}
\newcommand{\nr}{n^{(r)}}
\newcommand{\nm}{n^{(m)}}
\newcommand{\dnm}{\dot n^{(m)}}
\newcommand{\Omr}{\Omega_{(r)}}
\newcommand{\Omm}{\Omega_{(m)}}
\newcommand{\Omt}{\Omega_{(tot)}}
\newcommand{\trel}{\tau_{\hbox{\tiny{rel}}}}
\newcommand{\dtr}{\dot \tau_{\hbox{\tiny{rel}}}}
\newcommand{\kB}{k_{_B}}
\newcommand{\tH}{t_{_H}}
\newcommand{\Se}{S_{(e)}}
\newcommand{\dSe}{\dot S_{(e)}}
\newcommand{\rhow}{\rho_{\hbox{\tiny w}}}
\newcommand{\prw}{p_{\hbox{\tiny w}}}
\newcommand{\nw}{n_{\hbox{\tiny w}}}
\newcommand{\Tw}{T_{\hbox{\tiny w}}}
\newcommand{\mw}{m_{\hbox{\tiny w}}}
\newcommand{\Nw}{N_{\hbox{\tiny w}}}
\newcommand{\nuw}{\nu_{\hbox{\tiny w}}}
\newcommand{\Omw}{\Omega_{\hbox{\tiny w}}}
\newcommand{\rhobar}{\rho_{\hbox{\tiny b}}}
\newcommand{\pbar}{p_{\hbox{\tiny b}}}
\newcommand{\nbar}{n_{\hbox{\tiny b}}}
\newcommand{\mbar}{m_{\hbox{\tiny b}}}
\newcommand{\Nbar}{N_{\hbox{\tiny b}}}
\newcommand{\nubar}{\nu_{\hbox{\tiny b}}}
\newcommand{\Ombar}{\Omega_{\hbox{\tiny b}}}
\newcommand{\rhoel}{\rho_{\hbox{\tiny e}}}
\newcommand{\pel}{p_{\hbox{\tiny e}}}
\newcommand{\nel}{n_{\hbox{\tiny e}}}
\newcommand{\mel}{m_{\hbox{\tiny e}}}
\newcommand{\Nel}{N_{\hbox{\tiny e}}}
\newcommand{\nuel}{\nu_{\hbox{\tiny e}}}
\newcommand{\rhog}{\rho_{\hbox{\tiny $\gamma$}}}
\newcommand{\pg}{p_{\hbox{\tiny $\gamma$}}}
\newcommand{\ngo}{n_{\hbox{\tiny $\gamma$}}}
\newcommand{\Ng}{N_{\hbox{\tiny $\gamma$}}}

\begin{document}

\title{A qualitative and numerical study of the matter-radiation interaction 
in Kantowski-Sachs cosmologies}

\author{Alan A. Coley$^*$, Antonio Sarmiento G.$^\dagger$ and Roberto A. 
Sussman$^\ddagger$}

\affiliation{$^*$ Department of Mathematics and Statistics, Dalhousie 
University, Halifax, Nova Scotia B3H 3J5, Canada,
\\
$^\dagger$ Instituto de Matem\'aticas, U. Cuernavaca, UNAM, Av. Universidad 
s/n, 62200 Morelos, M\'exico.
\\
$^\ddagger$ Instituto de Ciencias Nucleares, Apartado Postal 70543, UNAM,
M\'exico D. F., 04510 M\'exico.}
\date{\today}

\begin{abstract}
\hskip3cm{\bf TO APPEAR IN THE PHYSICAL REVIEW D}
\vskip0.3cm
We examine, from both a qualitative and a numerical point of view, the 
evolution of Kantowski--Sachs cosmological models whose source is a mixture of 
a gas of weakly interacting massive particles (WIMP's) and a radiative gas 
made up of a ``tightly coupled'' mixture of electrons, baryons and photons. 
Our analysis is valid from the end of nucleosynthesis up to the duration of 
radiative interactions ($10^6$ K $> \ T \ > \ 4 \times 10^3$ K). In this 
cosmic era annihilation processes are negligible, while the WIMP's only 
interact gravitationally with the radiative gas and the latter behaves as a 
single dissipative fluid that can be studied within a hydrodynamical 
framework. Applying the full transport equations of Extended Irreversible 
Thermodynamics, coupled with the field and balance equations, we obtain a set 
of governing equations that becomes an autonomous system of ordinary 
differential equations once the shear viscosity relaxation time, $\trel$, is 
specified. Assuming that $\trel $ is proportional to the Hubble time, the 
qualitative analysis indicates that models begin in the radiation dominated 
epoch close to an isotropic equilibrium point (saddle). We show how the form 
of $\trel$ governs the relaxation timescale of the models towards an 
equilibrium photon entropy, leading to ``near-Eckart'' and transient regimes  
associated with ``abrupt'' and ``smooth'' relaxation processes, respectively. 
Assuming the WIMP particle to be a super-symmetric neutralino with mass $ \ 
\mw \sim 100$ GeV, the numerical analysis reveals that a physically plausible 
evolution, compatible with a stable equilibrium state and with observed bounds 
on CMB anisotropies and neutralino abundance, is only possible for models 
characterized by initial conditions associated with nearly zero spatial 
curvature and total initial energy density very close to unity. An expression 
for the relaxation time, complying with physical requirements, is obtained in 
terms of the dynamical equations. It is also shown that the ``truncated'' 
transport equation does not give rise to acceptable physics.
\end{abstract}

\pacs{04.40.Nr, 05.70.Ln}

\maketitle

\section {Introduction}

The radiative era of cosmic evolution extends from the end of cosmological
nucleosynthesis to the decoupling of baryonic matter and radiation, covering 
the temperature range  $4 \times 10^3 - 10^6$ K (roughly between 1 eV and 1
keV). During this period  cosmic matter can be
described\cite{kotu,padma,peacock} as a mixture of two  main non--interacting
components:  one, a non-relativistic and collisionless  gas of WIMP's (cold
dark matter, CDM), the other a tightly coupled mixture of non-relativistic
baryons, electrons and ultra-relativistic matter  (``radiation'', {\it i. e.},
photons and neutrinos). The standard approach to  the radiative era consists of
using a  Friedmann-Lema{\^\i}tre-Robertson-Walker (FLRW) space-time 
background~\cite{kotu,padma,peacock} whose sources are described either by 
equilibrium kinetic theory~\cite{bern}, gauge invariant 
perturbations~\cite{efsta}, or by hydrodynamical models~\cite{ct,rdm,pm} 
which in general, fail to incorporate a physically plausible description of 
the matter-radiation interaction since they assume a full thermodynamical 
equilibrium throughout the evolution. Since the tight coupling between 
electrons, baryonic matter and radiation follows from various processes of 
radiative interaction~\cite{wei,sw,peeb}, mostly involving photons and 
electrons, we can ignore the non-interacting neutrinos and assume the 
baryon--electron--photon mixture to evolve with a common temperature (local 
thermal equilibrium) and to behave as a single fluid, the ``radiative 
fluid''. This fluid must be dissipative in order to provide an adequate 
macroscopic model for these interactions~\cite{wei}. Ideally, all dissipative 
fluxes (heat flux, bulk and shear viscosities) should be taken into 
consideration in the study of this tight coupling. However, in order to deal 
with a mathematically tractable problem, while still aiming at a physically 
interesting generalization of previous work, we shall study the case in which 
only shear viscosity is present. Bulk viscosity is not significant in the 
temperature ranges we are considering~\cite{wei}, and, although neglecting 
the contribution of heat flux carries physical limitations, this is 
compensated by the ensuing mathematical simplification of the field and 
transport equations. This approach has already been tested in various known 
and new exact solutions~\cite{susstr}.

The simplest class of metrics allowing for anisotropic shear viscous stresses
are the Kantowski-Sachs cosmologies~\cite{ksmh,krs}, characterized by a
4-dimensional isometric group. As the source of space-time we consider a
momentum-energy tensor made of CDM (the WIMP gas) and the dissipative 
radiative fluid whose anisotropic pressure can be identified with shear 
viscous pressure. Considering the WIMP particle to be the supersymmetric 
lightest neutralino, with mass $100$ GeV~\cite{Ellis,Report,Torrente}, we can 
safely assume that throughout the radiative era these WIMP's are 
non-relativistic, collisionless and only interact gravitationally with 
radiation and baryonic matter. It is also reasonable to assume, for the 
prevailing temperatures of this era, that the pressure of the WIMP gas and 
the internal energy of the baryons and electrons are negligible in comparison 
with the radiation equilibrium pressure. Although the radiative era is 
dominated by radiation, the rest mass-energy density of the WIMP's is not 
negligible and dominates that of the baryons and electrons, hence the full 
source, CDM plus radiative fluid, can be well approximated by a momentum 
energy tensor in which CDM provides the bulk of the rest mass energy 
(``matter'', which ends up dominating the whole dynamics), while the photons 
(``radiation'') provide the bulk of thermal and dissipative effects. The 
shear viscosity associated with this source must satisfy appropriate 
constitutive and transport equations from irreversible thermodynamics that 
comply with causality and stability~\cite{hl1,hl2,wi,wjs}; these 
thermodynamical theories are known generically as Extended Irreversible 
Thermodynamics (EIT)~\cite{ddj,jcl,hl3}. The application of such theories to 
particular physical systems requires phenomenological coefficients, like the 
coefficient of shear viscosity, to be provided by Kinetic Theory. In 
particular, for the  tight coupling of electrons, baryonic matter and 
radiation and its associated photon-electron interaction, the coefficients 
corresponding to the ``radiative gas'' model~\cite{wei,jcl,uwi,dd1,dd2} 
should be employed. The entropy production must be positive definite and the 
relaxation time of shear viscosity must be a positive and monotonously 
increasing function, somehow related to the collision times of the radiative 
processes associated with the radiative era. All these timescales must 
overtake the Hubble expansion time as baryonic matter and radiation decouple. 

The paper is organized as follows: Sections II to IV present and discuss the
field equations of Kantowski-Sachs geometry for a mixture of CDM and a 
radiative fluid, the application of Extended Irreversible Thermodynamics and 
the appropriate set of equations of state for the models, as well as the 
evolution equations for the geometric and state variables. The dynamical 
analysis is carried on in Section V by defining a set of normalized 
variables, which then leads to a self-consistent and well-behaved autonomous 
system of ordinary differential (governing evolution) equations. From the 
qualitative analysis, we identify a saddle point associated with a radiation 
dominated FLRW cosmology and contained in the invariant set $ \ \chi=0, \ $ 
associated with the flat Bianchi I model. We argue that initial conditions 
must be defined near this point. In Section VI we discuss various assumptions 
on the form of the relaxation time for shear viscosity, these assumptions 
lead to the identification of a ``near-Eckart'' and transient regimes, 
respectively associated with a swift and slow rate of transiency. The effects 
of using a ``truncated'' transport equation are discussed qualitatively in 
Section VII, while Section VIII deals with the numerical analysis of the 
models bearing in mind the qualitative results obtained in previous sections. 
The main result that follows from the qualitative and numerical analysis is 
the fact that a physically plausible evolution is possible only for: (a) 
initial conditions and evolution close to $ \ \chi=0; \ $ and (b) using the 
full (not truncated) shear viscosity transport equation of EIT. A detailed 
discussion and summary of these results is provided in Section \ref{d_&_c}.

\section{Kantowski-Sachs cosmologies with anisotropic stresses}

The simplest non-FLRW cosmological metric allowing for anisotropic pressure is
that of Kantowski-Sachs (KS) models:
\be
ds^2 = - c^2dt^2 + A^{2}(t) dr^2 + B^2(t) \left[ d \theta^2 + \sin^2(\theta) 
d \phi^2 \right]. \label{ksmetric}
\ee
For a co-moving 4-velocity $ \ u^a, \ $ the expansion scalar and shear tensor
associated with the metric (\ref{ksmetric}) are:
\ba
\Theta \ = \ \frac{\dot A}{A} \ + \ \frac{2\dot B}{B}, \qquad \quad 
\sigma^a\,_b \ = \ {\bf{\hbox{diag}}}\left[0,-2\sigma,\sigma,\sigma \right],
\nonumber\\
\sigma \ \equiv \ \frac{1}{3} \left(\frac{\dot B}{B} \ - \ \frac{\dot A}{A}
\right), \qquad \qquad \qquad \label{kindefs}
\ea
where a dot denotes the derivative with respect to proper time of fundamental
observers, which for the KS metric (\ref{ksmetric}) in co-moving coordinates 
is given by $ \ t, \ $ ({\it i. e.}, $ \ \dot{A} = A_{,t} = u^a A_{,a}$). We 
consider as the source of (\ref{ksmetric}), the following stress-energy 
tensor:
\be
T^{ab} \ = \ \rho u^au^b \ + \ ph^{ab} \ + \ \Pi^{ab}, \label{EMtensor}
\ee
where $ \ h^{ab} = c^{-2} u^a u^b + g^{ab} \ $ and $ \ \Pi^a\,_b \ $ is the 
anisotropic pressure tensor satisfying $ \ \Pi_{ab} u^b = \Pi^a\,_a = 0. \ $ 
The most general form of this tensor for the metric (\ref{ksmetric}) is:
\be
\Pi^a\,_b \ = \ {\bf{\hbox{diag}}} \ \left[ \ 0, \ -2P, \ P, \ P \ \right], 
\label{Pi}
\ee
where $ \ P = P(t) \ $ is an arbitrary function to be determined by the field
equations and subjected to an evolution law for a given physical model 
associated with (\ref{ksmetric}) and (\ref{EMtensor}). The field equations 
then become:
\ba
\kappa \rho \ = \ - G^t\,_t \ = \ \frac{\dot B^2}{B^2} \ + \ \frac{2 \dot B}{B}
\frac{\dot A}{A} \ + \ \frac{1}{B^2}, \qquad \qquad 
\label{feq_rho}
\\
\nonumber\\
3 \kappa p \ = \ 2 \ G^\theta\,_\theta \ + \ G^r\,_r \ = \qquad \qquad \qquad 
\qquad \qquad \nonumber\\
- \frac{\dot B^2}{B^2} \ - \ \frac{2 \ddot A}{A} \ - \ \frac{4\ddot B}{B} \ 
- \frac{2 \dot B}{B}\frac{\dot A}{A} \ - \ \frac{1}{B^2}, \label{feq_p}
\\
\nonumber\\
3 \kappa P \ = \ G^\theta\,_\theta \ - \ G^r\,_r \ = \qquad \qquad \qquad 
\qquad \qquad \ \ \nonumber\\
\ \frac{\dot B^2}{B^2} \ - \ \frac{\ddot A}{A} \ + \ \frac{\ddot B}{B} \ - \ 
\frac{\dot B}{B} \frac{\dot A}{A} \ + \ \frac{1}{B^2}, \label{feq_P}
\ea
where $ \ \kappa = 8 \pi G/c^2, \ $ while the energy balance is given by:
\be
\dot \rho \ + \ (\rho \ + \ p) \ \Theta \ + \ 6 \ \sigma \ P \ = \ 0.
\label{Ebalance}
\ee

\subsection{Mixture of cold dark matter and a radiative fluid}

We will assume that the stress-energy tensor (\ref{EMtensor}) corresponds to 
a mixture of a non-relativistic gas of WIMP's and a radiative fluid with 
shear viscosity corresponding to a ``tightly coupled'' mixture of photons,
electrons and baryons sharing a common temperature $ \ T. \ $ Hence, $ \ 
\rho \ $ and $ \ p \ $ in eqn. (\ref{EMtensor}) are the total mass-energy 
density and equilibrium pressure given by:
\ba
\rho \ = \ \rhow \ + \ \rhobar \ + \ \rhoel \ + \  \rhog, \nonumber\\
\qquad p \ = \ \prw \ + \ \pbar \ + \ \pel \ + \ \pg, \label{mixture_dec}\\
\qquad  \pg \ = \ \rhog/3, \qquad \qquad \nonumber
\ea
with the subindices $ \ \gamma, \ w, \ b, \ $ and $ \ e \ $ denoting photons, 
WIMP's, baryons and electrons, respectively. The three latter components 
satisfy each the equation of state of a non-relativistic ideal gas:
\ba
\rhow \ = \ (\mw c^2 + 3 \kB \Tw/2) \ \nw, \quad \ \ \prw \ = \ \nw \ \kB 
\Tw, \quad \nonumber\\
\rhobar \ = \ (\mbar c^2 + 3 \kB T/2) \ \nbar, \qquad \pbar \ = \ 
\nbar \ \kB T, \qquad \label{nonrel}\\
\rhoel \ = \ (\mel c^2 + 3 \kB T/2) \ \nel, \qquad \pel \ = \ \nel 
\ \kB T, \qquad \ \nonumber
\ea
where $ \ \mw, \ \mbar, \ \mel \ $ are the respective particle masses of the 
WIMP's, the baryons (a proton mass) and the electrons, $ \ \kB \ $ is 
Boltzmann's constant, $ \ \Tw \ $is the temperature of the WIMP gas and $
\  T \ $ is the common temperature of the radiative mixture. During the 
radiative era creation/annihilation processes cease to be significant and so 
the particle number densities, $ \ \nw, \ \nbar, \ \nel, \ $ satisfy 
conservation laws of the form:
\be
\dot n \ + \ n\Theta \ = \ 0,\qquad \hbox{with}\qquad n \ = \ \nw, \ \nbar, 
\ \nel \label{n_Gamma}
\ee
which can be integrated, leading to
\be
n \ = \ \frac{N}{AB^2}, \qquad \hbox{with}\qquad N \ = \ \Nw, \ \Nbar, \ 
\Nel \label{nm}
\ee
where $ \ \Nw, \ \Nbar, \ \Nel \ $ are the constant ({\it i. e.}, conserved) 
number of WIMP's, baryons and electrons, respectively. 

The radiation component of the radiative fluid can be given either in terms 
of: 
\par
($i$) the Stefan-Boltzmann law:
\be
\rhog^{_{{sb}}} \ = \ a \ T^4, \qquad \qquad \pg^{_{{sb}}} 
\ = \ a \ T^4/3, \label{sb_law}
\ee
where $ \ a \equiv \pi^2 \kB^4/(15 \hbar^3 c^3)$ is Stefan-Boltzmann constant, 
or 
\par
($ii$) an ultra-relativistic ideal gas:
\be
\rhog^{_{ig}} \ = \ 3 \ \ngo \ \kB \ T, \qquad \qquad 
\pg^{_{ig}} \ = \ \ngo \ \kB \ T, \label{ig_law}
\ee
where $ \ \ngo \ $ is the number density of ultra-relativistic particles, 
subjected to a balance law analogous to (\ref{n_Gamma}) and given by an 
expression similar to (\ref{nm}) with the conserved photon number $ \ 
\Ng$.

In order to simplify eqn. (\ref{mixture_dec}), we can examine the ratios of 
particle numbers and rest mass densities of the different particle 
components. Considering the currently estimated~\cite{padma} ratio of photons 
per baryon, we have:
\be
\nubar \ \equiv \ \frac{\Nbar}{\Ng} \ = \ \frac{\nbar}{\ngo} \ 
\simeq \ 2.67 \times 10^{-8}\,\Ombar\,h^2,
\label{nu}
\ee
where $ \ \Ombar \ $ is the baryon abundance today (roughly $ \ 0.04 \pm 
0.01$) and $ \ h \simeq 0.7 \ $ is the adimensional Hubble 
factor~\cite{lan}. Regarding the WIMP gas, if we assume that it is made
up of the lightest supersymmetric  neutralinos with $ \ \mw \sim 100$ 
GeV~\cite{Ellis,Report,Torrente}, we have that \cite{Report}:
\be \nuw \ \equiv \ \frac{\Nw}{\Ng} \ = \ \frac{\nw}{\ngo} \ \simeq 
\ 2.82\times 10^{-8} \ \Omw \ h^2. \label{nuchi}
\ee
where $ \ \Omw \simeq 0.3 \pm 0.1 \ $ is the neutralino (CDM) abundance today. 
Using eqns. (\ref{nonrel}), (\ref{nm}) and (\ref{ig_law}), we can rewrite 
(\ref{mixture_dec}) as:
\ba 
\rho \ = \ \mw c^2 \nw \left(1+\frac{\mbar \nubar}{\mw \nuw} + \frac{\mel 
\nuel}{\mw \nuw} \right) + \qquad \qquad \quad \nonumber\\
\qquad \quad 3 \ngo \kB T \ \left(1+\frac{\nubar}{2}+\frac{\nuel}{2}+
\frac{\nuw \Tw}{2 T} \right),\nonumber\\
p \ = \ \ngo \kB T \left(1+\nubar+\nuel+\frac{\nuw \Tw}{T} \right), \qquad 
\label{mixture_dec2}
\ea
Bearing in mind that for electrons $ \ \Nel \sim \Nbar, \ $ so that $ \
\nuel \sim \nubar, \ $ while $ \ \mel \ll \mbar \approx 1$ GeV, $ \ \nuw
\ll 1, \ $  and $ \ \nuel \approx \nubar \ll 1, \ $ then,
\ba 
\frac{\mbar \nubar}{\mw \nuw} \simeq 10^{-2} \frac{\Ombar}{\Omw} \ll 1, 
\qquad \quad \frac{\mel \nuel}{\mw \nuw} \simeq 10^{-5} \ \frac{\Ombar}{\Omw} 
\ll 1,\nonumber
\ea 
while for the temperature range $ \ 4 \times 10^3$ K $< T < 10^6$ K, we have:
\be 
0.013 \ \alt \  \frac{\mw \ c^2 \ \nw}{\ngo \ \kB \ T} \ \alt \ 3.25,
\label{comparison}
\ee
showing that the radiative era is initially radiation dominated but rest mass 
energy density is not negligible and ends up becoming dominant. Therefore, 
even if $ \ \Tw/T \ $ in eqn. (\ref{mixture_dec2}) is not negligible, we 
have: 
\ba 
\rho \ \simeq \ \mw c^2 \nw \ + \ 3 \ngo \kB T, \qquad \ \ p \ \simeq \ \ngo 
\kB T,\label{eqstate0}
\ea
The same type of approximation can be obtained if we use the 
Stefan-Boltzmann law (\ref{sb_law}), since the ratio of pressures in eqns. 
(\ref{nonrel}) and (\ref{sb_law}): $ \ \pg^{_{{sb}}}/\pbar = 
a T^3/[3 \nbar \kB], \ $ is proportional to $ \ \nubar \ $ (likewise for 
WIMP's). Therefore, the mixture of a gas of WIMP's and a radiative fluid can 
be accurately described in the desired temperature range by the approximated 
equation of state:
\be
\rho \ = \ m c^2 \nm \ + \ \rho^{(r)}, \qquad p \ = \ p^{(r)} \ = \ 
\rho^{(r)}/3\nonumber\\ 
\ee
with
\be 
m \ \equiv \ \mw, \qquad \nm \ \equiv \ \nw, \qquad p^{(r)} \ \equiv \ \pg,
\label{eqstate}
\ee
where $ \ \rho^{(r)} \ $ follows from either one of eqns. (\ref{sb_law}) or
(\ref{ig_law}), hence $ \ \nr = \ngo \ $ and $ \ \rho^{(m)} = \mw c^2 
\nw. \ $ For the remaining of this paper, the superindices $ \ {}^{(r)} \ $ 
and $ \ {}^{(m)} \ $ will refer to quantities associated with photons 
(``radiation'') and WIMP's (``matter''), respectively. From eqns. 
(\ref{feq_rho}) and (\ref{feq_p}), the equation of state (\ref{eqstate}) can 
be given as the following constraint:
\be
\frac{ \ddot A}{A} + \frac{2 \ddot B}{B} + \frac{2 \dot A}{A}\frac{\dot B}{B} 
+ \frac{ \dot B^2}{B^2} + \frac{1}{B^2} - \frac{\kappa m c^2 \nm}{2}  \ = \ 0.
\label{eqstate2}
\ee

\section{Extended Irreversible Thermodynamics}

If the source (\ref{EMtensor}) is meant to describe a mixture of 
non-relativistic CDM and a radiative fluid (as argued in previous sections), 
the anisotropic pressure must be identified with a shear viscous stress of 
the latter fluid and must be compatible with a suitable thermodynamical 
formalism. We shall consider the so called ``Extended Irreversible 
Thermodynamics'' (EIT)~\cite{wi,wjs,ddj,jcl}, a theory complying with 
causality and stability requirements~\cite{hl3} and supported by the Kinetic 
Theory of gases, Information Theory and by the Theory of Hydrodynamical 
Fluctuations~\cite{jcl}. When shear viscosity is the only dissipative agent, 
the corresponding generalized entropy current, $ \ S^a, \ $ obeying the 
usual balance law with non-negative divergence, and up to second order in 
$ \ \Pi^{ab}, \ $ takes the form:
\be
S^a \ = \ n S u^a, \qquad \qquad S \ = \ \Se \ - \ \frac{ \alpha \Pi_{cd} 
\Pi^{cd}}{2 n T}, \label{Sdef}
\ee
where $ \ S \ $ is the entropy per particle, $ \ n \ = \ \nr \ + \ \nm \ = 
\ (1 + \nuw) \ \nr \ $ is the total particle number density ($\approx \nr$),
$ \ \alpha \ $ is a phenomenological coefficient to be specified later and
$ \ \Se \ $ is defined by the equilibrium Gibbs equation:
\be
n T \dSe \ = \ \dot \rho \ - \ (\rho \ + \ p^{(r)}) \frac{\dot n}{n} \ = \ 3 
\dot p \ + \ 4 p \Theta \ = \ - \ \sigma_{ab} \Pi^{ab}, \label{eqGibbs}
\ee
where we have used eqns. (\ref{n_Gamma}) and (\ref{eqstate}) to eliminate 
$\dot \rho. \ $ Fulfillment of the second law of thermodynamics requires $ 
\ S^a \ _{;a} \geq 0, \ $ which from the definition of $ \ S^a \ $ and $ \ 
S \ $ in (\ref{Sdef}) leads to:
\be
\dot S \ \geq \ 0, \label{dot_S1}
\ee
together with
\be
\alpha \ = \ \frac{\trel}{2 \ \eta}, \label{alpha}
\ee
and the evolution equation of the viscous pressure, {\it i. e.}, the 
transport equation~\cite{hl3,matr}:
$$
\trel \dot \Pi_{cd} h^c_a h^d_b \ + \ \Pi_{ab} \left[1 + \frac{\epsilon_0 \eta 
T}{2} \left( \frac{\trel}{T \eta} u^c \right)_{;c} \right] \ + \ 2 \eta 
\sigma_{ab} = 
$$
$$
\trel \dot \Pi_{cd} h^c_a h^d_b \ + \ \Pi_{ab} \left[1 - \frac{\epsilon_0 
\trel}{2} \left(\frac{\dot T}{T} + \frac{\dot\eta}{\eta} - \frac{\dtr}{\trel} 
- \Theta \right) \right] \ +
$$
\be
\qquad \qquad \qquad \qquad \qquad 2 \eta \sigma_{ab} \ = \ 0,\label{transp_eq}
\ee
where $ \ \eta, \ T, \ \trel, \ \sigma_{a b}, \ \Pi_{a b} \ $ are the
coefficient of shear viscosity, the temperature, the relaxation time, and 
the shear and shear viscosity tensors, respectively. The parameter 
$ \ \epsilon_0 \ $ can take only two  values: $ \ \epsilon_0 = 1 \ $ 
(``full'' transport equation) and $ \ \epsilon_0 = 0 \ $ (``truncated'' 
transport equation, also known as the Israel-Stewart 
equation~\cite{wi,wjs,jcl}), while Eckart's non-causal transport equation 
follows by setting $ \ \trel = 0. \ $ The coefficient of shear viscosity 
as well as other related quantities can be obtained by a  variety of 
means~\cite{wei} including Kinetic Theory, Statistical Mechanics or 
both~\cite{uwi,dd1,dd2,zim}. Unless specifically stated otherwise, we shall 
consider only  the full transport equation $ \ \epsilon_0 = 1. \ $ We will 
discuss the implications of the truncated equation ($\epsilon_0 = 0$) in 
Section \ref{truncated}. Evaluating $ \ \dot S \ $ from (\ref{Sdef}) and 
using eqns. (\ref{eqGibbs}), (\ref{alpha}), and (\ref{transp_eq}) we 
obtain:
\be
\dot S \ = \ \frac{\Pi_{ab} \ \Pi^{ab}}{2 \ \eta  \ n \ T} \ = \ \frac{2 \ 
[ \ \Se \ - \ S \ ]}{\trel} \ \geq \ 0, \label{dot_S2_full}
\ee
where we have used eqn. (\ref{Sdef}) and $ \ \epsilon_0 = 1 \ $ in eqn. 
(\ref{transp_eq}). Notice how $ \ \dot S \ $ takes a very simple form, 
illustrating the role of $ \ \trel \ $ as the reference timescale 
associated with the entropy change from $ \ S \ $ to $ \ \Se > S. \ $ 
Also, the second law of thermodynamics, ({\it i. e.}, (\ref{dot_S1})) is 
fulfilled if $ \ \eta \geq 0 \ $ or, equivalently, if $ \ \trel \ > \ 0$ 
and $ \ \Se \ \geq  S \ $ hold. Another important requirement that follows 
from the second law of thermodynamics and the stability of equilibrium 
states is that $ \ S \ $ be a convex functional, {\it i. e.}, $ \ \delta^2 
\ S < 0. \ $ For the KS models all quantities depend only on time and so a 
necessary (but not sufficient) condition is given by $ \ \ddot S \ < \ 0, \ $ 
which leads to:
\be
\left( 1 + \frac{\dtr}{2} \right) \ \Pi_{ab} \ \Pi^{ab} \ + \ 2 \ 
\eta \ \sigma_{ab} \ \Pi^{ab} \ > \ 0. \label{c_dot_tau}
\ee
For the applicability of the general relations, eqns. (\ref{Sdef}), 
(\ref{dot_S1}), (\ref{alpha}), (\ref{transp_eq}), (\ref{dot_S2_full}) and 
(\ref{c_dot_tau}) to the models we are concerned, we must impose the 
equation of state (\ref{eqstate}), with  either one of the choices eqn. 
(\ref{sb_law}) or eqn. (\ref{ig_law}). As a physical reference to infer 
the form that the coefficients $ \ \trel, \  \eta, \ \alpha \ $ may take 
for the radiative fluid, consider the``radiative  gas'' model  associated 
with the photon-electron interaction \cite{wei,ddj,jcl,uwi}, characterized 
by $ \ p, \ \rho \ $ complying with the equations of state discussed in 
section III. For the  radiative gas the forms of $ \ \eta, \ \alpha \ $ in 
terms of the relaxation time of the dissipative process, $ \ \trel \ $, 
are:
\be
\eta_{_{(rg)}} \ = \ \frac{4}{5} \ p^{(r)} \ \trel, \qquad \qquad 
\alpha_{_{(rg)}} \ = \  \frac{5}{8 \ p^{(r)}}, \label{rg}
\ee
where $ \ p^{(r)} \ $ is either $ \ aT^4/3 \ $ or $ \  \nr \kB T \ $ 
(eqns. (\ref{sb_law}) or (\ref{ig_law}), respectively) and the subscript 
``$(rg)$'' emphasizes that these quantities are specific to the radiative 
gas. Applying eqn. (\ref{rg}) into eqns. (\ref{Sdef}) and (\ref{eqGibbs}), 
we get for the entropy per particle:
\be
S \ = \ \Se \ - \  \frac{15 \ P^2}{8 \ p^{(r)} \  \nr \ T}, \label{Sdef2}
\ee
\be 
\dSe \ = \ - \ \frac{3 \ \sigma \ P}{ \nr \ T}, \label{dot_Se}
\ee
where we have neglected the entropy of the non-re\-lativ\-istic matter, so 
that $ \ \nm + \nr = (1 + \nuw) \ \nr \approx $ $\nr, \ $ and $ \ S \ $ is 
approximately the photon entropy (this is justified because the number of 
photons is so much larger than that of WIMP's, baryons and electrons). 
The transport equation (\ref{transp_eq}) becomes:
\be
\dot P \ + \ \frac{8}{5} p^{(r)} \sigma \ + \ \left( \frac{4}{3} \Theta \ + \ 
\frac{1}{\trel} \right) P \ + \ \frac{\lambda \sigma}{p^{(r)}} P^2 \ = \ 0,
\label{transp_eq2}
\ee
where $ \ \lambda \equiv 1 + 1/(2 \lambda_0) \ $ and $ \ \lambda_0 = 1/2$ or 
$2 \ $ (for $ \ p^{(r)} \ $ given by the Stefan-Boltzmann law (\ref{sb_law}) 
or the ideal gas law (\ref{ig_law}), respectively); also condition 
(\ref{dot_S2_full}) takes the form:
\be
\dot S \ = \ \frac{15 \ P^2}{4 \ p^{(r)} \ \ \nr \ T \ \trel} \ = \ 
\frac{2 \ [ \ \Se \ - \ S \ ]}{\trel} \ \geq \ 0. \label{dot_S3}
\ee
Notice that $ \ S < \Se \ $ and $ \ \trel > 0 \ $ must hold in order for 
$ \ \dot S > 0 \ $ to be satisfied, while (\ref{eqGibbs}) implies that 
we must have $ \ \sigma \ P \ < 0 \ $ in order that $ \ \dSe \ > 0. \ 
$ Regarding the interpretation of $ \ \Se, \ $ if we assume $ \ \dSe \ > 0 \ 
$\footnote{Notice that $\dSe  \ > 0$ is a sufficient but not necessary 
condition for $\dot S > 0$. Under the framework of Extended Irreversible 
Thermodynamics, the latter is the physicaly relevant condition.} and the 
Steffan-Boltzmann law (\ref{sb_law}), eqn. (\ref{eqGibbs}) then yields:
\be 
\Se \ = \ \frac{4 \ a \ T^3}{3 \ \nr}, \label{Se}
\ee  
a function that is only constant (equilibrium photon entropy) if $ \ P 
= 0. \ $ This constant is given explicitly by the black body
formulae~\cite{lan}: 
$$ 
\nr|_{_{P=0}} \ = \ \frac{30 \zeta(3) a T^3}{\kB \pi^4} \quad \Rightarrow 
\qquad \qquad \qquad \qquad \qquad
$$
\be
\qquad \qquad \qquad \qquad \Se|_{_{P=0}} \ = \ \frac{2 \pi^4 \kB}{45 
\zeta(3)} \ \approx \ 3.60 \ \kB.\label{Seeq}
\ee
where $ \ \zeta \ $ is the Riemann zeta function. However, if we 
characterize the evolution to equilibrium as $ \ P \to 0, \ $ then, as 
this evolution proceeds, $ \ \Se \ $ in eqn. (\ref{Se}) must tend to 
the constant entropy given in (\ref{Seeq}). Hence, we can identify eqn. 
(\ref{Seeq}) as the equilibrium state associated with eqn. 
(\ref{dot_S2_full}) and eqn. (\ref{dot_S3}), attained as $ \ P \to 0 \ $ 
and $ \ S \to \Se \ $ ({\it i. e.} as the radiation relaxes) in the 
timescale provided by the relaxation time $ \ \trel. \ $ For the ideal 
gas law, equation (\ref{eqGibbs}) does not yield eqn. (\ref{Se}), but 
$ \ \Se \ \propto \ \kB\ln (T^3/\nr), \ $ an expression that coincides 
with eqns. (\ref{Se}) and (\ref{Seeq}) only in equilibrium (if $ \ \dot 
\Se \ = 0 $). However, since both EIT and Eckart's theory assume near 
equilibrium states, quantities like $ \ P^2 \ $ and $ \ \sigma P, \ $, 
appearing in eqns. (\ref{Sdef2}) to (\ref{dot_S3}) must be small, hence 
the ratio $ \ T^3/\nr \ $ is nearly constant and so we can also assume 
that $\ \Se \ $ given by eqn. (\ref{Se}) is approximately valid for the 
ideal gas law. 

Conditions (\ref{dot_S3}) and $ \ \delta^2 \ S < 0 \ $ associated with
eqn. (\ref{c_dot_tau}) must be satisfied by any self consistent 
thermodynamical system. The importance of these conditions will become 
evident when discussing the numerical integration of the evolution 
equations. The relaxation time, $ \ \trel, \ $ is qualitatively 
analogous to and larger than the mean  collision time between particles 
and it may, in principle, be estimated by  collision integrals provided 
the interaction potential is known. Since we are concerned with mixtures 
of baryons, electrons and photons evolving in the temperature range $ \ 
4 \times 10^3$ K $< T < 10^6$ K (from the end of cosmic nucleosynthesis 
to decoupling), convenient references for comparing $ \ \trel \ $ are the 
collision times associated with Compton and Thompson 
scatterings~\cite{zp}:
\be
t_c \ = \ \frac{\ \mel \ c^2}{\kB T} \ t_\gamma, \label{t_c}
\ee
\be
t_\gamma \ = \ \frac{1}{2 c \sigma_{_T} \nbar} \left[ {1 + \left( {1 + 
\frac{4 h^3 \nbar e^{B_0/\kB T}}{\left( 2 \pi \mel \kB T \right)^{3/2}}} 
\right)^{1/2}} \right],\label{t_gamma}
\ee
where $ \ \sigma _{_T}, \ B_0, \ \mel, \ $ and $ \ h \ $ are the 
Thompson scattering cross section ($6.65 \times 10^{-25}$ cm$^2$), the 
hydrogen atom binding energy ($13.6$ eV), the electron mass and Planck's 
constant, respectively. Equation (\ref{t_gamma}) is obtained from the 
number density of free electrons provided by the Saha equation. Notice 
that we are using the baryon number density, $ \ \nbar, \ $ and not the 
number density of WIMPS, $ \ \nm \ $. For higher temperatures in the range 
of interest, Compton scattering is the most efficient radiative process 
keeping baryonic matter and radiation tightly coupled, though it is no 
longer effective in lower and intermediate temperature ranges  ($T \ < \ 
10^4$ K). The photon-electron interaction of the radiative era requires 
that microscopic collision times $ \ t_\gamma, \ t_c, \ $ as well as 
$ \ \trel, \ $ be much smaller than the timescale of cosmic expansion given 
by the Hubble time, approximately $ \ \tH \ \equiv \ 3/\Theta. \ $ For the 
lower temperature range of the radiative era, just before recombination, 
Thomson scattering becomes the dominant radiative process, so that the 
decoupling of baryonic matter and radiation can be associated with the 
condition $ \ t_\gamma \ = \ \tH, \ $ which should be approximately 
equivalent to $ \ \trel \ = \ \tH. \ $

\section{Evolution Equations}

Since we need to determine a self-consistent set of ordinary differential
equations governing the KS models, it is convenient to express the field
equations and eqn. (\ref{eqstate2}) in terms of $ \ \Theta \ $ and $ \ 
\sigma \ $ by eliminating $ \ \dot A, \ \dot B \ $ from (\ref{kindefs}):
\be
\frac{\dot A}{A} \ = \ \frac{\Theta}{3} \ - \ 2\sigma, \qquad \qquad 
\frac{\dot B}{B} \ = \ \frac{\Theta}{3} \ + \ \sigma,\label{kindefs2}
\ee
which leads to:
\ba
\kappa \rho \ = \ \frac{\Theta^2}{3} \ - \ 3 \sigma^2 \ + \ \frac{1}{B^2}, 
\qquad \qquad \label{feq_rho2}
\\
\kappa p^{(r)} \ = \ - \frac{\Theta^2}{3} \ - \ \frac{2\dot\Theta}{3} \ - \ 3 
\sigma^2 \ - \ \frac{1}{3B^2}, \label{feq_p2}
\\
\kappa P \ = \ \dot \sigma \ + \ \sigma \Theta \ + \ \frac{1}{3  B^2}, \qquad 
\qquad \label{feq_P2}
\ea
while, using eqns. (\ref{feq_rho2}) and (\ref{feq_p2}), the equation of 
state (\ref{eqstate2}) becomes:
\be
\dot \Theta \ + \ \frac{2 \ \Theta^2}{3} \ + \ 3 \ \sigma^2 \ + \ 
\frac{1}{B^2} \ -  \ \frac{\kappa \ m \ c^2 \ \nm }{2} \ = \ 0. 
\label{eqstate3}
\ee
We can eliminate $ \ B \ $ from the equations above with the help of 
equation (\ref{feq_rho2}) (which becomes a constraint) and the second of 
equations (\ref{kindefs2}) (the evolution equation for $ \ B$). Using 
equation (\ref{eqstate}), equations (\ref{feq_P2}) and (\ref{eqstate3}) 
then become:
\be
\dot \sigma \ + \ \sigma^2 \ + \ \sigma \Theta \ - \ \frac{\Theta^2}{9} \ + \ 
\frac{\kappa}{3} m c^2 \nm \ + \ \kappa p^{(r)} \ - \ \kappa P \ = \ 0,
\label{dot_sigma}
\ee
\be
\dot \Theta \ + \ \frac{\Theta^2}{3} \ + \ 6 \ \sigma^2 \ + \ \frac{\kappa}
{2} \ m \ c^2 \ \nm \ + \ 3 \ \kappa \ p^{(r)}  \ = \ 0, \label{dot_Theta}
\ee
which are the evolution equations for $ \ \sigma \ $ and $ \ \Theta. \ $ 
Since we are assuming particle number conservation, an evolution equation 
for non-relativistic particle number density (the WIMP's) follows from 
equation (\ref{n_Gamma}):
\be
\dnm \ + \ \nm \ \Theta \ = \ 0, \label{dot_n}
\ee
a conservation law satisfied also by $ \ \nr = \nm/\nuw \ $ (if using the 
ideal gas law (\ref{ig_law})). Another evolution equation is provided by
(\ref{Ebalance}), which applied to the equation of state (\ref{eqstate}) 
and using $ \ \sigma_{ab} \ \Pi^{ab} \ = \  6 \ \sigma \ P \ $ yields:
\be
\dot p^{(r)} \ + \frac{4}{3} \ p^{(r)} \ \Theta \ + \ 2 \ \sigma \ P \ = \ 
0, \label{dot_p}
\ee
becoming the evolution equation for $p^{(r)}$. The transport equation
(\ref{transp_eq2}) derived in the previous section, namely: 
\be 
\dot P \ + \ \frac{8}{5} p^{(r)} \sigma \ + \ \left( \frac{4 \Theta}{3} \ + \ 
\frac{1}{\trel} \right) P \ + \ \frac{\lambda \sigma}{p^{(r)}} P^2 \ = \ 0,
\label{dot_P}
\ee
is the evolution equation for the shear stress $ \ P. \ $ In addition to these 
evolution laws, eqn. (\ref{dot_S3}) can be thought of as an evolution equation 
for $ \ S, \ $ while we can transform eqn. (\ref{dot_p}) into an evolution 
equation for $ \ T \ $ by using either (\ref{sb_law}) or (\ref{ig_law}), 
leading to:
\be
\frac{\dot T}{T} \ + \ \frac{\Theta}{3} \ +\ \frac{\sigma \ P}{\lambda_0 \ 
p^{(r)}} \ = \ 0. \label{dot_T}
\ee
Equations (\ref{dot_sigma}), (\ref{dot_Theta}), (\ref{dot_n}), 
(\ref{dot_p}), and (\ref{dot_P}) represent a self-consistent and closed 
system of first order ODE's for $ \ \nm, \ P, \ p^{(r)}, \ \sigma, \ 
\Theta. \ $ Notice that this set of evolution equations is fully 
determined if $\ \trel \ $ is known. In the cosmological context $ \ 
\trel \ $ might be proportional to the timescale defined by the 
expansion scalar (approximately the Hubble time):
\be
\trel \ \propto \ \tH \ \equiv \ 3/\Theta. \label{trel_prop_H1}
\ee
Alternatively, and depending on the temperature and energy range one is
considering, $ \ \trel \ $ could be identified as proportional to a 
collision time (say, Thomson or Compton scattering) given by equations 
(\ref{t_c}) or (\ref{t_gamma}), \ {\it i. e.},
\be
\trel \ \propto \ t_\gamma(n,T), \qquad \qquad \trel \ \propto \ 
t_c(n,T),
\ee
Since we can assume in equation (\ref{dot_P}) two different equations of 
state for the radiation component, strictly speaking, equations 
(\ref{dot_sigma})-(\ref{dot_P}) constitute two different systems of 
evolution equations parameterized by the two possible values of 
$ \ \lambda_0 \ $ and $ \ \lambda $:
\vskip0.3cm
\noindent{--Stefan-Boltzmann law:}
\be
3 p^{(r)} \ = \ a T^4, \quad \hbox{with} \quad \lambda_0 = 2 \quad \hbox{and} 
\quad \lambda = 5/4, \quad \ \nonumber
\ee
\noindent{--Ideal gas law:}
\be
\quad \ p^{(r)} \ = \ \nr \kB T, \quad \hbox{with} \ \ \lambda_0 = 1/2 \ \ 
\hbox{and} \ \ \lambda = 2. \ \label{lambdas}
\ee

\section{Dynamical Analysis}\label{dyn_an}

\subsection{The governing equations.}

Let us define:
\be
Q \ \equiv \ \frac{P}{p^{(r)}}, \label{Qdef}
\ee
a ratio that is related (from eqn. (\ref{Sdef2})) to the deviation from 
equilibrium:
\be
\Se \ - \ S \ = \ \frac{15}{8} \ \lambda_0 \ Q^2,\nonumber
\ee
where:
\be 
\lambda_0 \ = \ \frac{p^{(r)}}{\nr T} = \left\{
\begin{matrix}
{\frac{3}{4} \ \Se, \quad \hbox{Stefan-Boltzmann law} \hfill}\\
\label{Sdef4}\\
{\quad \kB, \qquad \quad \ \hbox{Ideal Gas law} \quad } \end{matrix} \right.
\ee
with $ \ \Se \ $ given by eqn. (\ref{Se}). Therefore, $ \ Q \ $ must be 
a small quantity for the cosmic times we are interested in. Note that 
when $ \ P \ $ (and hence $ \ Q$) is zero, the shear vanishes and the 
Kantowski-Sachs model reduces to an isotropic FLRW model. In principle, 
$ \ P \ $ (and hence $ \ Q$) can be positive or negative, but since $ \ 
P \ $ represents a viscous pressure it should be negative in the 
expanding regime. We  shall focus henceforth on the case in which $Q$ is
negative. The energy conditions imply that $ \ - 1 \le Q \le 1/2. \ $ In 
addition, on physical grounds we expect the second term in equation 
(\ref{Ebalance}) to dominate the third term in this equation, which is 
satisfied whenever $ \ \sigma^2 Q^2/(\Theta/3)^2 < 4. \ $ However, since 
physically we expect $ \ Q^2 \ $ to be small, these constraints are 
satisfied handily.

We introduce now the new normalized variables:
\ba
\Sigma \ \equiv \ \frac{\sigma}{\Theta/3}, \qquad \qquad \Omm \ \equiv \ 
\frac{\kappa m c^2 \nm}{\Theta^2/3}, \nonumber\\
\Omr \ \equiv \ \frac{\kappa p^{(r)}}{(\Theta/3)^2}, \qquad \qquad \qquad
\label{newvars}
\ea
a definition that is motivated by the fact that $ \ \Theta/3 \ $ is 
approximately the Hubble expansion factor $ \ H = \Theta/3 + \sigma_{ab} 
n^a n^b \ $ for unit space-like vectors defined by \ $n^an_a = 1, \ u^a 
n_a = 0, \ $ hence $ \ \Omm \ $ and $ \ \Omr \ $ are approximately 
equivalent to the observational parameters (the $ \ \Omega$'s) for the 
CDM and radiation contents of the mixture. We also define (for $ \ 
\Theta > 0$) the new independent variable:
\be
\frac{d}{dt} \ = \  \frac{\Theta}{3} \ \frac{d}{d\tau} \qquad \Rightarrow 
\qquad \tau \ = \ \int{\frac{\Theta}{3} \ dt}, \label{newtime}
\ee
the evolution equations (\ref{dot_sigma}) to (\ref{dot_P}) become:
\be
\Omm' \ = \ - \ \Omm \left[ \ 1 \ - \ 4 \Sigma^2 \ - \ \Omm \ - \ 2 \Omr 
\ \right], \label{EqOmega}
\ee
\be
Q' \ = \ - \ \frac{8}{5} \ \Sigma \ - \ \frac{Q}{\trel \ \Theta/3} \ - \ 
(\lambda - 2) \ \Sigma \ Q^2. \label{EqQ}
\ee
\be
\Omr' \ = \ - \Omr \left[2 \ + \ 2 \Sigma \ (Q - 2 \Sigma) \ - \ \Omm \ - \ 2 
\Omr \right], \label{EqOmegar}
\ee
\ba
\Sigma' \ = \ \left( 1 - 2 \Sigma \right) \left( 1 - \Sigma^2 \right) \ - \ 
\Omm \left( 2 - \Sigma \right)/2 \ - \quad \nonumber\\
\Omr \left( 1 - Q -\Sigma \right), \label{EqSigma}
\ea
where we have used
\be
\frac{3 \ \dot \Theta}{\Theta^2} \ = \ \frac{\Theta '}{\Theta} \ = \ \ - 
\ 1 \ - \ 2 \ \Sigma^2 \ - \ \frac{\Omm}{2} \ - \ \Omr, \label{EqTheta}
\ee
and a prime denotes differentiation with respect to $ \ \tau. \ $ Note 
that this last equation implies that $ \ \Theta \ $ is monotonically 
decreasing. The constraint (\ref{eqstate3}) becomes:
\be
\chi \ \equiv \ 1 \ - \ \Sigma^2 \ - \ \Omm \ - \ \Omr \ = \ - \ \frac{3}
{B^2 \ \Theta^2}. \label{eqstate4}
\ee
We note that:
\ba
\chi' \ = \  \chi \left[ \ 2 \Sigma \ (2 \Sigma \ - \ 1) \ + \ \Omm \ + 
\ 2 \Omr \ \right]. \nonumber
\ea
Clearly $ \ \chi = 0 \ $ is an invariant set of the above differential 
equations, which corresponds to the Bianchi I (zero curvature) sub-case.

Eventually, the models re-collapse and $ \ \Theta \ $ changes sign. At 
the point of maximum expansion (when $ \ \Theta = 0$) the variables above 
diverge and the normalized equations are no longer valid. However, for 
the times we are interested in, in the expanding phase far from 
re-collapse, the above variables and equations are valid. Indeed, in 
principle we can use the above system to follow the evolution of the 
models all the way back to the big bang. From equation (\ref{dot_Theta}) 
we can see that in this regime the curvature is small and that the 
variables $ \ \Sigma^2, \ \Omm, \ \Omr \ $ are well-behaved. Compact 
variables can be defined by normalizing with $ \ \Theta^2 \ + B^{-2} \ $ 
(instead of $ \ \Theta^2$) that are valid for all times~\cite{CG}; 
however, the physical assumptions used here are not valid at later 
times.

In addition to equations (\ref{EqOmega}) to (\ref{EqSigma}), evolution
equations for $ \ n, \ T, \ \Se \ $ and $ \ S \ $ follow by using the 
variables defined in eqns. (\ref{newvars}) and (\ref{newtime}) in 
equations (\ref{dot_n}), (\ref{dot_T}), (\ref{dot_Se}) and 
(\ref{dot_S3}):
\ba
n' \ = \ - 3 \ n, \qquad \Rightarrow \qquad \qquad \qquad \nonumber\\
n(\tau) = n_i e^{- 3 \tau}, \qquad \hbox{where:} \quad n = \nm \quad \hbox{or} 
\quad \nr, \label{Eqn}
\ea
\be
T' \ = \ - T \ \left[1 + \frac{\Sigma \ Q}{\lambda_0} \right], 
\label{EqT}
\ee
\be
\Se' \ = \ - 3 \ \lambda_0 \ \Sigma \ Q, \label{EqSe}
\ee
\be
S ' \ = \ \frac{2 \ \tH }{\trel} \ \left[ { \ \Se \ - \ S} \ \right] \ = \ 
\frac{15 \ \tH }{4 \ \trel} \ \lambda_0 \ Q^2, 
\label{EqS}
\ee
where $ \ \lambda_0 \ $ is given by eqn. (\ref{Sdef4}), the subindex 
$ \ i \ $ denotes evaluation at an initial time $ \ \tau = \tau_i, \ $ 
and $ \ \tH = 3/\Theta, \ $ is approximately the Hubble time which 
follows from eqn. (\ref{EqTheta}) as:
\be
\tH \ = \ \frac{3}{\Theta_i} \exp \left\{ \int{ \left[ 1 + 2 \Sigma^2 + 
\frac{\Omm}{2} + \Omr \right] \ d\tau} \right\}, \label{tH}
\ee
while the relation between physical time $t$ and $\tau$ follows from
(\ref{newtime}) and (\ref{EqTheta}):
\be
t \ = \ \int{\tH \ d \tau \ }, \label{t_vs_tau}
\ee
where $ \ \tH \ $ is given by (\ref{tH}) above.

\subsection{Qualitative Properties}

Consider the dynamical implications of assuming that $\trel$ is given by 
eqn. (\ref{trel_prop_H1}), namely:
\be
\trel \ = \ \frac{\gamma_0}{\Theta/3} \ = \ \gamma_0 \ \tH, 
\label{trel_prop_H2}
\ee
where $ \ \gamma_0 \ > 0 \ $ is a constant. For the range we are interested 
in, $ \ \gamma_0 \leq 1. \ $ Equation (\ref{EqQ}) becomes:
\be
Q' \ = \ - \ \frac{8 \Sigma}{5} \ - \frac{Q}{\gamma_0} \ - \ (\lambda - 2) \ 
\Sigma \ Q^2, \label{EqQ2}
\ee
and so equations (\ref{EqOmega}), (\ref{EqOmegar}), (\ref{EqSigma}) and
(\ref{EqQ2}) now constitute a closed four-dimensional system of first order
autonomous differential equations for $ \ (\Omr, \ \Omm, \ \Sigma, \ Q)$, a
system that depends on the value of the constant parameter $ \ \gamma_0. \ 
$ Moreover, from equation (\ref{eqstate4}) and the above discussion, in the 
regime we are interested in $ \ \Omm, \ \Omr, \ $ and $ \ \Sigma \ $ are 
bounded and physical conditions imply that $ \ - 1 \le  Q \le 1/2. \ $ 
Consequently a local analysis of the stability of the equilibrium points of 
this system will provide useful dynamical information.

Setting the right-hand-side of equation (\ref{EqQ2}) to zero we obtain:
\be
(\lambda - 2) \ \Sigma \ Q^2 \ + \ \frac{Q}{\gamma_0} \ + \ \frac{8 \Sigma}{5}
\ = \ 0, \label{EqQ3}
\ee
which is a quadratic equation for $ \ Q \ $ if $ \ \Sigma \ $ is given and 
$ \ \lambda \ne 2$.
\vskip0.3cm
\noindent{\em Stefan-Boltzman law.} ($\lambda=5/4$):

\noindent The equilibrium points at finite values are (note that all such 
points are given below; however, $(\Omm, \Omr, \Sigma^2)$ are not necessarily 
bounded by unity):

\begin{enumerate}
\item[(i)] $\Omm = 0 = \Omr$, $ \ \ $ six points:
\subitem{(ia)} $\Sigma = 1/2 \ $ and 

$\qquad \qquad \quad Q_{\pm} = 4 \ [3 \gamma_0 \pm \sqrt{9 \gamma_0^2 + 2/15} 
\ ]/(9 \gamma_0^2);$
\subitem{(ib)} $\Sigma = 1 \ $ and 

$\qquad \qquad \quad Q_{\pm} = 2 \ [3 \gamma_0 \pm \sqrt{9 \gamma_0^2 + 8/15} 
\ ]/(9 \gamma_0^2);$
\subitem{(ic)} $\Sigma = - 1 \ $ and 

$\qquad \qquad \quad Q_{\pm} = - 2 \ [3 \gamma_0 \pm \sqrt{9 \gamma_0^2 + 
8/15} \ ]/(9 \gamma_0^2).$

\item[(ii)] $\Omm = 0, \ \Omr \ne 0, \ \ $ four points:
\subitem{(iia)} $\Omr = - (12 \gamma_0 + 17/15)/(9 \gamma_0^2), \ $ and 

$\qquad \qquad \quad \Sigma_{\pm} = Q_{\pm} = \pm 2 \sqrt{3 \gamma_0 + 8/15}/
(3 \gamma_0);$
\subitem{(iib)} $\Omr{}_{\pm} = - \frac{(4+3\gamma_0)[2+\gamma_0(1 \pm u)]}
{\gamma_0[2+3\gamma_0(1 \mp u)]},$

$\qquad \qquad \quad \Sigma_{\pm} = (4 + 3 \gamma_0)/[2 + 3 \gamma_0 
(1 \mp u)], \ $ and 

$\qquad \qquad \quad Q_{\pm} = (2 \pm 3 \gamma_0 u)/(4 + 3 \gamma_0), \ $ 
where 

$\qquad \qquad \quad u = 2 \sqrt{1 + (8 \gamma_0/5)(4 + 3 \gamma_0)}/
(3 \gamma_0);$
\subitem{} $\Sigma = 0 = Q, \ \Omr = 1$ is a particular solution.
\item[(iii)] $\Omr = 0, \ \Omm \ne 0, \ \ $ one particular solution:
\subitem{} $\Sigma = 0 = Q, \ $ and $ \ \Omm=1.$
\item[(iv)] $\Omm \ne 0, \ \Omr \ne 0$, two points:
\subitem{} $\Omm = (40+79 /\gamma_0)/(64 \gamma_0),\qquad \qquad \quad $

$\qquad \qquad \quad \Omr = - 15(8 + 3 \gamma_0)/(128 \gamma_0),$

$\qquad \qquad \quad \Sigma = \pm \sqrt{5(8 + 3 \gamma_0)/(2 \gamma_0)}/8, 
\quad $ and
 
$\qquad \qquad \quad Q_{\pm} = \mp 4 \sqrt{2(8 + 3 \gamma_0)/(45 \gamma_0)}.$
\end{enumerate}

The two equilibrium points given by (ia), can be shown to lead to a value 
for $ \ Q \ $ which is unphysical in the sense that its magnitude is much 
too large (in fact, when $ \ \Omm = 0 = \Omr, \ \ Q \ $ would be expected 
to vanish). The equilibrium points (iv) are also unphysical: since clearly 
$ \ \Sigma^2 \le 1/4 \ $ and we get from (iv) that $ \ \Sigma^2 Q^2 = 1/4, 
\ $ this would imply that $ \ Q^2 \ge 1$ ! At the equilibrium points (ii) 
when $ \ \Omr \ne 1 \ $ we have that $ \ \Sigma = 20 Q/[\gamma_0(15Q^2 - 
32)], \ $ which leads to a quartic equation for Q; however, for physical 
values of the parameter $ \ \gamma_0, \ $ this equation has no real roots 
and hence no solutions of physical interest. The two equilibrium points 
given by the particular solution of (ii) and by (iii), corresponding to 
FLRW models, namely $(\Omr, \Omm, \Sigma, Q)$ given by $(1,0,0,0)$ and 
$(0,1,0,0)$, can easily be shown to be saddles. The equilibrium points 
(ib,c), namely $ \ (\Omr, \Omm, \Sigma, Q) = (0, 0, \pm 1, Q_{\pm}), \ $ 
which belong to the invariant set $\chi = 0$ ({\it i. e.,} correspond to 
a Bianchi I model with no curvature) have eigenvalues $ \ 3, \ - 1 / 
\gamma_0 + 3\lambda^2 \Sigma Q/2, \ 2 - 2\Sigma Q, \ 4 - \Sigma. \ $. In 
this case, the eigenvalue $ \ - 1/\gamma_0 + 3 \lambda^2 \Sigma Q/2 \ $ 
can only be positive if we take the 'positive square root' of eqn. (37), 
({\it i. e.,} $Q = Q_+$). However, for $ \ \{\lambda^2, \gamma_0 \} \le 1, 
\ $ {\it i. e.,} $ \ \lambda^2\,\gamma_0 \le 1, \ $ it follows that the 
eigenvalue $ \ 2 - 2 \Sigma Q_+ \ $ can never be positive. Consequently, 
this equilibrium point can not be a source.

We note that there are no sinks at finite values. However, this is to be
expected since the models evolve toward maximum expansion at which the
variables become unbounded. The models subsequently re-collapse.

\noindent{\em Ideal Gas case} ($\lambda = 2$):

Equations (\ref{EqOmega}), (\ref{EqOmegar}) and (\ref{EqSigma}) remain
unchanged and the governing equation (\ref{EqQ2}) becomes:
\be
Q' \ = \ - \ \frac{8 \ \Sigma}{5} \ - \ \frac{Q}{\gamma_0}. 
\label{EqQ4}
\ee
At an equilibrium point we immediately see that:
\be
Q \ = \ - \ \frac{8 \ \gamma_0 \ \Sigma}{5}. \label{Qval}
\ee
However, there are no major qualitative changes in the analysis. In 
particular, there are fewer equilibrium points:

\begin{enumerate}
\item[(i)] $\Omm = 0 = \Omr$, three points:
\subitem{(ia)} $\Sigma=1/2 \ $ and $ \ Q = -4 \gamma_0/ 5 ;$
\subitem{(ib)} $\Sigma = 1 \ $ and  $ \ Q = -8 \gamma_0/ 5;$
\subitem{(ic)} $\Sigma = - 1 \ $ and $ \ Q = 8 \gamma_0/ 5.$
\item[(ii)] $\Omm = 0, \ \Omr \ne 0$:
\subitem{} $\Sigma = 0 = Q, \ \ \Omr = 1$ is a particular solution; 
otherwise,
\subitem{} A quadratic equation for $\Sigma$ which gives another two 
equilibrium points.
\item[(iii)] $\Omr = 0, \ \Omm \ne 0$, one particular solution:
\subitem{} $\Sigma = 0 = Q, \ $ and $\Omm = 1.$
\item[(iv)] $\Omm \ne 0, \ \Omr \ne 0$, two points that imply:
\subitem{} $\Sigma^2 \ = \ 5/(16\gamma_0) \ \ge \ 5/16 \ $ (for
$\gamma_0 \le 1$), which leads to $ \ \Omm + 2\Omr  \ = \ 1 - 5 / 
(4\gamma_0) \ < \ 0$.
\end{enumerate}

\section{Assumptions on the relaxation time}

In the general dynamical system, eqns. (\ref{EqOmega}) -- (\ref{EqSigma}), 
the relaxation time $ \ \trel \ $ needs to be specified in order for the 
governing system of equations to be closed. For the qualitative dynamical 
analysis we have assumed (\ref{trel_prop_H2}), so that $ \ 1/(\trel\Theta) 
\ $ is a constant, since otherwise we would either not be dealing with an 
autonomous system, or would be looking at an autonomous but much more 
difficult dynamical system. However, equation (\ref{trel_prop_H2}) is a 
simplifying assumption that cannot be supported by thermodynamical 
arguments, perhaps a more realistic assumption would be to consider 
instead:
\be
\trel \ = \  \frac{\gamma(\tau)}{\Theta/3} \ = \ \gamma(\tau) \ \tH
\label{trel_prop_H3}
\ee
so that equations (\ref{EqQ}) and (\ref{EqS}) become
\be
Q' \ = \ - \ \frac{8 \ \Sigma}{5} \ - \ \frac{Q}{\gamma(\tau)} \ - \ (\lambda 
- 2) \ \Sigma \ Q^2. \label{EqQgamma}
\ee
\be
S' \ = \ \frac{2}{\gamma(\tau)} \left[ { \ \Se \ - \ S} \ \right], 
\label{EqS2}
\ee
where $ \ \gamma = \gamma(\tau) \ $ is a function that could be suitably 
adjusted so that $ \ \trel \ $ has a form that is qualitatively analogous 
to that of microscopic timescales like eqn. (\ref{t_c}) or (\ref{t_gamma}), 
timescales that are physically relevant for the matter source under 
consideration. The ratio $ \ 3/(\trel\Theta) = \tH/\trel = 1/\gamma \ $ 
should provide a comparison of the timescale for the relaxation (transient) 
effects in the radiative fluid with the timescale of cosmic expansion. 
Hence this ratio should approach unity as baryonic matter and radiation 
decouple, so that $ \gamma \approx 1$ should be a consistent choice for 
near decoupling conditions, while $ \ \gamma \gg 1 \ $ or $ \ \gamma \ll 
1, \ $ correspond to after decoupling (late times) and much before 
decoupling (earlier times). Ideally, we should obtain $ \ \trel \ $ from 
collision integrals associated with each of the various radiative 
processes occurring in the radiative era, but such an endeavor would merit 
a separate paper by itself and will not be attempted here. Instead, we will
consider $ \ \trel \ $ as an ``effective'' relaxation time, encompassing 
the different radiative processes. We will examine the non-transient limit 
(or near-Eckart regime) and use the dynamical equations themselves in order 
to suggest a suitable form for $\trel$. The discussion of this section will 
be complemented and tested numerically in section \ref{numerics} (see also 
figures 5 and 6).

\subsection{The ``near-Eckart'' and the transient regimes.}
\label{regimes}

In order to examine the relaxation process as $ \ S \to \Se, \ $ we will 
assume that $ \ \gamma(\tau) \ $ in (\ref{trel_prop_H3}) is a smooth 
function so that we can always expand it in the form $ \ \gamma 
\approx \gamma(0) + \gamma'(0) \ \tau + \gamma''(0) \ \tau^2/2. \ $ 
Therefore, at early times $ \ \tau \approx 0 \ $ we can always associate 
the constant $ \ \gamma_0 \ $ in eqn. (\ref{trel_prop_H2}) as $ \ \gamma_0 
= \gamma(0), \ $ so that the corresponding form of $ \ \trel \ $ is 
approximately correct at least near $ \ \tau = 0. \ $ Also, at early times 
we must have $ \ \gamma(0) < 1, \ $ and so $ \ 1/\gamma_0 \gg 1, \ $ while 
$ \ \Sigma(0) \ $ and $ \ Q(0) \ $ are necessarily small quantities. Hence 
the terms $ \ \Sigma \ $ and $ \ \Sigma Q^2 \ $ in eqn. (\ref{EqQgamma}) 
will be much smaller than $ \ Q(0)/\gamma_0 \ $ and so we have that near 
$ \ \tau = 0, \ Q' \approx - Q/\gamma_0, \ $ so that:
\ba
Q \ \approx \ Q(0) \ \exp \left( \frac{- \tau}{\gamma_0} \right), \qquad 
\nonumber\\
S \ \approx \ \Se \ - \ \frac{15}{8} \ \lambda_0 \ Q^2(0) \ \exp \left( 
\frac{-2 \tau}{\gamma_0}\right), \label{near_t0}
\ea
where $ \ \lambda_0 \ $ is given by eqn. (\ref{Sdef4}). Since the
process of relaxation to equilibrium can be characterized as the decay of the
dissipative flux $ \ Q\to 0 \ $ as $\ S \ $ grows and asymptotically approaches
the  equilibrium state given by eqn. (\ref{Seeq}), the numerical value of
$ \ \gamma_0 \ $ in (\ref{near_t0}) may be  interpreted as a measure of the
``rate of transiency'' in terms of how  fast or slow the system accomplishes
this relaxation in comparison with the timescale provided  by $ \ \tH. \ $ We
can then identify two  possible situations
\bi
\item{The near-Eckart regime}. If $ \ \gamma_0 \ll 1, \ $ then eqn. 
(\ref{near_t0}) indicates a very fast relaxation with a very abrupt decay 
of $ \ Q \ $ and $ \ S \ $ to $ \ \Se. \ $ In the very limit $ \ \gamma_0 
\to 0 \ $ (so that $ \ \trel \to 0 \ $ as well) we have $ \ Q \to Q(0) \ 
\delta(\tau), \ $ so that the relaxation is infinitely fast, in agreement 
with the non-causal nature of Eckart's classical theory. The fast 
relaxation associated with a very small $ \ \gamma_0 \ $ implies a very 
short duration of the relaxation process (small $ \ \trel$), indicating 
that the approximation $ \ \gamma = \gamma_0, \ $ as well as the 
expressions in eqn. (\ref{near_t0}) for $ \ Q(\tau) \ $ and $ \ S(\tau), 
\ $ are approximately valid for the whole evolution time. Hence, since 
$ \ Q \ $ practically vanishes very quickly, we have for most of the 
evolution time that $ \ S\approx \Se, \ $ agreeing with the fact that in 
Eckart's theory the entropy is rigorously and unambiguously given by its 
equilibrium form (the hypothesis of ``local equilibrium'', see section 
1.3.1 of~\cite{jcl}). The near-Eckart regime is appropriate to describe 
a given radiative process for which microscopic timescales are much 
smaller than the cosmological expansion timescale $ \ \tH. \ $ See 
figures 5.
\item{The transient regime.} If $ \ \gamma_0 < 1 \ $ but $ \ \gamma_0 
\approx O(10^{-1}) - O(1), \ $ then $ \ Q \ $ and $ \ S \ $ also decay 
but the relaxation process is much slower, hence the term ``transient''. 
In this case, the expressions in eqn. (\ref{near_t0}) and the 
assumption $ \ \gamma = \gamma_0 \ $ are only good approximations for 
$ \ \gamma, \ Q(\tau) \ $ and $ \ S(\tau) \ $ near $ \ \tau=0. \ $ In 
general, we must use:
\ba
Q \ \approx \ Q(0) \ \exp\left[-\int{\frac{d\tau}{\gamma(\tau)}}\right], \qquad
\nonumber\\
S \ \approx \ \Se - \frac{15}{8} \ \lambda_0 \ Q^2(0) \ \exp \left[ - \int{
\frac{2 \ d\tau}{\gamma(\tau)}}\ \right], \label{not_near_t0}
\ea
leading to eqn. (\ref{EqS2}) and to:
\be
S \ '' \ \approx \ \Se'' \ - \frac{2 \ (1 + \gamma') \ (\Se-S)}
{\gamma^2},
\ee
so that the fulfillment of $ \ S' > 0 \ $ and $ \ S'' < 0 \ $ can be 
examined in terms of $ \ \gamma(\tau) \ $ and $ \ \Se. \ $ Sufficient 
(but not necessary) conditions follow by demanding that $ \ \gamma \ $ 
is a monotonically increasing function ($\gamma' > 0$) and $ \ \Se' > 0, 
\ \Se'' < 0. \ $ Another condition on $ \ \gamma \ $ is furnished by 
eqn. (\ref{c_dot_tau}), which together with eqns. (\ref{rg}), (\ref{Qdef}), 
(\ref{newvars}), (\ref{newtime}), (\ref{EqTheta}), and (\ref{trel_prop_H3}) 
yield
\ba
0 \ < \ 2 + \gamma' Q^2 \ + \qquad \qquad \qquad \qquad \qquad \quad 
\nonumber\\
\gamma \left[\frac{16}{5} \Sigma Q - \left( 1 + 2 \Sigma^2 + \frac{\Omm}{2} + 
\Omr \right) Q^2 \right], 
\label{c_dot_tau2}
\ea
a condition that should be tested numerically for any given choice of $ \ 
\trel$.

The relaxation time $ \ \trel \ $ can be approximated by a given $ \ 
\gamma = \gamma_0, \ $ but as the decoupling era is reached, $ \ \gamma \ 
$ must increase to $ \ O(1) \ $ allowing for $ \ \trel \ $ to overtake $ \
 \tH$.
\ei

\subsection{Dynamical relaxation times.}\label{dynrel}

Let us consider the following ansatz:
\be
\frac{1}{\gamma} \ = \ - \frac{8}{5} \ \frac{\Sigma}{Q} \ \left[ \ 1 + 
\zeta(\tau) \ \right], \label{eqzeta}
\ee
which provides an exact relation for the relaxation time in eqn. 
(\ref{trel_prop_H3}). In regimes in which the Eckart theory is a good 
approximation, we can assume $ \ \gamma \approx \gamma_0 \ll 1, \ $ so that 
eqn. (\ref{eqzeta}) implies $ \ Q \propto \Sigma \ $ but $ \ |\Sigma/Q| \gg 
1 \ $ and $ \ \zeta \ $ is 'small'. Such regimes would correspond to a 
radiative process that takes place in timescales smaller than a mean 
collision time, thus decaying very fast to equilibrium. In case we wish to 
consider processes taking place on timescales comparable and larger than 
main collision times, then transient effects are important and the 
near-Eckart regime is  no longer appropriate. In particular, we can 
construct an expression for $ \ \trel \ $ that acts as an `effective' 
relaxation time that encompasses the relaxation times for the main 
radiative processes acting in the radiative era under consideration. Having 
this idea in mind, a reasonable and more general expression for $ \ \zeta \
 $ can be obtained from the dynamical equations themselves. Consider the
conditions discussed above for a near-Eckart regime, and assume that:
\be
Q \ = \ - \mu_0 \ \Sigma, \label{Q_Sigma}
\ee
where $ \ \mu_0 \ $ is a non-negative constant. We obtain an expression 
analogous to eqn. (\ref{eqzeta}) by substituting eqn, (\ref{Q_Sigma}) into 
the (full EIT) evolution equations (\ref{EqQ}), and (\ref{EqSigma}), 
leading to the consistency requirement:
\be
\frac{1}{\gamma} \ = \ \frac{1}{\gamma_0} \ + \ a_0 \Omm \ - \ b_0 \Sigma^2 \ 
+ \ \left(1 - \mu_0 - \frac{1}{\Sigma}\right) \chi, 
\label{gamma_eq}
\ee
where
\ba
\frac{1}{\gamma_0} \ = \ 1 + \mu_0 + \frac{8}{5\mu_0} \ \ge \ 1 + 4 
\sqrt{\frac{2}{5}} \ > \ 1, \nonumber\\
a_0 \ \equiv \ \frac{1}{2} - \mu_0, \qquad b_0 \ = \ 1 + (\lambda-3) \mu_0,
\label{constants}
\ea
and $ \ \chi \ $ is the curvature term $ \ \chi \equiv 1 - \Omm - \Omr - 
\Sigma^2 \ $ given by eqn. (\ref{eqstate4}), a term that is very small and 
can be neglected. This is an extremely simple dynamical relation for the 
relaxation time, and for a wide range of conditions it might be a very 
good approximation. In addition, it has some important physical properties. 
For early times in which $ \ \chi, \ \chi/\Sigma, \ \Omm, \ \Sigma^2 \ $ 
are very small, we have a near-Eckart regime associated with $ \ \gamma 
\approx \gamma_0 \ll 1, \ $ as expected. This is appropriate for the 
Compton scattering, the dominant radiative process in the early part of 
the period under consideration, a process that quickly thermalizes and 
ceases to be effective. At later times, as $ \ \Omm \ $ increases toward 
a value of order unity at recombination (and to a lesser extent, the
curvature term also grows), the constants $ \ \gamma_0, \ a_0, \ b_0 \ $ 
in eqn. (\ref{constants}), as well as the initial conditions, can be 
selected in such a way that $ \ \gamma, \ $, given by (\ref{gamma_eq}), 
increases sufficiently as to allow $ \ \trel \ $ to overtake $ \ \tH \ $ 
and to approach the characteristic timescale of Thomson scattering in 
eqn. (\ref{t_gamma}). We show in section \ref{numerics} that adequate 
parameters and initial conditions can be found so that $ \ \trel \ $ 
associated with eqn. (\ref{gamma_eq}) has the expected behavior (see 
figures 6).

\section{Truncated Theory}\label{truncated}

The discussion so far has been based on the full transport equation. In 
order to appreciate the effect of considering the truncated transport 
equation, it is useful to rewrite equation (\ref{transp_eq}) with $ \ 
\eta \ $ and $ \ \trel \ $ given by eqns. (\ref{rg}) and 
(\ref{trel_prop_H3}) in terms of $ \ Q \ $ and $ \ \Sigma, \ $ thus 
allowing the full and truncated equations to appear jointly. This yields:
\be
Q' \ = \ - \ \frac{8 \ \Sigma }{5} \ - \ \left[ \frac{1}{\gamma(\tau)} - 4(1 - 
\epsilon_0) \right] Q \ - \ (\lambda \epsilon_0 - 2) \Sigma Q^2.
\label{EqQjoint}
\ee
where the full and truncated theories are respectively  given by $ \ 
\epsilon_0 = 1 \ $ and $ \ \epsilon_0 = 0. \ $ Assuming a transient 
regime, instead of eqn. (\ref{near_t0}) we have near $ \ \tau = 0$:
\ba
Q \ \approx \ Q(0) \ \exp \left[ -\frac{\tau}{\gamma_0} + 4(1-\epsilon_0) 
\ \tau \right], \qquad \qquad \nonumber\\
S \approx \Se - \frac{15}{8} \lambda_0 Q^2(0) \exp \left\{2 \left[-\frac{\tau}
{\gamma_0} + 4(1 - \epsilon_0) \tau \right] \right\}, \label{near_t0tr}
\ea
so that using the truncated transport equation ($\epsilon_0 = 0$) 
introduces a large linear term ($ \propto 4 \tau $) that is absent in the 
full theory. This linear term could change dramatically the form of $ \ Q
 \ $ and the relaxation of $ \ S \ $ to $ \ \Se. \ $ As we show below, 
the truncated transport might lead to $ \ Q \ \Sigma \ $ being positive 
(implying that $ \ \dSe < 0$), so even at this level (early times) 
there might be problems of consistency in the truncated approach of EIT.

A comprehensive analysis in the case of the truncated theory, similar to 
that presented in Section V, can be  undertaken. The evolution equations 
(\ref{EqOmega}), (\ref{EqOmegar}), and  (\ref{EqSigma}) remain unchanged, 
while the evolution equation (\ref{EqQ}) for $ \ Q \ $ must be replaced 
by the truncated equation that follows by setting $ \ \epsilon_0 = 0 \ $ 
in eqn. (\ref{EqQjoint})
\be
Q' \ = \ - \ \frac{8 \ \Sigma}{5} \ - \ \left( \frac{1}{\gamma_0} \ - \ 4 
\right) \ Q \ + \ 2 \ \Sigma \ Q^2. \label{EqTr2}
\ee
At an equilibrium point:
\be
2 \ \Sigma \ Q^2 \ - \ \left(\frac{1}{\gamma_0} \ - \ 4 \right) \ Q \ - 
\ \frac{8 \ \Sigma}{5} \ = \ 0. \label{crit_pts}
\ee
However, we immediately note from equation (\ref{crit_pts}) that close to 
an equilibrium point for small $ \ \Sigma \ $ and $ \ Q \ $,
\be
\Sigma \simeq  \frac{5}{8} \left(4 - \frac{1}{\gamma_0}  \right) Q,
\label{A1}
\ee
so that for $ \ \gamma_0 \geq \frac{1}{4} \ $ we have that $ \ \Sigma \ $ 
and $ \ Q \ $ have the same sign (unlike the non-truncated theory case), 
and hence it is immediately clear that there will be a different 
qualitative dynamical behavior in the truncated theory.

In particular, close to the FLRW equilibrium point $ \ (\Omr, \Omm, \Sigma, 
Q) = (1, 0, 0, 0) \ $ (which we will use to determine the initial 
conditions in our numerical  analysis in the following section), from a 
calculation of the corresponding  eigenvalues we have that:
\be
\Omm \propto e^\tau, \qquad \qquad (1-\Omr) \propto e^{2\tau}, \label{A2}
\ee
\be
\Sigma, \ Q \ \propto \ e^{\alpha_{+} \tau} + \beta e^{\alpha_{-} \tau}, 
\label{A3}
\ee
from equations (\ref{EqOmega}) - (\ref{EqSigma}) in the non-truncated case, 
where $ \ \beta \ $ is a constant, and
\be
\alpha_{\pm} = \frac{1}{2 \gamma_0} \left[ \gamma_0 - 1 \pm \sqrt{ \left(
\gamma_0 - 1 \right)^2 - 4 \gamma_0 \left( 1 + 8 \gamma_0/5 \right)} \right].
\label{A4}
\ee
For physical values $ \ \gamma_0 \leq 1, \ $ both $ \ \alpha_{\pm} \ $ have 
negative real parts and as noted earlier, this FLRW equilibrium point is a 
saddle $ \ (\mbox{e. g., for} \ \gamma_0 = 1/2, \ \alpha_{\pm} = - 1/2 \pm i 
\sqrt{67/20})$.

In the truncated case we have that eqns. (\ref{A2}) and (\ref{A3}) are 
satisfied close to the FLRW equilibrium point, but now with
\be
\alpha_{\pm}^{\hbox{\tiny{tr}}} = \frac{1}{2 \gamma_0} \left[ 5 \gamma_0 - 1 
\pm \sqrt{ \left( 5 \gamma_0 - 1 \right)^2 + 4 \gamma_0 \left( 12 \gamma_0/5 - 
1 \right) } \right]. \label{A5}
\ee
We first note that for $ \ \gamma_0 \geq 1/5, \ $ at least one of the $ \ 
\alpha^{\mbox{\tiny{tr}}} \ $ has a positive real part which leads to a 
change of stability (indeed, for $ \ 1/5 \leq \gamma_0 \leq 5/12, \ $ this 
equilibrium point is a source !).  Physically, this means that in the 
truncated case $ \ \Sigma \ $ and $ \ Q \ $ in equation (\ref{A3}) have a  
growing  mode, and hence their magnitudes increase leading to a breakdown 
in the physical model (and  the time period for which the assumptions are 
valid). This breakdown is seen in the numerics (figures 7). As a comparison, 
for $ \ \gamma_0 = 1/2, \quad \alpha_{\pm} \simeq \frac{3}{2}(1 \pm 1.1); \ $ 
the  growing mode which evolves approximately as $ \ e^{3 \tau}, \ $ leads to 
a rapid increase in the magnitudes of $ \ \Sigma \ $ and $ \ Q \ $ and the 
models fail after a relatively short cosmological time (see figures 7).

\section{Numerical Integration}\label{numerics}

From the dynamical analysis carried out in section \ref{dyn_an}, there are 
no sources (at finite values) in the physical regime; that is, during the 
time period for which the various physical assumptions used here are valid 
there are no past attractors. It is reasonable to assume as conditions 
prevailing at the beginning of the regime we are interested in, that the 
universe is approximately  isotropic and spatially homogeneous ({\it e. 
g.}, almost FLRW with $ \ |Q| \ll 1 \ $ and $ \ \Sigma \ll 1$) and that 
the radiation component is dominant. These assumptions are consistent 
with current observations, and we are also assuming that some mechanism 
(such as, for example, inflation) at early times has driven the universe 
toward this configuration. Thus the model is close to the particular 
solution mentioned in (ii) with $ \ (\Omr,\Omm,\Sigma,Q) \ $ given by $ \ 
(1,0,0,0). \ $ As noted above, this equilibrium point is a saddle, however 
it is a 'stronger' attractor than other saddles in that it has more 
positive eigenvalues. Also, this equilibrium point lies in the invariant 
subspace $ \ \chi=0, \ $ associated with the zero curvature Bianchi I 
sub-case. Since $ \ (\Omr,\Omm,\Sigma,Q) \ = (1,0,0,0) \ $ is an 
equilibrium point lying in an invariant set, if a model is driven toward 
this point in phase space it can stay an arbitrarily long time close to 
this point ({\it i. e.}, the universe can spend an extended period close 
to this FLRW model). The universe will then eventually begin to evolve 
away from this configuration and from the invariant set as time evolves. 
Therefore, we shall assume initial conditions for our numerical 
integration based on the fact that the universe starts evolving from 
situations close to this equilibrium point.

\subsection{Initial conditions.}

Since we must have (for physical reasons) $ \ Q < 0 \ $ and $ \ Q \Sigma <
0, \ $ initial conditions close to $ \ (\Omr,\Omm,\Sigma,Q) = (1,0,0,0) \ 
$ can be given by:
\ba
\Omr(0) \ = \ 1 - \epsilon, \qquad \Omm(0) \ = \ \epsilon - \delta \ > \ 0,
\nonumber\\ 
\Sigma(0) \ = \ \Sigma_i \ > \ 0, \qquad Q(0) \ = \ Q_i \ < \ 0, \qquad 
\label{initconds}
\ea
where $ \ \epsilon, \ \delta, \ \Sigma_i, \ Q_i \ $ are real small 
constants. The initial values of ``total Omega'': $ \ \Omt = \Omm + \Omr \ 
$ and $ \ \chi \ $ given by eqn. (\ref{eqstate4}) are:
\be 
\Omt(0) \ = \ 1 - \delta, \qquad \qquad \chi(0) \ = \ \delta - (\Sigma_i)^2, 
\label{initconds2}
\ee
so that the value of $ \ \delta \ $ reflects the deviation of $ \ \Omt \ $ 
from unity, while the deviation from the invariant set $ \ \chi = 0 \ $ 
depends on $ \ \delta \ $ and $ \ (\Sigma_i)^2$.

\bi
\item{The factor $ \ \gamma$}. 

As discussed earlier, during the interactive range we are interested in, 
the various timescales must satisfy
\be
\tH \ > \ \trel \ > \ t_c \ > \ t_\gamma, \label{init_tscales1}
\ee
where $ \ \tH=3/\Theta, \ \trel \ $ follows from eqn. (\ref{trel_prop_H3}) 
and $ \ t_c, \ t_\gamma \ $ are given by eqns. (\ref{t_c}) and 
(\ref{t_gamma}). Considering that $ \ (\Theta_i/3)^2 \approx (1/3) \kappa \ a 
\ T_i^4 \ $ and inserting the constants appearing in eqns. (\ref{t_c}) and 
(\ref{t_gamma}), we arrive at the following initial values:
\ba 
\log_{10}(\tH)|_i \ \approx \ 8.8, \qquad \log_{10}(t_c)|_i \ \approx \ 7.6, 
\nonumber\\
\log_{10}(t_\gamma)|_i \ \approx \ 3.4, \qquad \qquad \label{init_tscales2}
\ea
thus suggesting the range
\be 
0.06 \ < \ \gamma_i \ < \ 1 \ \label{init_gamma0}
\ee
for the initial value $ \ \gamma(0) = \gamma_i \ $\footnote{We distinguish 
between the constant initial value $ \ \gamma_i \ $ and the case in which 
$ \ \gamma = \gamma_0 \ $ for all the evolution period.}. 

\noindent{If assuming that $ \ \gamma = \gamma_0 \ $ for all times, then 
the near-Eckart regime can be associated with $ \ \gamma_0 \leq 
O(10^{-2}), \ $ while a transient regime follows by setting $ \ \gamma_0 
\approx O(1) < 1$.}

\item{The constants $ \ \Sigma_i \ $ and $ \ Q_i$}.

From evaluation of eqn. (\ref{Sdef4}) at $ \ \tau=0, \ $ the constant $ \ 
Q_i^2 \ $ is proportional to  the initial deviation of the photon entropy 
from its equilibrium value:
\be 
\frac{15 \ \lambda_0}{8 \ \Se} \ Q_i^2 \ \approx \ Q_i^2 \ \approx \ 1 \ - 
\ \frac{S(0)}{\Se(0)}, \label{e4vals}
\ee
where we have used eqns. (\ref{Se}) and (\ref{Seeq}) so that $ \ \lambda_0/
\Se \approx O(1), \ $ with $ \ \lambda_0 \ $ given by eqn. (\ref{Sdef4}). 
Both  numbers $ \ \Sigma_i, \ Q_i \ $ are initial ratios of off-equilibrium 
and anisotropic  variables ($\Pi_{ab}, \ \sigma_{ab}$) with respect to 
equilibrium and isotropic  variables ($p^{(r)}, \ \Theta/3$). Since we are 
choosing initial conditions  close to a saddle point associated with near 
FLRW conditions and we must assume  near thermal equilibrium, then $ \ 
\Sigma_i, \ |Q_i| \ $ must necessarily be small  numbers ($\ll 1$). A 
maximal bound on $ \ |Q_i| \ $ and $ \ |\Sigma_i| \ $ can be fixed from 
CMB observations~\cite{mes} making it reasonable to take
\be 
|Q_i| < 0.01 - 0.1, \qquad |\Sigma_i| < 0.001 - 0.01, 
\label{e4_cmb}
\ee
However, we will comment further ahead on the sensitivity of the functions 
to these initial values.

\item{The constants $ \ \epsilon \ $ and $ \ \delta$}.

The values for $ \ \epsilon \ $ and $ \ \delta \ $ are  restricted by the 
ratio of photons to WIMP's. Considering the neutralino as the WIMP 
particle with $ \ m \sim 100$ GeV, using the ideal gas law and eqn. 
(\ref{comparison}) yields:
\be 
\frac{\epsilon-\delta}{1-\epsilon} = \frac{\Omm(0)}{\Omr(0)} = 
\frac{m c^2}{3 \kB T_i} \ \nuw \approx 0.013, \label{e1_e2}
\ee
leading to the following constraint  
\be 
\delta \ \approx \ 1.013 \ \epsilon \ - \ 0.013, 
\label{e1e2}
\ee
that must be satisfied by all initial conditions compatible with $ \ T_i 
\approx 10^6$ K and with the observational constraints $ \ \Omw \sim 0.3 
\pm 0.1 \ $ and $ \ h \simeq 0.7. \ $ Further restrictions on $ \ 
\epsilon \ $ and $ \ \delta \ $ follow by demanding that $ \ \Omr \ $ 
decreases and $ \ \Omm \ $ increases at the initial time $ \ \tau = 0. \ 
$ From the expressions for $ \ \Omm' \ $ and $ \ \Omr' \ $ in the 
differential equations, these  conditions imply:
$$
- 1 - 4 \Sigma_i^2 \ < \ - \epsilon - \delta \ < \ - 2 \ \Sigma_i |Q_i| - 4 \ 
\Sigma_i^2,
$$
so that $ \ \epsilon + \delta > 0 \ $ must hold, leading to the 
following minimal values of $ \ \epsilon \ $ and $ \ \delta:$
\be 
\epsilon \ > \ 0.0064,\qquad \delta \ > \ -0.0065.
\label{e11e22} 
\ee
The condition that $ \ \Omr \ $ decreases at $ \ \tau = 0, \ $ together 
with $ \ Q_i < 0 \ $ and $ \ \Sigma_i > 0, \ $ imply:
\be 
0 \ < \ \Sigma_i \ < \ \frac{1}{4} \sqrt {Q_i^2 + 4 ( \epsilon + \delta )} \ - 
\ \frac{|Q_i|}{4}. \label{e123}
\ee
\ei

\subsection{A physically plausible evolution and the range of validity 
of the models.}

It is important to specify a criterion in order to distinguish a 
physically plausible evolution for the models. We define such an 
evolution by the following conditions that must hold all along the 
range of validity discussed previously:

\begin{enumerate}

\item $ \ \Omm \ $ increases while $ \ \Omr \ $ decreases. The transition 
from a radiation- to a matter-dominated epoch occurs within the radiative 
era. However, the ratio $ \ \Omm/\Omr \ $ must remain finite in all the 
validity range.

\item $ \ S \ $ must be an increasing and convex function, tending 
asymptotically to the equilibrium photon entropy given by eqn. (\ref{Seeq}).

\item $ \ \dSe \ \propto \ - \Sigma Q\ $ must be very small. Idealy, we
should have $ \ \dSe \ > \ 0$, though this condition might fail to hold
as long as $ \ \dot S \ > \ 0 \ $ holds (see \cite{jcl} for examples).  

\item Initially, we have eqns. (\ref{init_tscales1}), (\ref{init_tscales2}), 
and (\ref{init_gamma0}), but then at later stages $ \ t_c \ $ (the Compton 
scattering timescale) is no longer relevant, while $ \ t_\gamma \ $ and $ \ 
\trel \ $ should overtake $ \ \tH, \ $ so that the baryon--photon decoupling 
is defined as $ \ t_\gamma = \tH \ $ and should occur at $ \ \tau = 
\tau_{_D} \ $ such that $ \ T(\tau_{_D}) = T_{_D} \simeq 4 \times 10^3$ K.
\end{enumerate}

We will not be concerned with the evolution of the models after the radiative 
era, since the assumptions regarding a hydrodynamic description of the 
radiative fluid break down. After the baryon--photon decoupling, an 
appropriate treatment of cosmic matter requires a different theoretical 
framework based on Kinetic Theory~\cite{ellisvan}.

Consider a transient regime ($\gamma \approx \gamma_0 \approx O(1) < 1$),
together with ``test'' initial conditions given by eqns. (\ref{initconds}) 
with $ \ \delta = 0, \ $ satisfying eqns. (\ref{e4_cmb}), (\ref{e1e2}), 
(\ref{e11e22}), and (\ref{e123}), hence $ \ \epsilon = 0.0128. \ $ For the
time being, $ \ \Sigma_i \ $ will be taken to be two orders of magnitude 
smaller than $ \ Q_i. \ $ This yields the following initial conditions 
lying very near $ \ \chi(0) = 0$:
\ba
\Omm(0) \ = \ 0.0128, \qquad \Omr(0) \ = \ 0.9872, \qquad \nonumber\\
\Sigma(0) \ = \ 0.001, \qquad Q(0) \ = \ - 0.1, \qquad \quad \nonumber\\ 
\hbox{so that:} \qquad \Omt(0) \ = \ 1, \qquad \chi(0) \ = \ - 10^{-6}. \qquad 
\label{trial_ic}
\ea
In order to illustrate how the different variables should behave in a 
physical evolution taking place in the appropriate timescale, we integrate 
the system of governing equations (\ref{EqOmega}), (\ref{EqOmegar}), 
(\ref{EqSigma}), and (\ref{EqQ2}) for (\ref{trial_ic}) and $ \ \gamma = 
\gamma_0 = 0.7. \ $ The results are displayed in figures 1 and 2. As 
shown by figures 1a, 1b, 1c and 1d, the functions $ \ \Omm, \ \Omr, \ 
\Sigma, \ Q, \ \Se \ $ and $ \ S \ $ comply with conditions 1, 2 and 3 of 
the physical evolution mentioned above (we shall discuss condition 4 in 
section VIII-E). Figure 2a depicts joint logarithmic plots of the various 
timescales $ \ t_\gamma, \ t_c, \ \tH, \ $ and physical time $ \ t, \ $ 
respectively given by eqns. (\ref{t_c}), (\ref{t_gamma}), (\ref{tH}) and
(\ref{t_vs_tau}), while the radiation temperature $ \ T \ $ is displayed 
in figure 2b. By comparing figures 1 and 2, it is evident that the range 
of validity of the models is roughly $ \ 0 \leq \tau < 6, \ $ 
corresponding to $ \ 10^6$ K $> T > 10^3$ K, with $ \ t \approx 10^5 \ $ 
years (the physical time for the radiative era), while the transition 
from radiation to matter dominance taking place at about $ \ \tau = 4 \ $ 
($T \approx 10^4$ K). Numerical integration of the governing equations 
for initial conditions different from (\ref{trial_ic}) might yield 
important qualitative changes in the state variables plotted in figures 1, 
like $ \ \Omm, \ \Omr \ $ or $ \ S, \ $ but not of those plotted in 
figures 2, such as $ \ T \ $ or the timescales (\ref{t_c}), (\ref{t_gamma}), 
(\ref{tH}), $ \ \trel \ $ or physical time $ \ t. \ $ Therefore, figures 2 
provide a general estimation of the range of validity of the models for a wide 
range of initial conditions. We will consider more general initial conditions 
in the following subsection.

\begin{figure}
\begin{center}
\leavevmode
\epsfxsize=2.6in
\epsffile{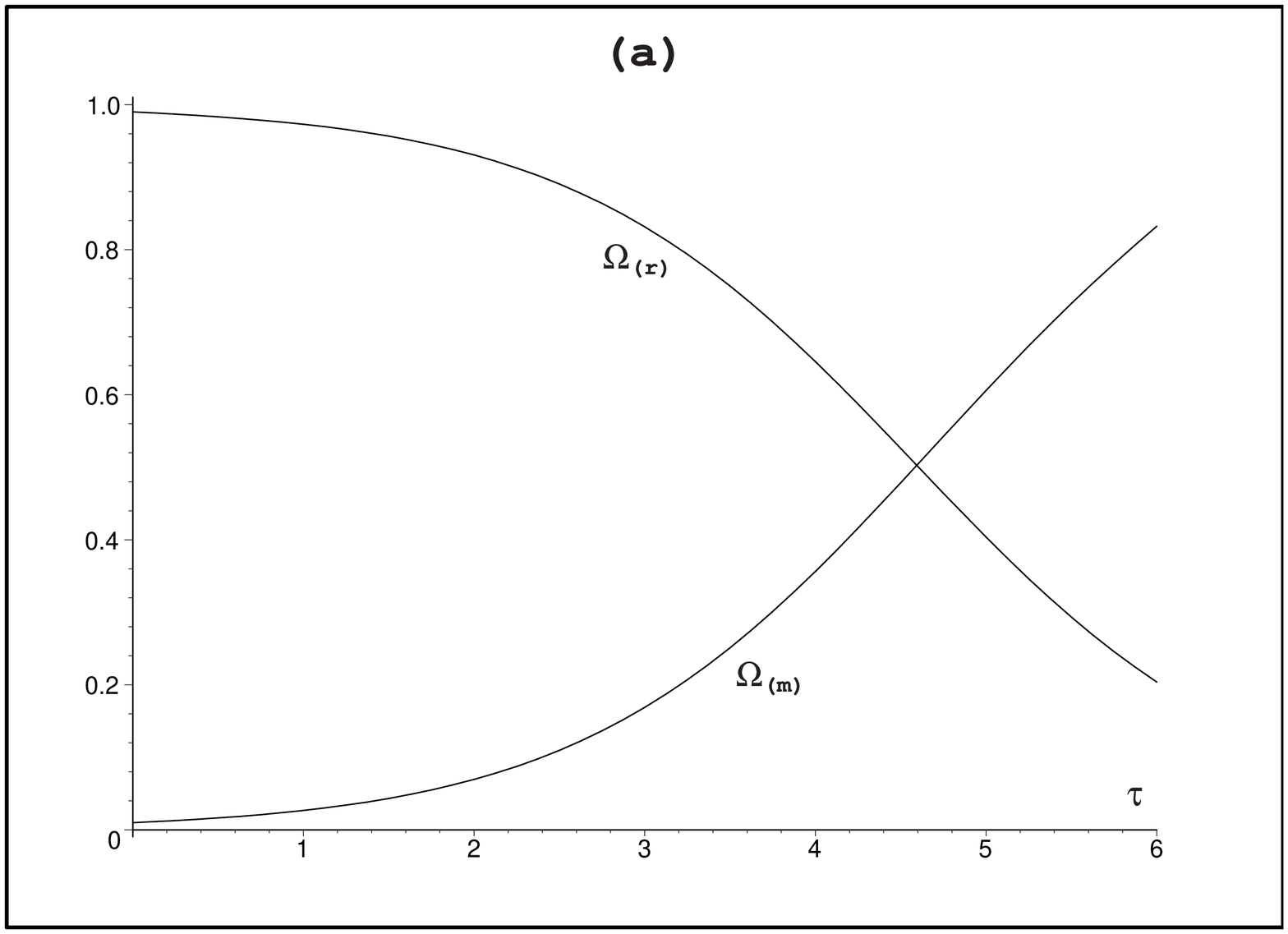}
\leavevmode
\epsfxsize=2.6in
\epsffile{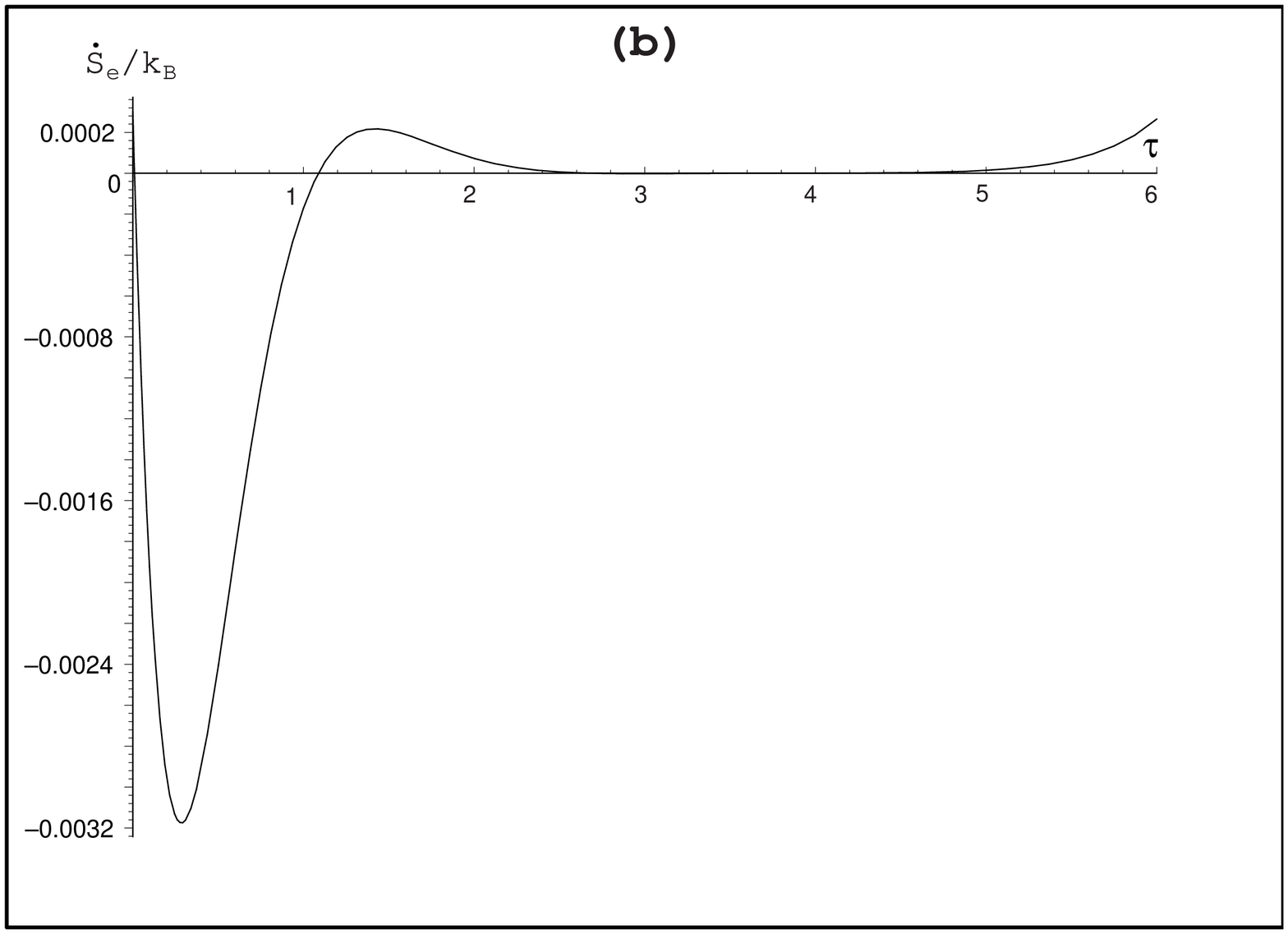}
\leavevmode
\epsfxsize=2.6in
\epsffile{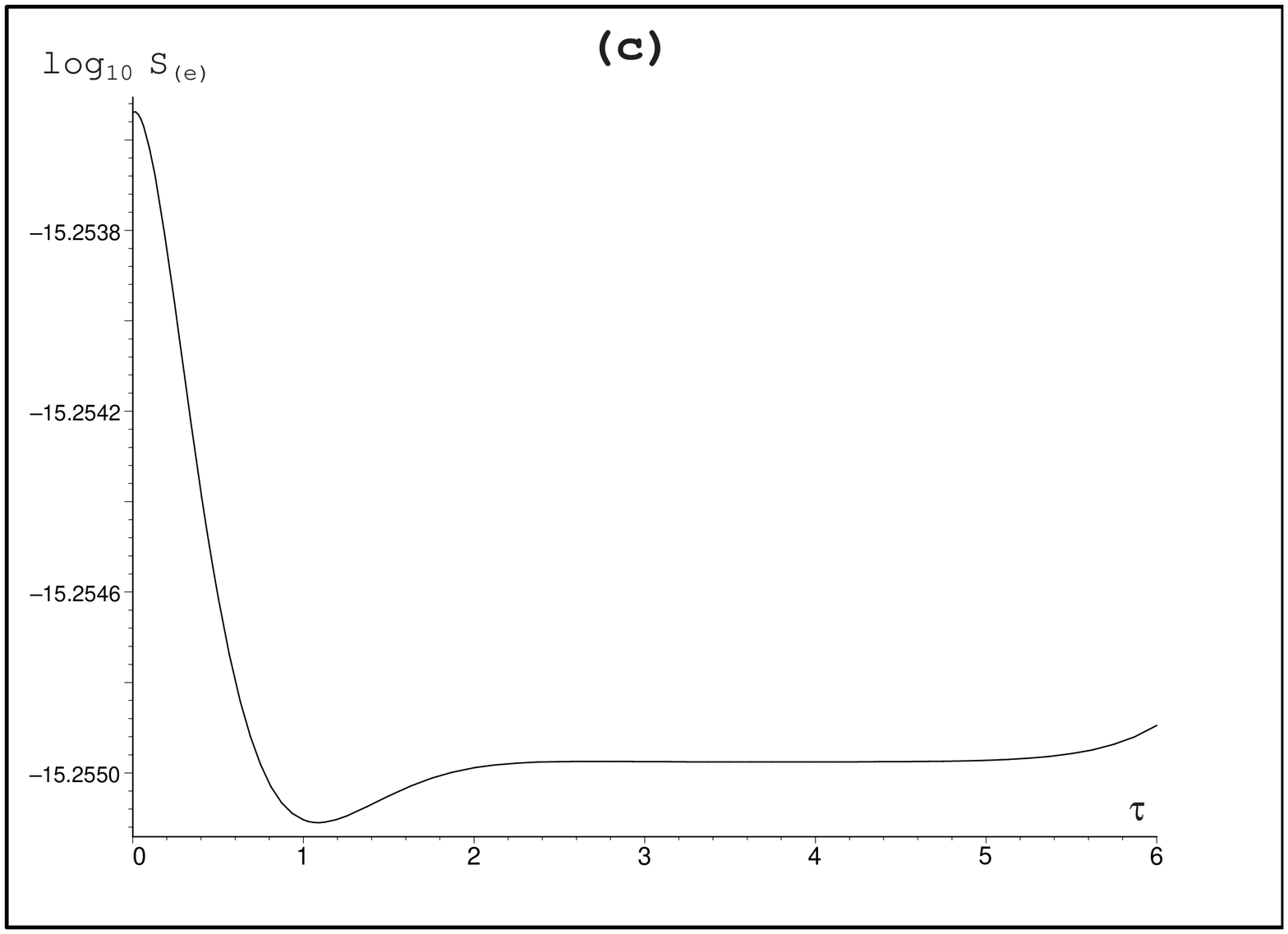}
\leavevmode
\epsfxsize=2.6in
\epsffile{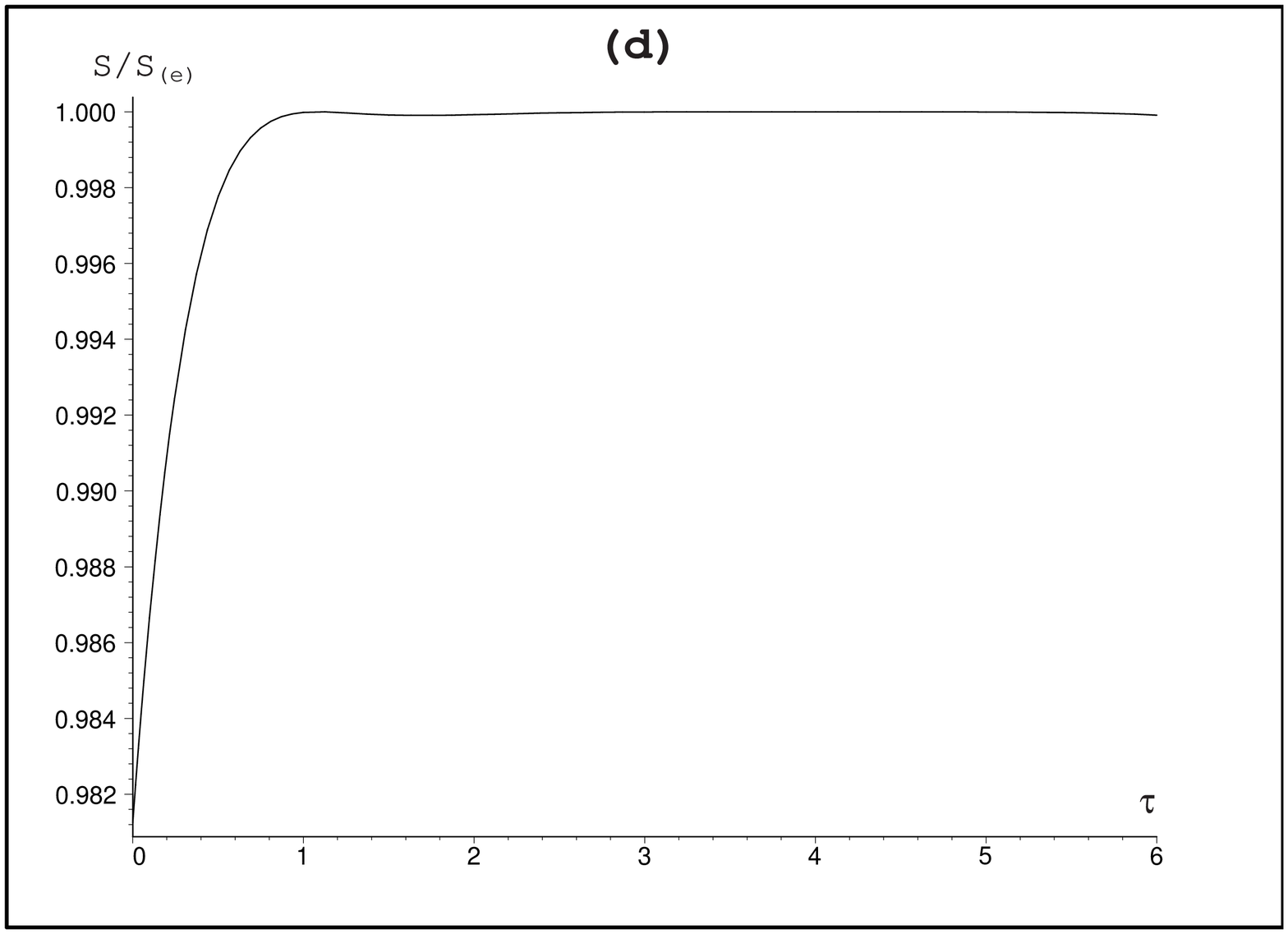}
\end{center}
\caption{${\hbox{A physically plausible evolution}}. \ \ $ These figures 
illustrate the fulfillment of conditions 1, 2 and 3 for a physically 
plausible evolution given in section VIII-B. The function $ \ S_{(e)} \ $ in 
(c) is given in cgs units and is almost equal to the equilibrium photon 
entropy. These figures (as well as those of figure 2) were obtained using 
initial conditions (102), with $ \ \tau_{\hbox{\tiny rel}}=0.7. \ $ Notice in 
figure 1b that $ \ \dot S_{(e)} \ $ becomes negative (though small) near the 
initial time. This behavior does not denote an unphysical situation since $\
\dot S \ > \ 0$ holds throughout the evolution.} \label{PPE}
\end{figure}

\begin{figure}
\begin{center}
\leavevmode
\epsfxsize=2.6in
\epsffile{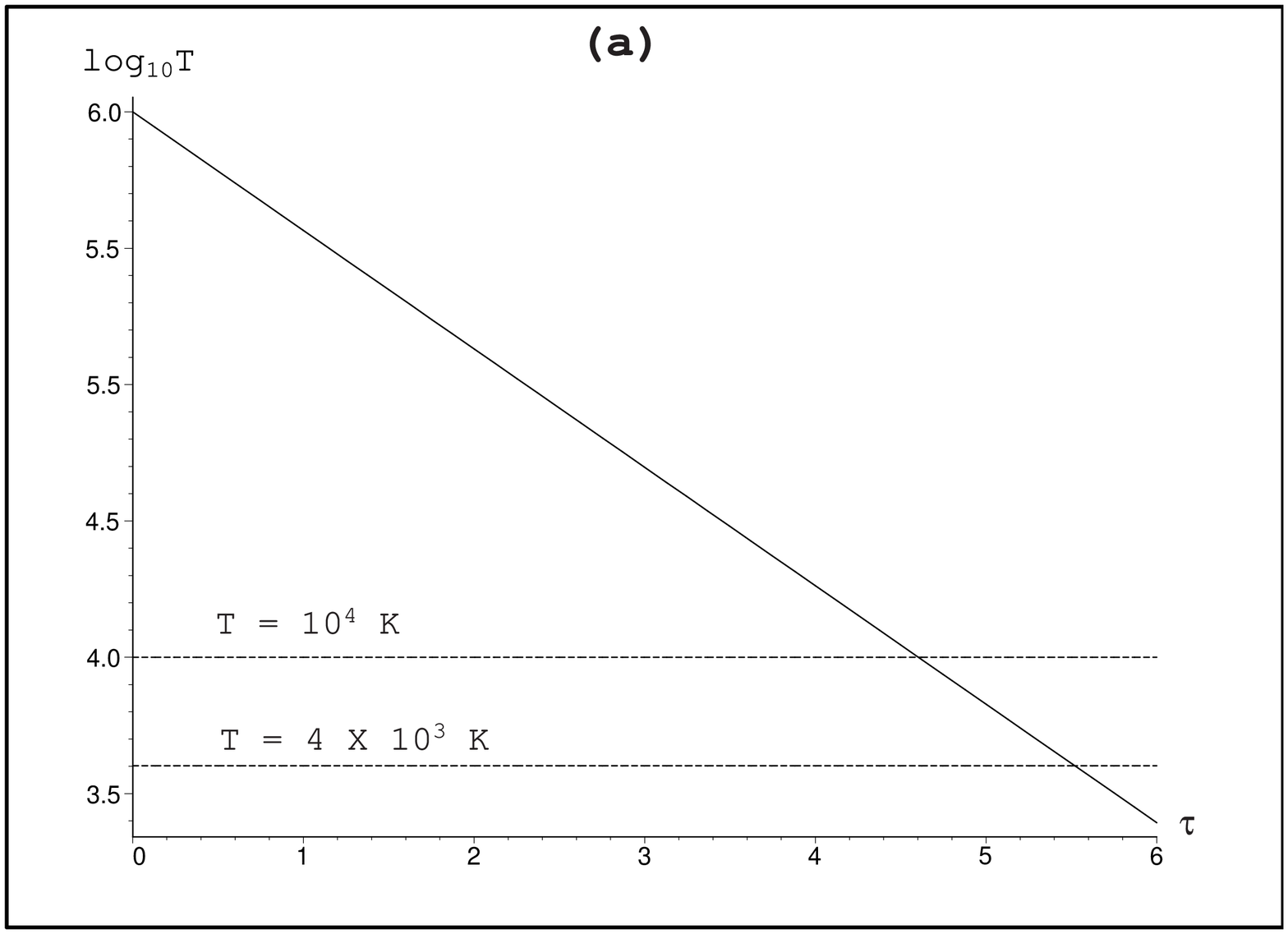}
\leavevmode
\epsfxsize=2.6in
\epsffile{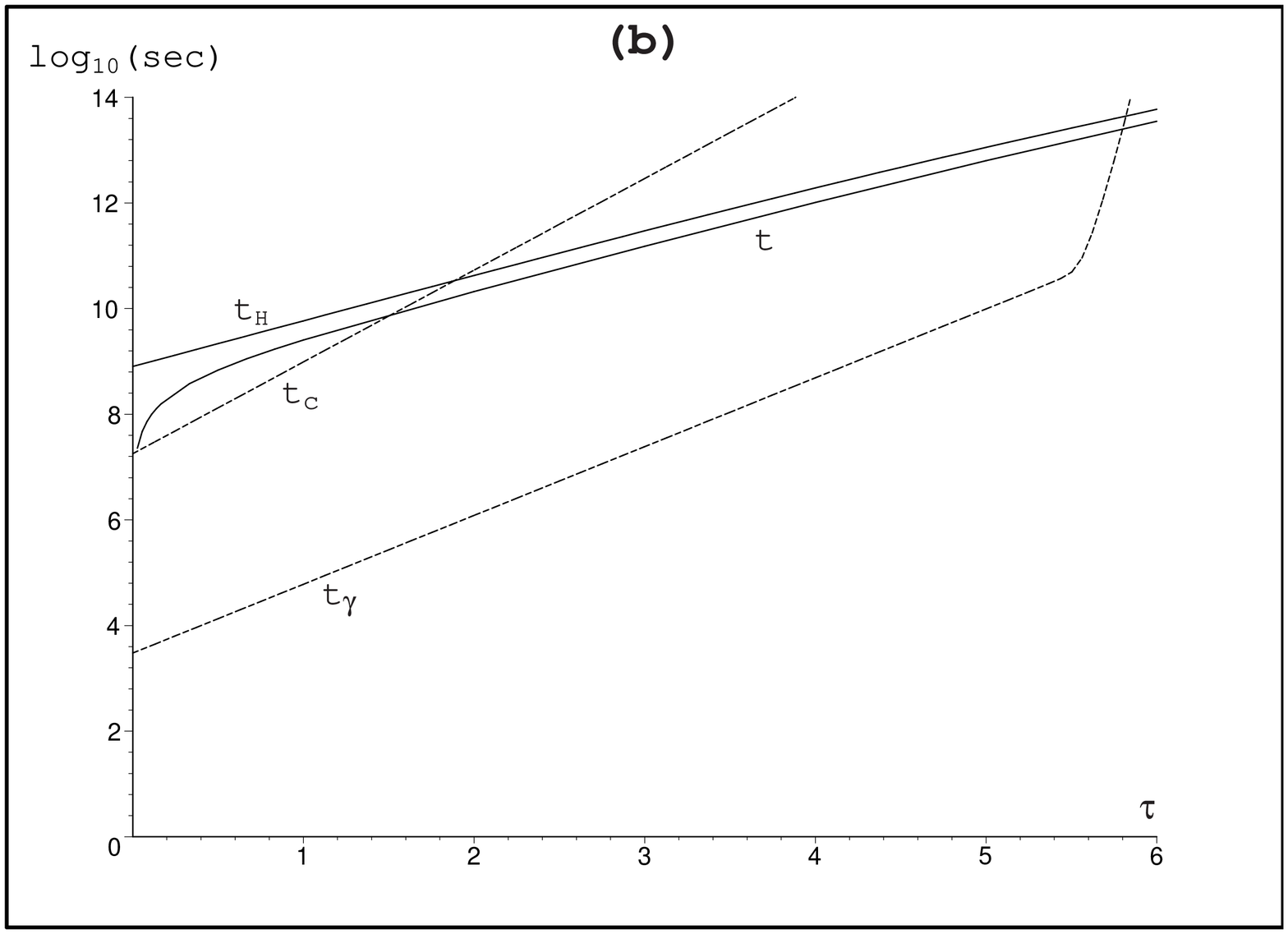}
\end{center}
\caption{${\hbox{Range of validity of the models.}} \ \ $ The range $ \ 0 
\leq \tau \leq 6 \ $ corresponds to the temperature range of the radiative 
era, between $ \ T_i = 10^6$ K and the baryon--radiation decoupling 
temperature $ \ 4 \times 10^3$ K (roughly at $ \ \tau = 5.5$). A logarithmic 
plot of the various timescales used in the paper (in seconds) are depicted 
in figure 2b: the Thomson and Compton scatterings ($t_\gamma, \ t_c$), the 
Hubble time $ \ t_{_{H}} \ $ and physical time $ \ t. \ $ Notice  how the 
decoupling temperature occurs at the same $ \ \tau \ $ as the condition $ \ 
t_\gamma = t_{_{H}}, \ $ while radiation-matter equality, $ \ \Omega_{(r)} 
= \Omega_{(m)} \ $ (see figure 1a) corresponds to $ \ T = 10^4 \ $ K 
(roughly $ \ \tau = 4.6.$)}
\end{figure}

\subsection{Sensitivity to deviations from $ \ \chi = 0$.}

We will test the effect of initial deviations from the invariant set $ \ 
\chi = 0 \ $ on state functions obtained by numerical integration of the 
governing equations. We consider initial conditions as in 
(\ref{initconds}), keeping $ \ \Sigma_i, \ $ and $ \ Q_i \ $ fixed, but 
now we take $ \ \delta \ne 0$:
\ba 
\begin{matrix}
{\Omm(0) \ = \ 0.0128 - \delta,}&{ \qquad \Omr(0) \ = \ 0.9872,}\\
{\Sigma(0) \ = \ 0.001,}&{ \qquad Q(0) \ = \ -0.1, \quad \ \ }\\
{\ \Omt(0)=1-\delta,}&{\qquad \chi(0) \ = \ \delta - 10^{-6}}, \label{delta_ic}
\end{matrix} 
\ea
Since $ \ \delta \ $ can be given in terms of $ \ \epsilon \ $ by eqn. 
(\ref{e1e2}) and $ \ \epsilon = 0.0128 \ $ corresponds to $ \ \delta = 0, \ $ 
testing values of $ \ \epsilon \ $ near $ \ \epsilon = 0.0128 \ $ determines 
the initial deviation from $ \ \Omt = 1 \ $ and $ \ \chi = 0. \ $ Notice that 
$ \ \delta \ $ can be positive or negative, respectively, for $ \ \epsilon > 
0.0128 \ $ or $ \ \epsilon < 0.0128, \ $ so that $ \ \chi(0) = 0 \ $ if $ \ 
\delta = \Sigma_i^2 = 10^{-6}, \ $ while $ \ \chi(0) \ $ is  positive/negative 
if $ \ \delta > \Sigma_i^2 = 10^{-6} \ $ or $ \ \delta < \Sigma_i^2 =  10^{-6} 
\ $ (though, from eqn. (\ref{eqstate4}), curvature has opposite sign to $ \ 
\chi$). We integrate numerically the governing equations for initial 
conditions (\ref{delta_ic}), $ \ \gamma = \gamma_0 = 0.7 \ $ (transient 
regime) and for  various values of $ \ \delta. \ $ The resulting forms of $ \ 
\Omm, \ \Omr, \ \Omt,$ and $ \ S \ $ are respectively plotted in figures 3a to 
3d (for positive curvature $ \ \chi < 0 \ $) and in figures 4a to 4d (for 
negative curvature $ \ \chi > 0$). These figures clearly show that a physical 
evolution is only possible for initial conditions that deviate very slightly 
from $ \ \chi(0) = 0 \ $ (less than $ \approx 10^{-5.5}$), leading to orbits 
that remain very close to the invariant set $ \ \chi = 0. \ $ If we fix $ \ 
\epsilon \ $ and $ \ \delta \ $ (for any combination of values compatible with 
(\ref{e1e2}), (\ref{e11e22}) and (\ref{e123})) and vary $ \ \Sigma_i, \ $ so 
that deviation from $ \ \chi(0) = 0 \ $ is governed by $ \ \Sigma_i, \ $ we 
obtain exactly the same behavior displayed by figures 3 and 4, leading to the 
same conclusion: a physical evolution is only possible for initial conditions 
for which $ \ |\chi(0)| \alt 10^{-5.5}$.

\begin{figure}
\begin{center}
\leavevmode
\epsfxsize=2.5in
\epsffile{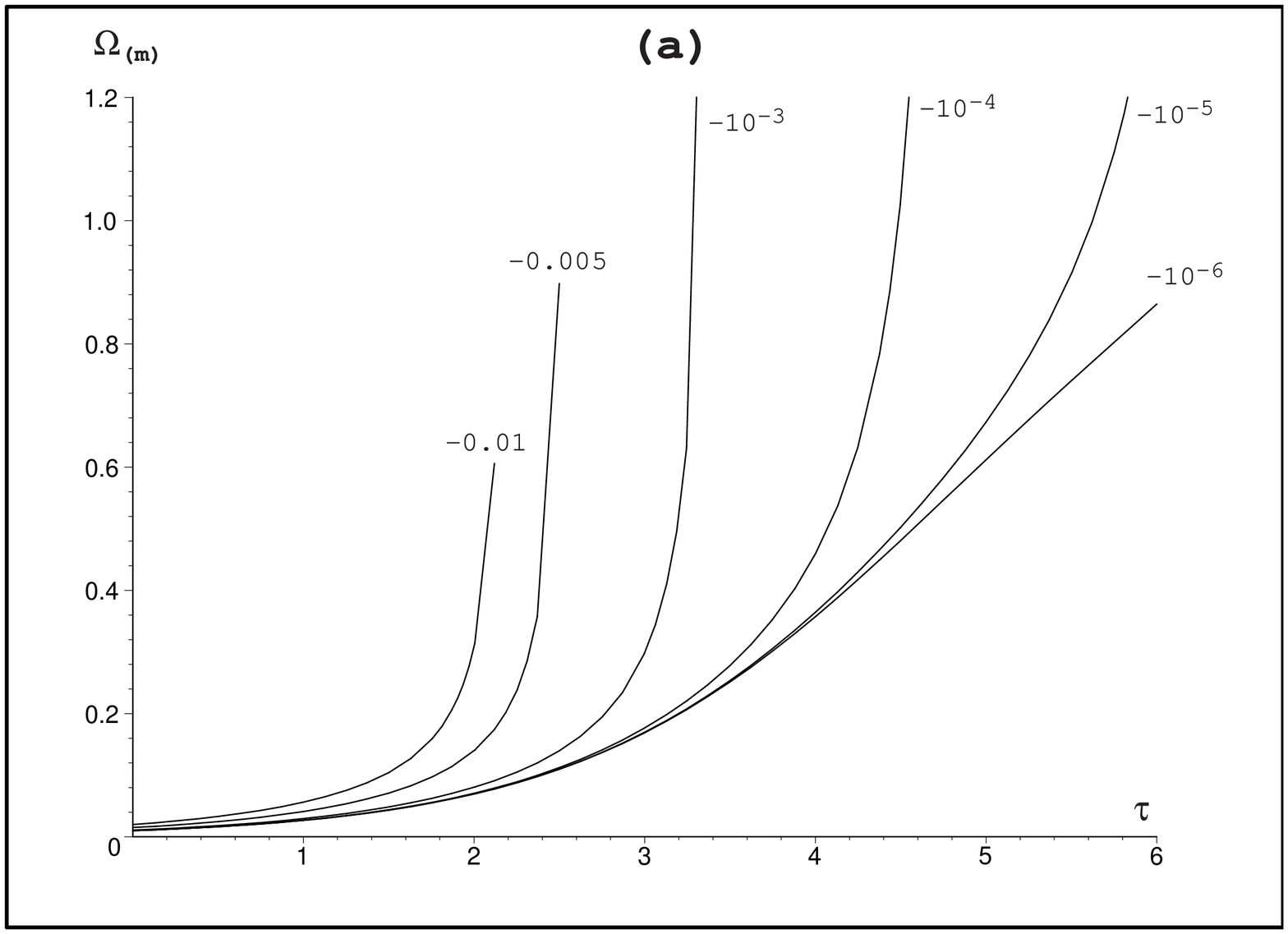}
\leavevmode
\epsfxsize=2.5in
\epsffile{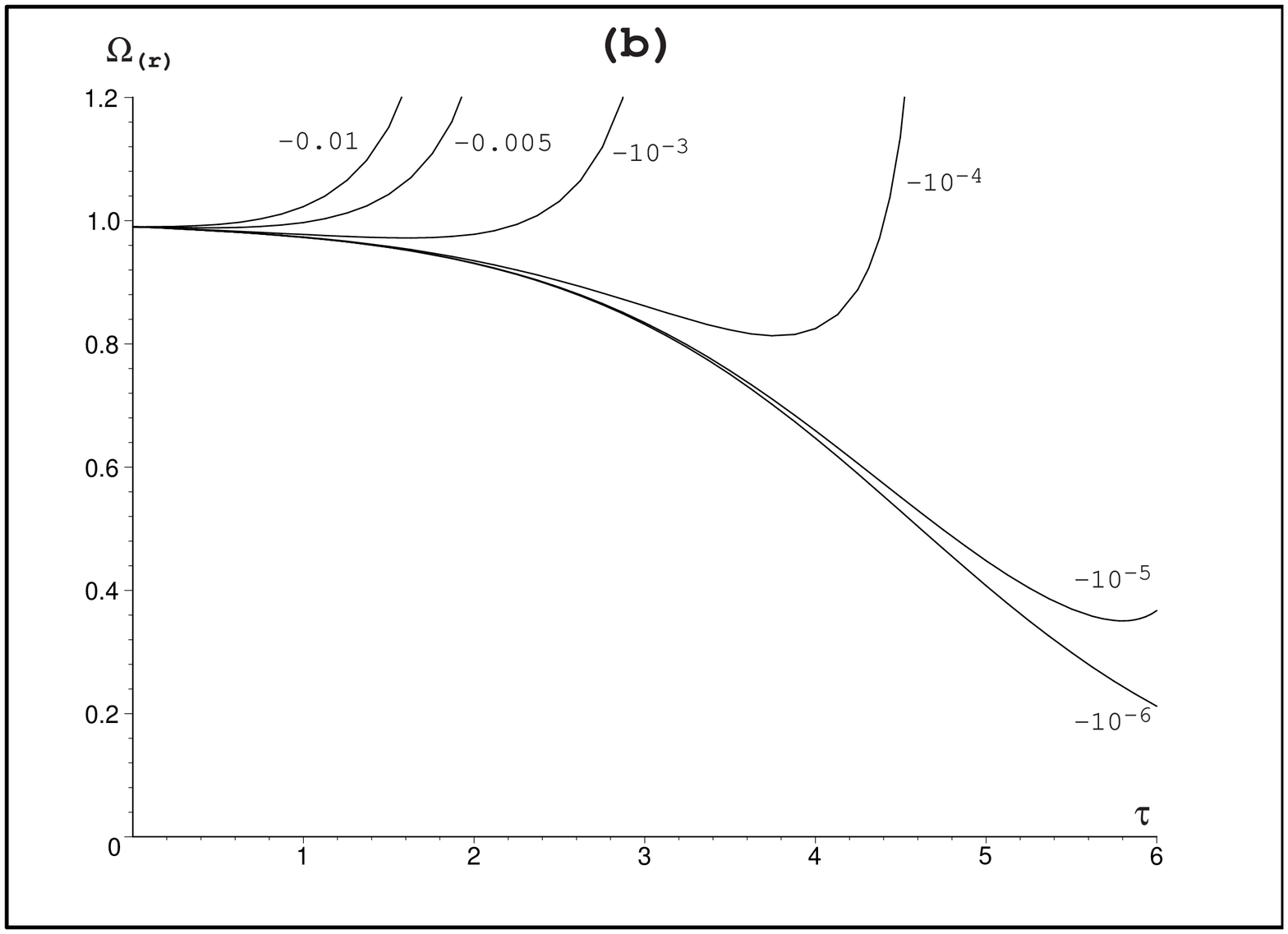}
\leavevmode
\epsfxsize=2.5in
\epsffile{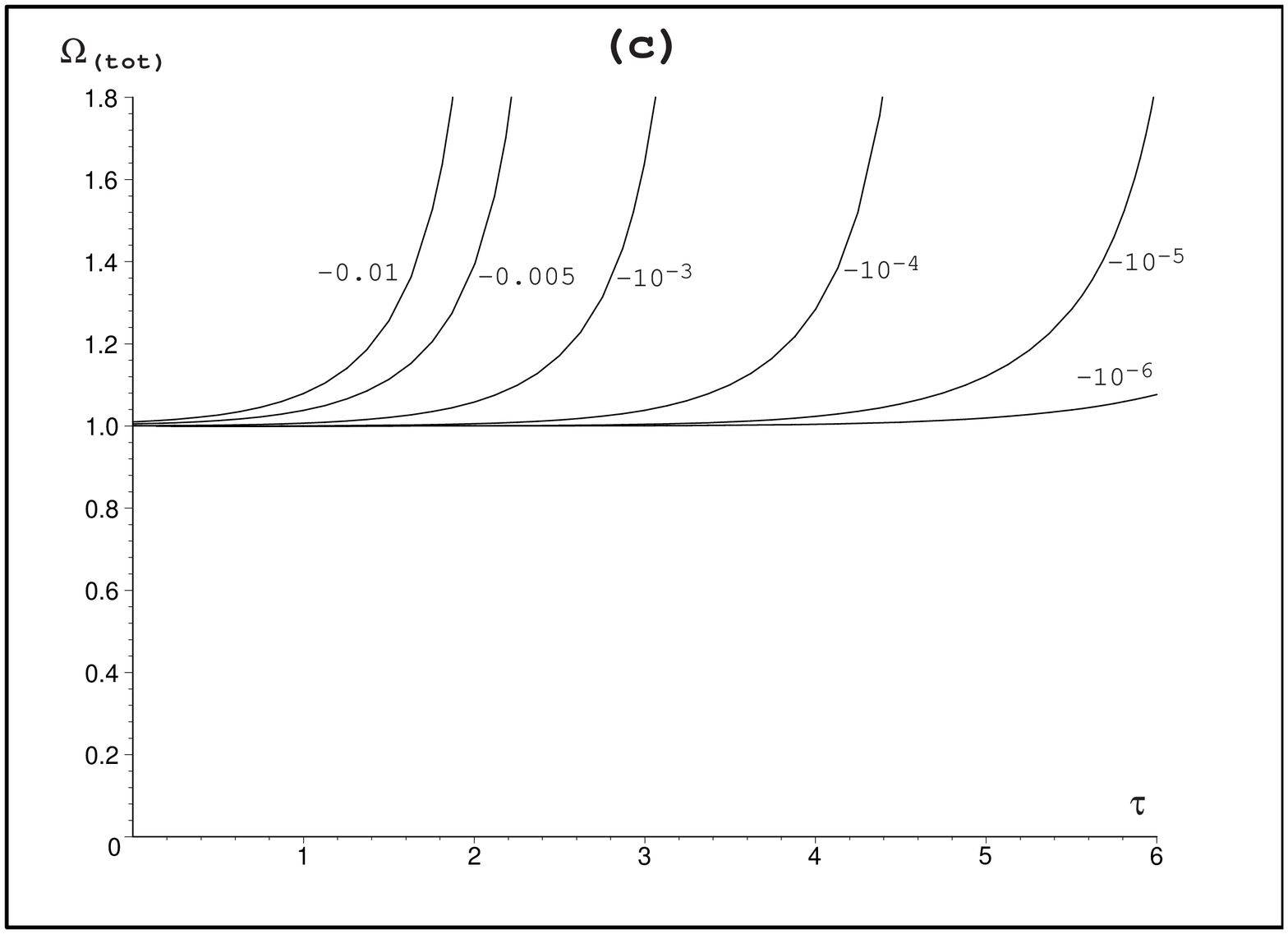}
\leavevmode
\epsfxsize=2.5in
\epsffile{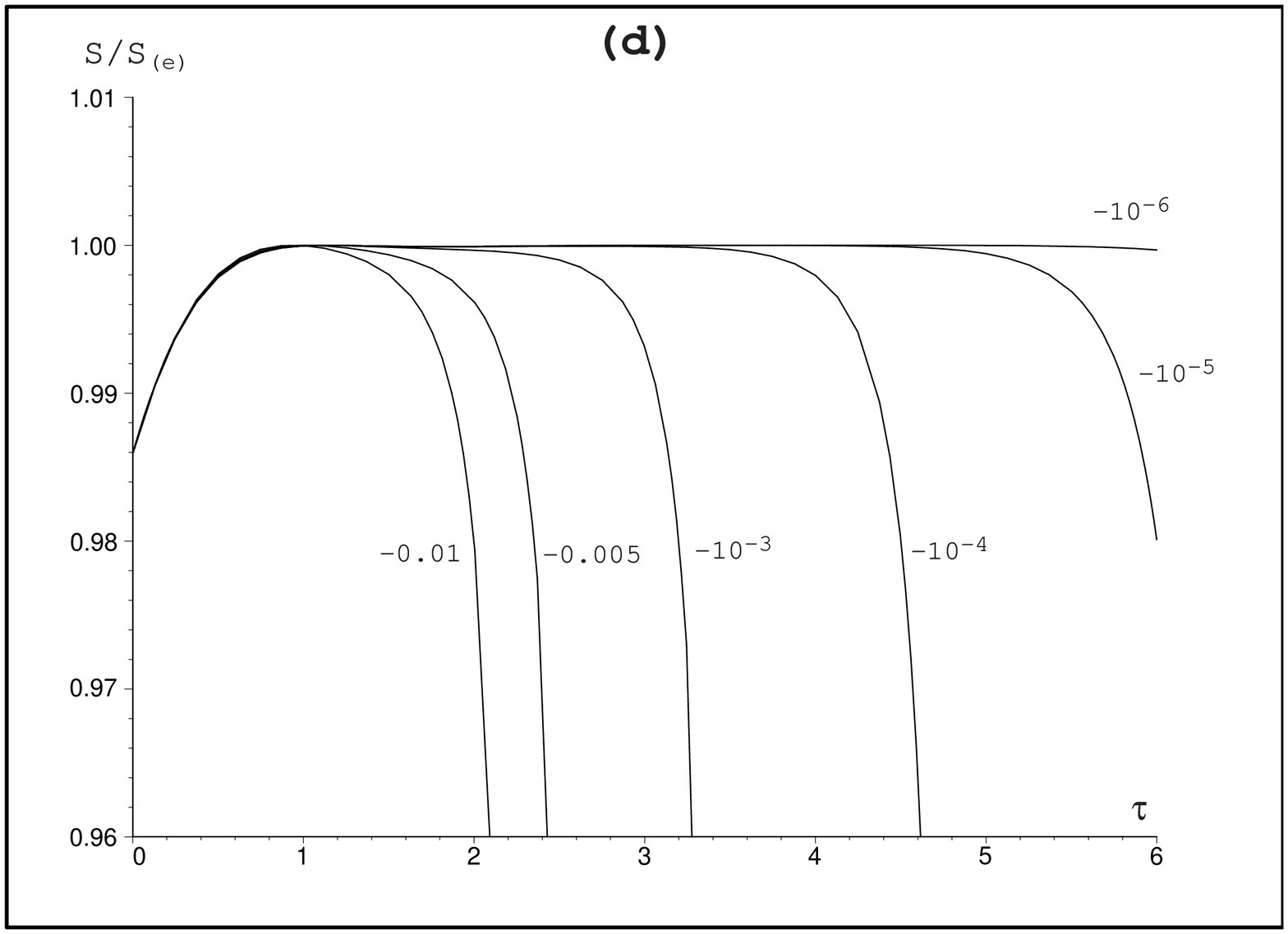}
\end{center}
\caption{${\hbox{Sensitivity to $ \ \chi(0) = 0, $ positive curvature.}} \ $ 
Figures display $ \ \Omm, \ \Omr, \ \Omt = \Omm + \Omr \ $ and $ \ S/\Se \ $ 
for initial conditions (103) with $ \ - 0.01 < \delta < 0; $ numerical values 
of $ \ \delta \ $ appear next to each curve. $ \ \Omm, \ \Omr, \ $ and $ \ 
\Omt \ $ branch upward and $ \ S/\Se \ $ downward; the branching time $ \ 
\tau \ $ decreases with increasing $ \ |\delta|. \ $ The branching up 
corresponds to $ \ \Theta \to 0 \ $, marking the re-collapsing stage of the 
models. A comparison of figures (a), (b), and (c) with (d), and with figures 
2a and 2b, shows that a physically plausible evolution for the duration of the 
radiative era ($0 < \tau  < 6$) and for the appropriate temperature ranges is 
only possible if $ \ |\delta|  < 10^{-5.5}.$} \label{initfunc_dens1}
\end{figure}

\begin{figure}
\begin{center}
\leavevmode
\epsfxsize=2.5in
\epsffile{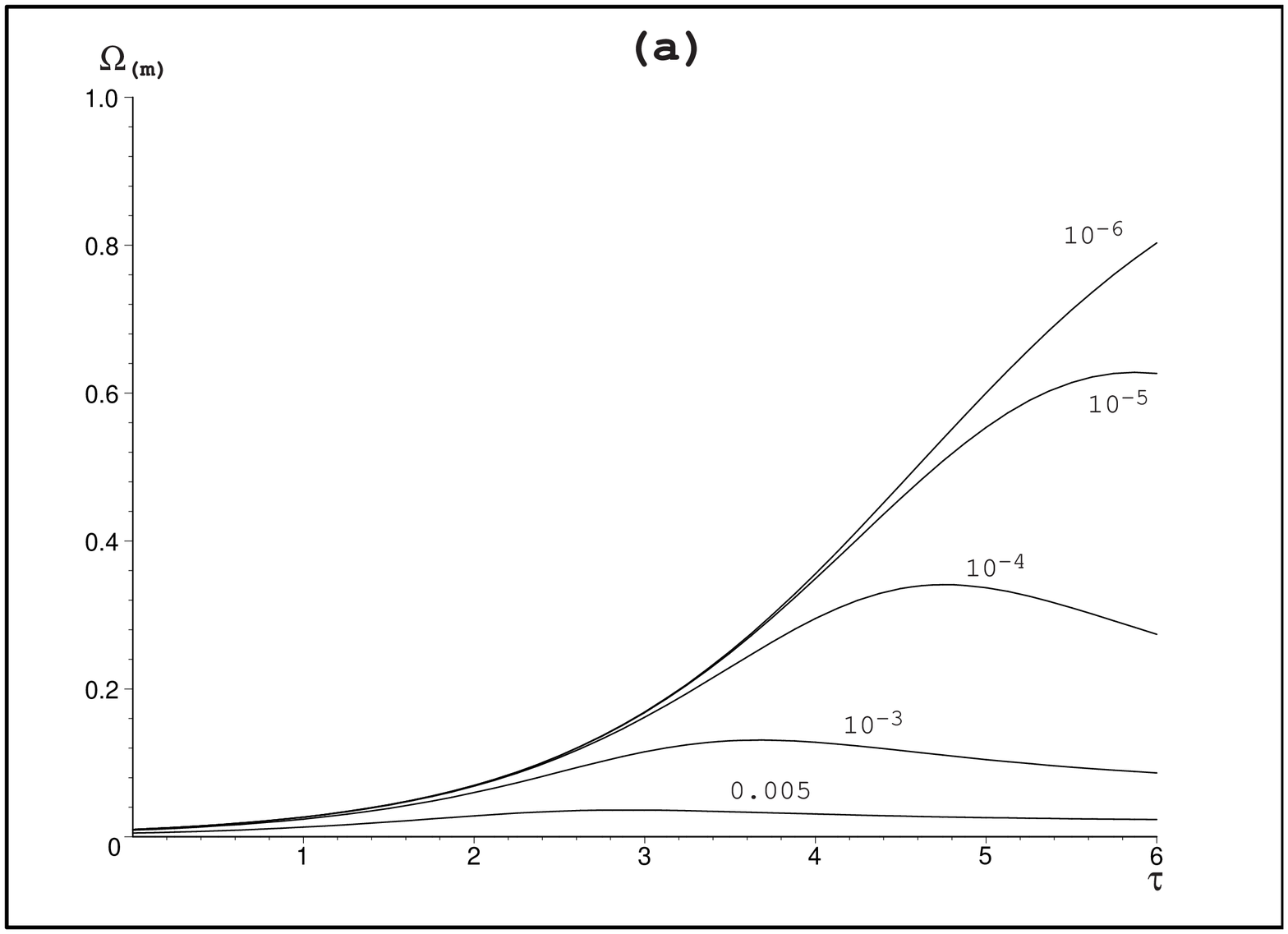}
\leavevmode
\epsfxsize=2.5in
\epsffile{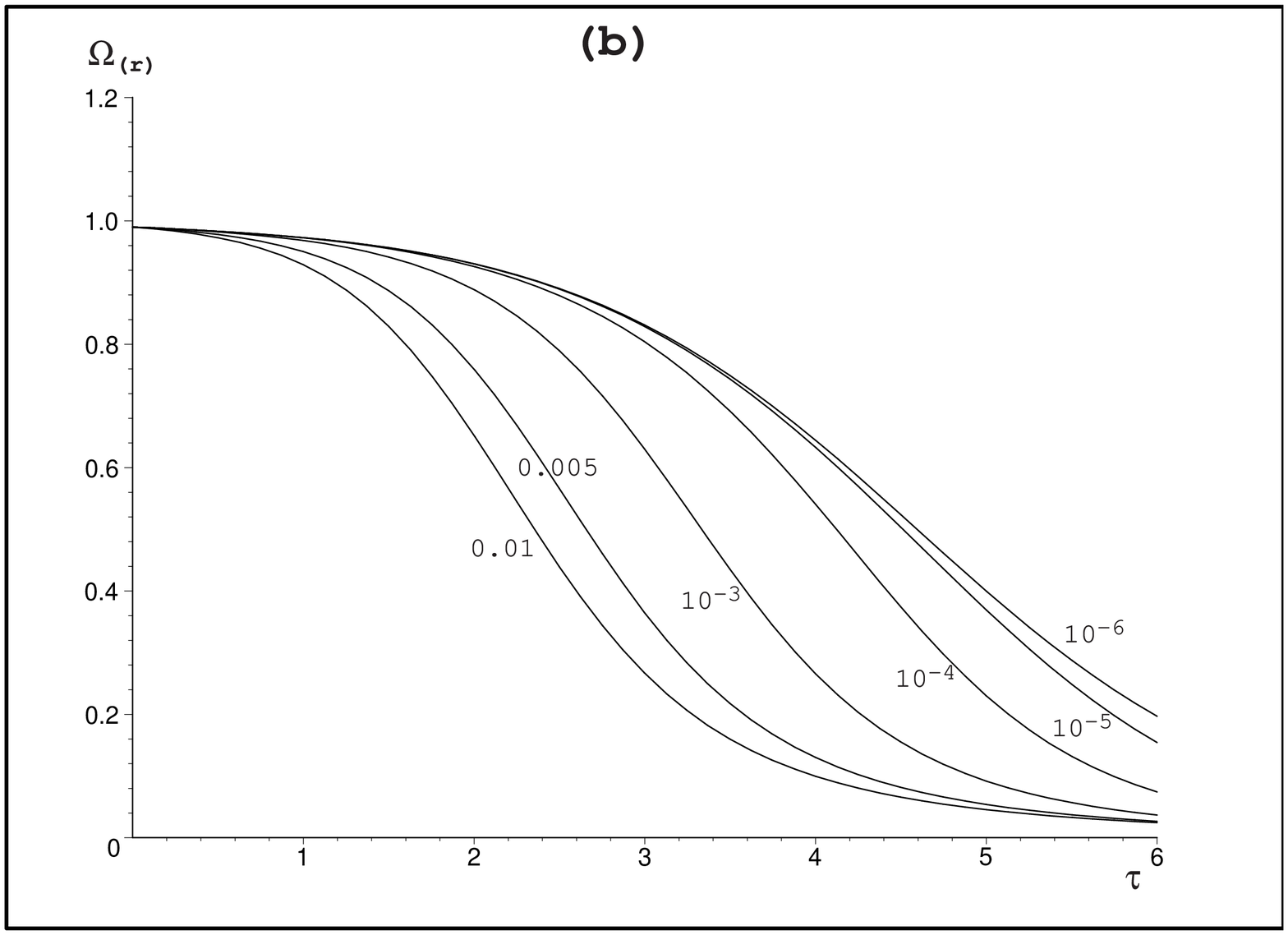}
\leavevmode
\epsfxsize=2.5in
\epsffile{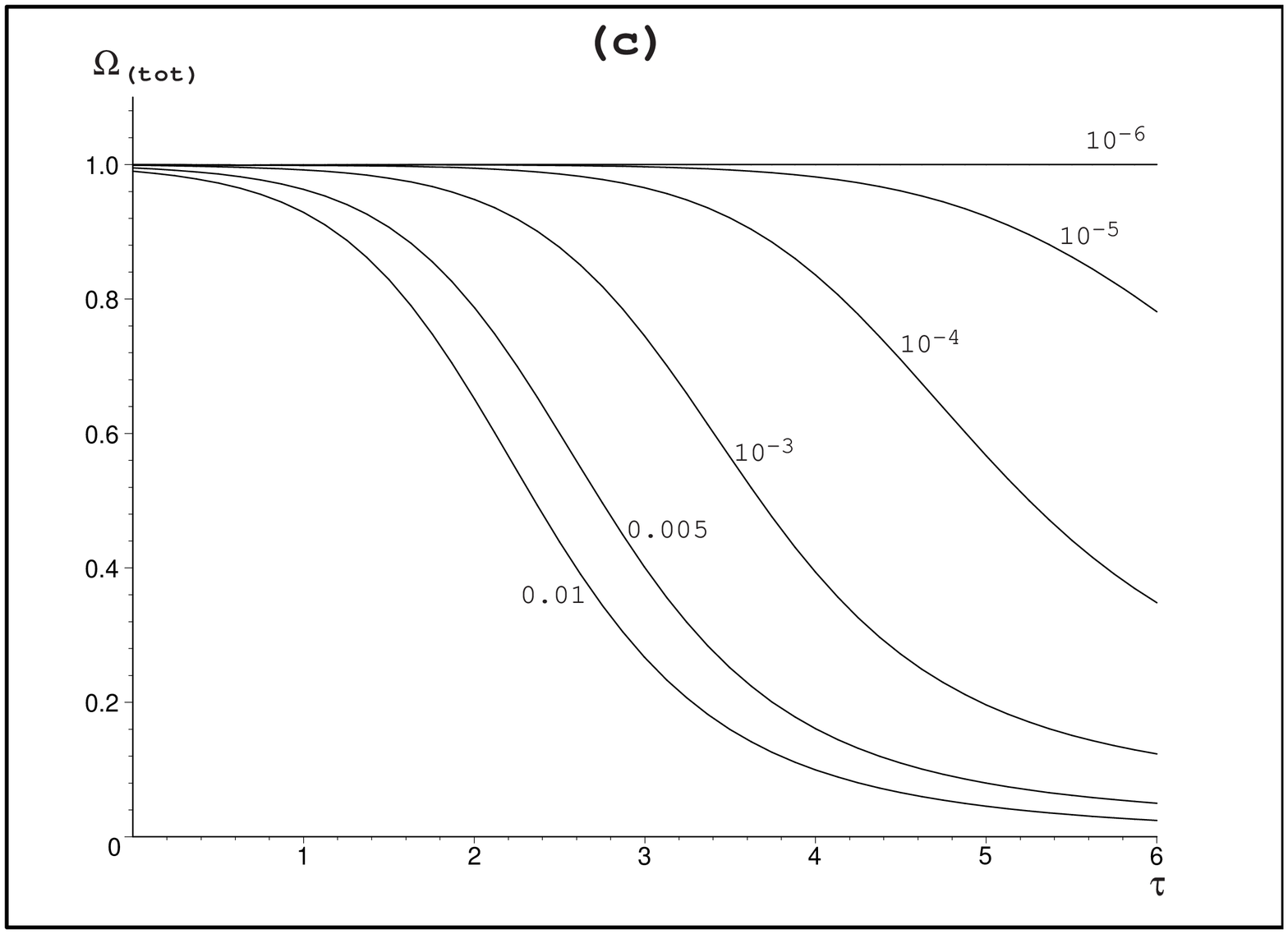}
\leavevmode
\epsfxsize=2.5in
\epsffile{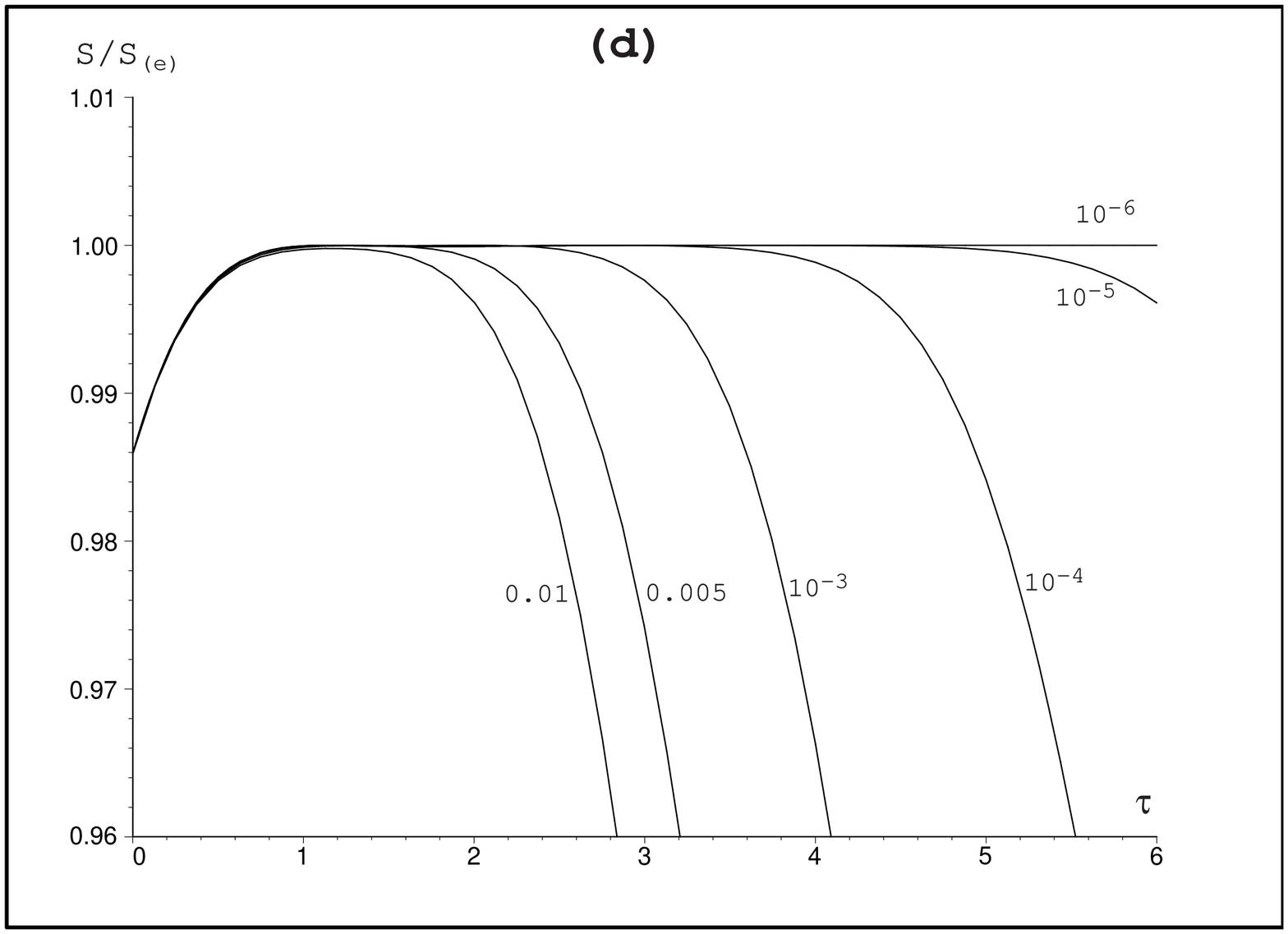}
\end{center}
\caption{${\hbox{Sensitivity to $ \ \chi(0) = 0, $ negative curvature.}} \ $ 
Analogues of figures 3 for $ \ \delta > 0. \ $ A similar branching of the 
functions $ \ \Omm, \ \Omr, \ \Omt, \ S/\Se, \ $ is observed even though the 
models do not re-collapse. Again, the function $ \ S/\Se \ $ satisfies the 
conditions for a physically plausible evolution only if initial conditions are 
given by $ \ \delta < 10^{-5.5}. \ $ Notice that the $ \ \Omm (\tau) \ $ 
curves in (a) with $ \ \delta \approx 10^{-5.5}, \ $ bend downward, so that 
for $ \ \tau \gg 6 \ $ they tend to the currently accepted values of present 
CDM abundance: $ \ \Omm \approx 0.3.$} \label{initfunc_dens2}
\end{figure}

\subsection{The near-Eckart regime}

We examine the near-Eckart regime by assuming $ \ \gamma = \gamma_0 = 0.001 
\ll 1, \ $ together with initial conditions (\ref{trial_ic}). As shown in 
figure 5a, the functions $ \ Q(\tau) \ $ and $ \ S(\tau) \ $ clearly have 
the form (\ref{near_t0}), indicating a quick relaxation in terms of an 
abrupt exponential decay (in about $ \ \tau \approx 0.01$). Figure 5b 
shows how the equilibrium entropy $ \ \Se \ $ tends to a constant value 
in about $ \ \tau \approx 0.02, \ $ a longer time than the relaxation of 
$ \ S/\Se, \ $ thus indicating that the effective relaxation time is 
provided by $ \ \Se, \ $ and $\ \Se'\ > \ 0 $ holds for all the evolution, in
agreement with the ``local equilibrium  hypothesis'' (see section
\ref{regimes}). Figure (5c) reveals how $ \  (5/8) \ Q/\Sigma \to \gamma_0 =
0.001, \ $ so that Eckart's transport  equation ($P + 2 \eta \sigma = 0 \ $
with $ \ \eta \ $ given by eqn.  (\ref{rg})) is approximately valid once the
quick relaxation is over. A  good approximation to the relaxation time in a
near-Eckart regime is given  by the assumption (\ref{trel_prop_H3}) with $ \
\gamma = - (5/8) \
\gamma_0 
\ (Q/\Sigma) \ $ and $ \ \gamma_0 = 0.001. \ $ Figure 5d illustrates that 
this is a correct assumption, since the obtained $ \ \trel \ $ overtakes 
$ \ \tH \ $ in the very short period that coincides with the duration of the 
relaxation process ($\tau \approx 0.02$).

The plots of the functions $ \ \Omm(\tau) \ $ and $ \ \Omr(\tau) \ $ are 
identical to those that would have resulted in the transient regime had we 
chosen initial conditions (\ref{trial_ic}) and a much larger value of 
$ \ \gamma_0. \ $ The function $ \ \Sigma(\tau) \ $ is affected, becoming 
almost constant for very small $ \ \gamma_0. \ $ This is reasonable, since 
eqn. (\ref{EqOmega}) does not contain $ \ Q, \ $ while eqns. 
(\ref{EqOmegar}) and (\ref{EqSigma}) do contain this function, but $ \ 
|Q(\tau)| \ $ and $ \ Q' \ $ decay very fast becoming almost zero for most 
of the time range and so the differential equations for the functions $ \ 
\Omm \ $ and $ \ \Omr \ $ (but not $ \ \Sigma$) are practically unaffected. 
The effect of varying $ \ \gamma_0 \ $ (as long as we have small values, 
$ \ < 0.01$) is simply to make the decay of $ \ Q \ $ and $ \ \Se' \ $ 
slightly more or less abrupt (depending on whether $ \ \gamma_0 \ $ is 
smaller or larger than $ \ 0.001$) and has no noticeable effect on 
numerical curves of other functions.

\begin{figure}
\begin{center}
\leavevmode
\epsfxsize=2.4in
\epsffile{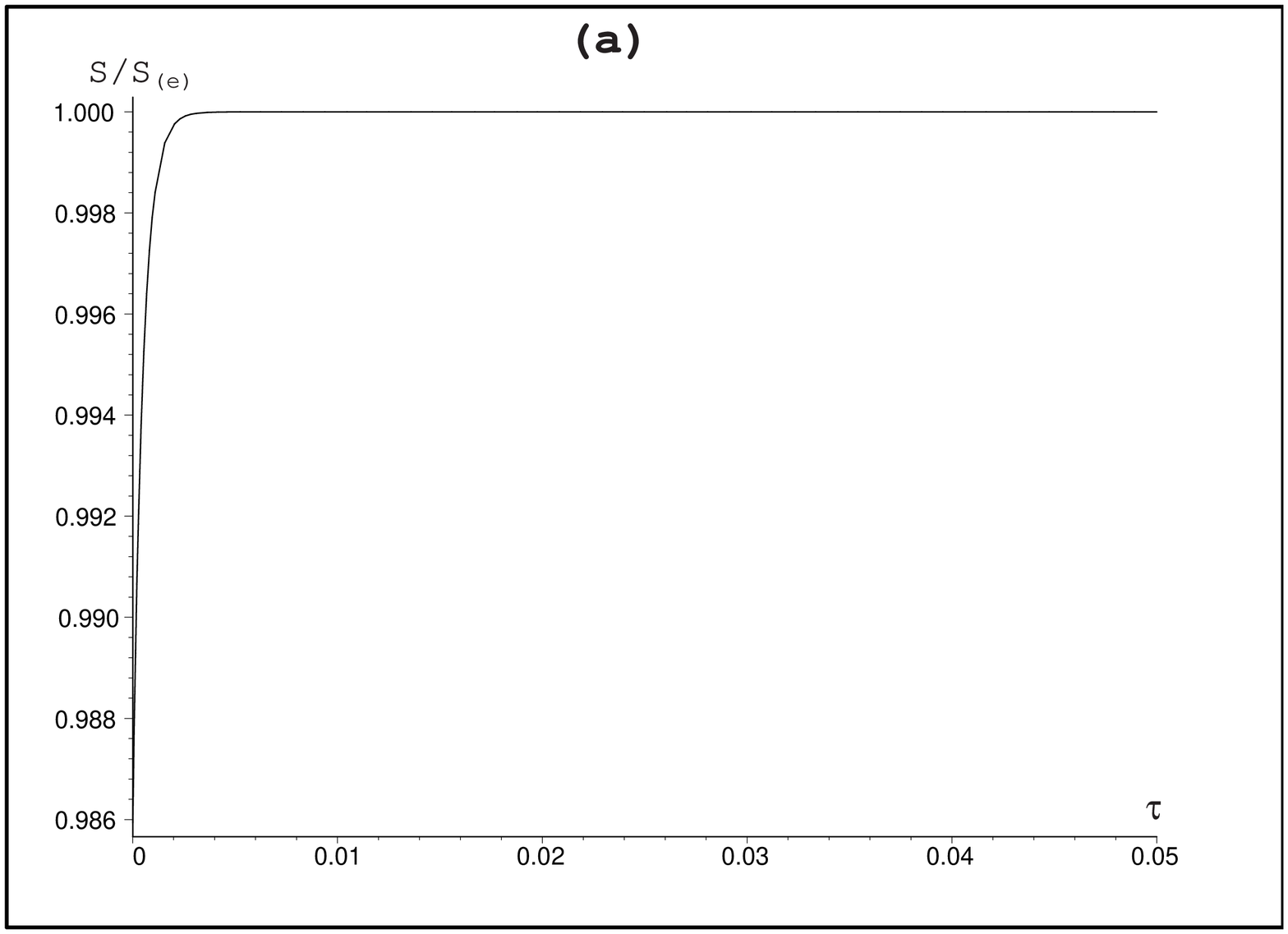}
\leavevmode
\epsfxsize=2.4in
\epsffile{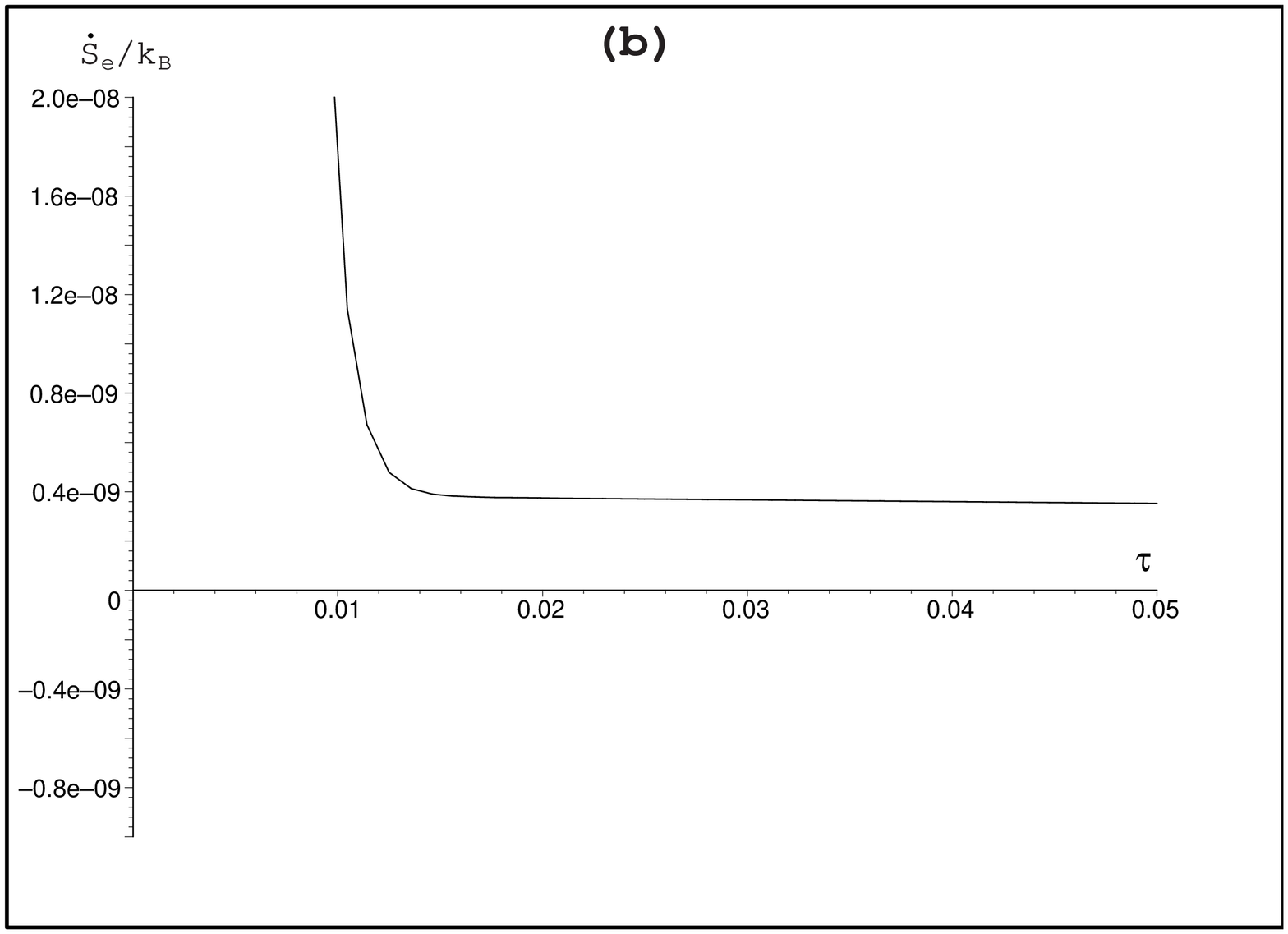}
\leavevmode
\epsfxsize=2.4in
\epsffile{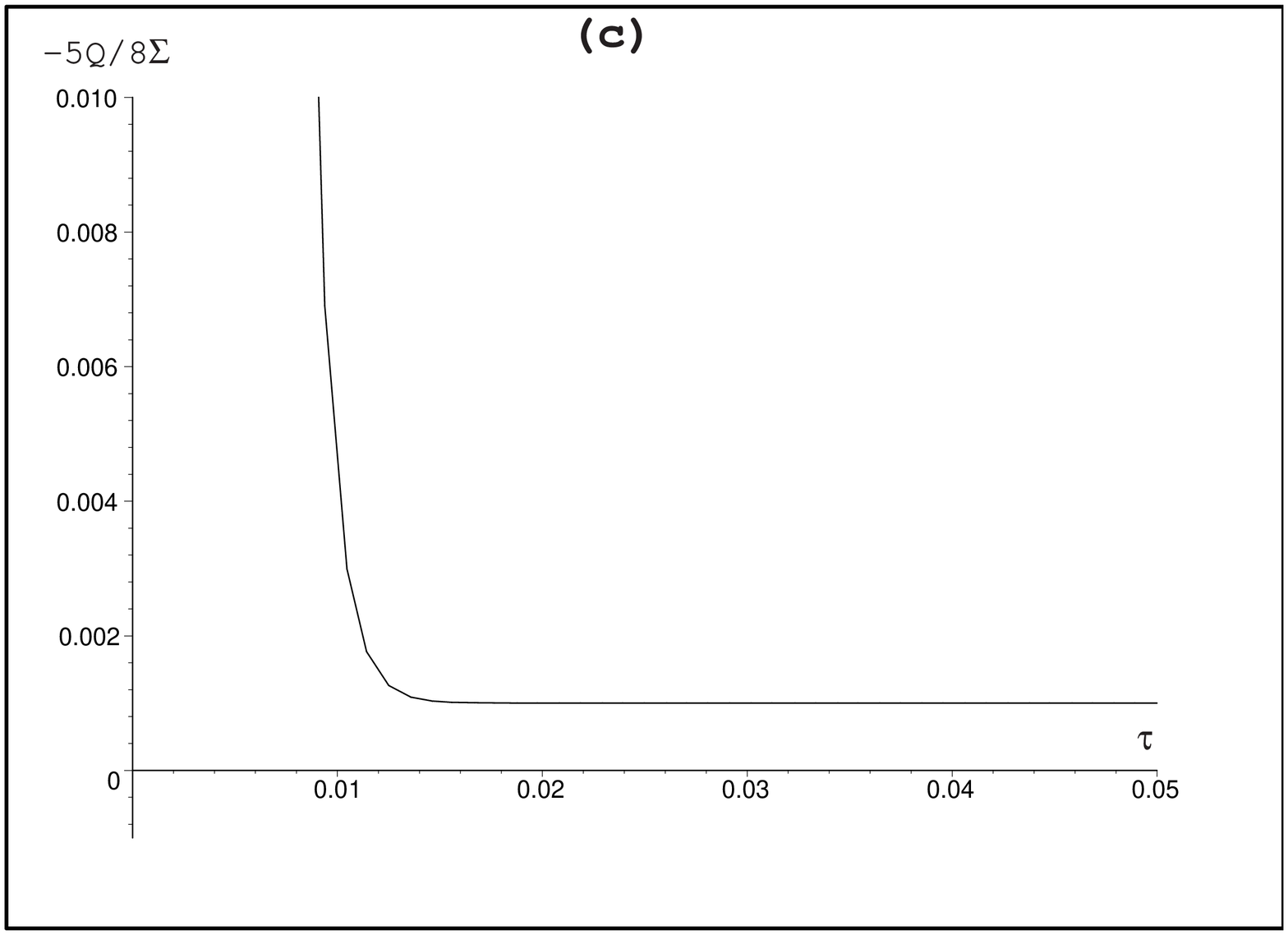}
\leavevmode
\epsfxsize=2.4in
\epsffile{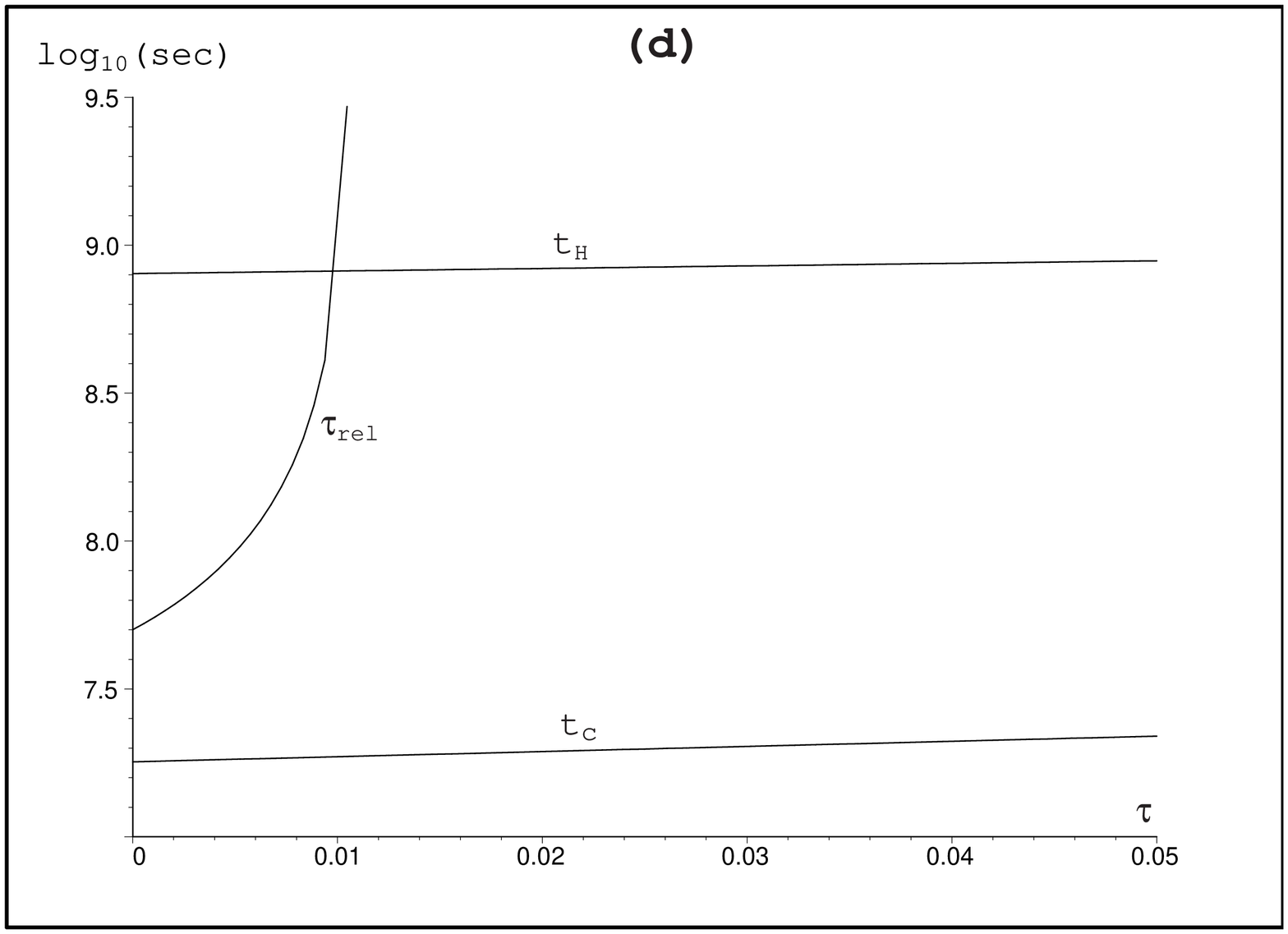}
\end{center}
\caption{${\hbox{``Near-Eckart'' regime}}.$ Func\-tions $S/\Se,$ $\dSe/\kB,$ 
and $ - (5/8)(Q/\Sigma)$ correspond to initial conditions (104) with $\gamma_0 
= 0.001. \ S/\Se$ in (a) relaxes much faster ($\tau \approx 0.002$) than in 
figures 1d, 3d or 4d, associated with the transient regime. The decay of $\Se$ 
in (b) takes longer ($\tau \approx 0.01$) than that of $S,$ hence $\Se$ is an 
adequate entropy function in agreement with the ``local equilibrium''
hypothesis characteristic of the Eckart regime ($\dSe > 0$ holds throughout 
the evolution). Panel (c) shows how $- (5/8)(Q/\Sigma) \to \gamma_0 = 0.001,$ 
indicating that Eckart's transport equation is approximately valid for $\tau > 
0.02.$ In panel (d) we used initial conditions (104) and $\gamma = - (5/8)
(Q/\Sigma) \gamma_0$ with $\gamma_0 = 0.001$ (instead of $\gamma = \gamma_0 = 
0.7$), which yields an excellent approximation to the relaxation parameter 
of a near-Eckart regime, overtaking $\tH$ at the relaxation timescale $\tau 
\approx 0.01;$ this timescale however, is too short in comparison with the 
relaxation timescale of the Compton scattering also shown in (d), hence this 
and other radiative processes must be studied within a transient regime.} 
\label{initfunc_test1}
\end{figure}

\subsection{Testing the relaxation times numerically.}

The assumption $ \ \gamma = \gamma_0 < 1 \ $ leads to a reasonable 
relaxation time only in the earlier stages of the evolution, while a 
choice $ \ \gamma = \gamma_0 > 1 \ $ might work for later conditions (near 
matter-radiation decoupling) but not for earlier times. This can be 
appreciated in figure 2b, since the relaxation times obtained for 
different values of $ \ \gamma = \gamma_0 \ $ would be curves parallel to 
$ \ \tH. \ $ As we mentioned before, $ \ \gamma_0 \ $ controls the rate of 
decay of $ \ Q \ $ and $ \ S, \ $ but it is still interesting to check if 
other functions are sensitive to changes in the numerical constant value 
of $ \ \gamma_0. \ $ Assuming initial conditions (\ref{trial_ic}) and 
integrating the dynamical system for various values of $ \ \gamma_0 > 0.1 
\ $ (thus excluding the non-transient zone) shows that the other functions 
(such as $ \ \Omm, \ \Omr \ $ or $ \ \Sigma$) are essentially identical to 
those shown in figures 1 to 4, being thus unaffected by the numerical value 
of $ \ \gamma_0$.

Since $ \ \gamma \ $ constant is not very realistic, we test now the 
expression for $ \ \trel \ $ that follows from equations (\ref{gamma_eq}) 
and (\ref{constants}) in section \ref{dynrel}. Ignoring the curvature term, 
we have that $\mu_0/\gamma$ equals
\be 
\frac{8}{5}+\mu_0(1+\mu_0)+\frac{\mu_0(1-2\mu_0)\Omm}{2}-\mu_0\left[1+
(\lambda-3)\mu_0\right]\Sigma^2,\label{gamma_dynrel}
\ee
It is evident that choosing a sufficiently small $ \ \mu_0 \ $ yields 
$ \ \gamma_i \approx 5 \mu_0/8 \ll 1, \ $ an initial value characteristic 
of a near-Eckart regime. As expected, numerical tests with $ \ \mu_0 \ll 1
 \ $ lead to practically identical curves as those corresponding to the 
near-Eckart regime (figures 5a, 5b and 5c). However, for $ \ \mu_0 > 1/2, \ 
$ the factor multiplying $ \ \Omm \ $ in the denominator of 
(\ref{gamma_dynrel}) can become negative for some $ \ \tau, \ $ thus 
opening the possibility that the denominator might become small and so $ \ 
\gamma \ $ might increase to $ \ \approx O(1) \ $ as $ \ \tau \ $ reaches 
later times related to the baryon--photon decoupling era. In order to 
explore this possibility, we integrated the  dynamical system for initial 
conditions (\ref{trial_ic}) and under the  assumption of $ \ \gamma \ $ 
given by eqn. (\ref{gamma_dynrel}), with various values of $ \ \mu_0 > 1/2. 
\ $ As shown by figure 6a, the choice $ \ \mu_0 > 15 \ $ leads to $ \ 
\gamma \ $ diverging as (approximately) $ \ \tau \to 7, \ $ this leads to 
$ \ \trel \ $ overtaking $ \ \tH \ $ (figure 6b) around $ \ \tau \approx 
6.8. \ $ It would have been nicer to have $ \ \trel \ $ overtaking $ \ \tH 
\ $ at an earlier time (say $ \ \tau \approx 5 \ - \ 5.5$) as required by 
condition 4 for a physically plausible evolution, but we feel that the 
form of $ \ \trel \ $ associated with (\ref{gamma_dynrel}) is a reasonable 
approximation to a physical relaxation time that acts as an ``effective'' 
relaxation time for the radiative era. Finally, another consequence of 
dealing with a more reasonable form for $ \ \trel \ $ is the fact that $ \ 
\dSe \propto - \Sigma \ Q \ $ being positive and very small (condition 
3 of a physically plausible evolution) is better satisfied than in the 
case of constant $ \ \gamma_0 \ $ (compare figures 1b and 6c).

\begin{figure}
\begin{center}
\leavevmode
\epsfxsize=2.7in
\epsffile{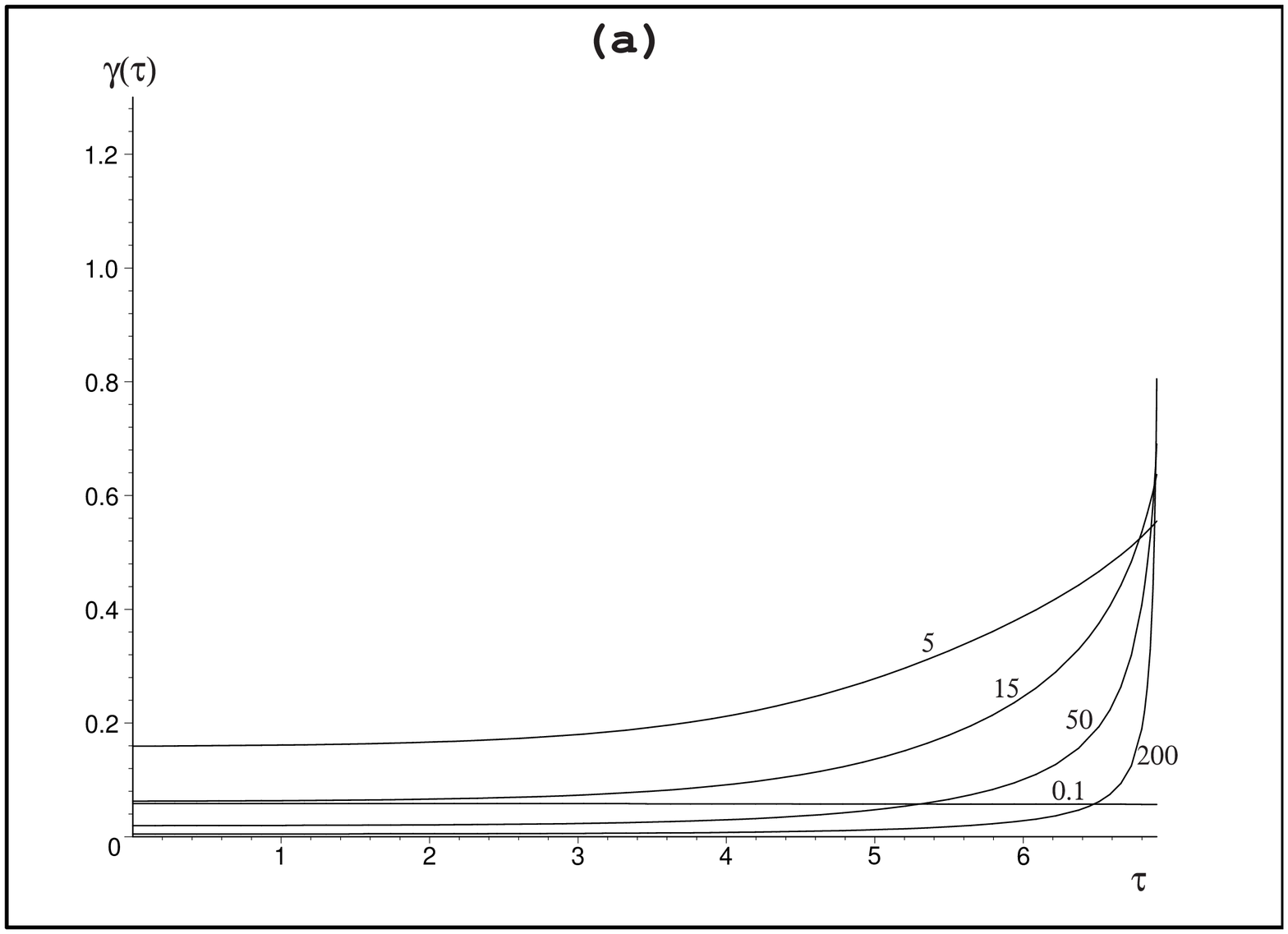}
\leavevmode
\epsfxsize=2.7in
\epsffile{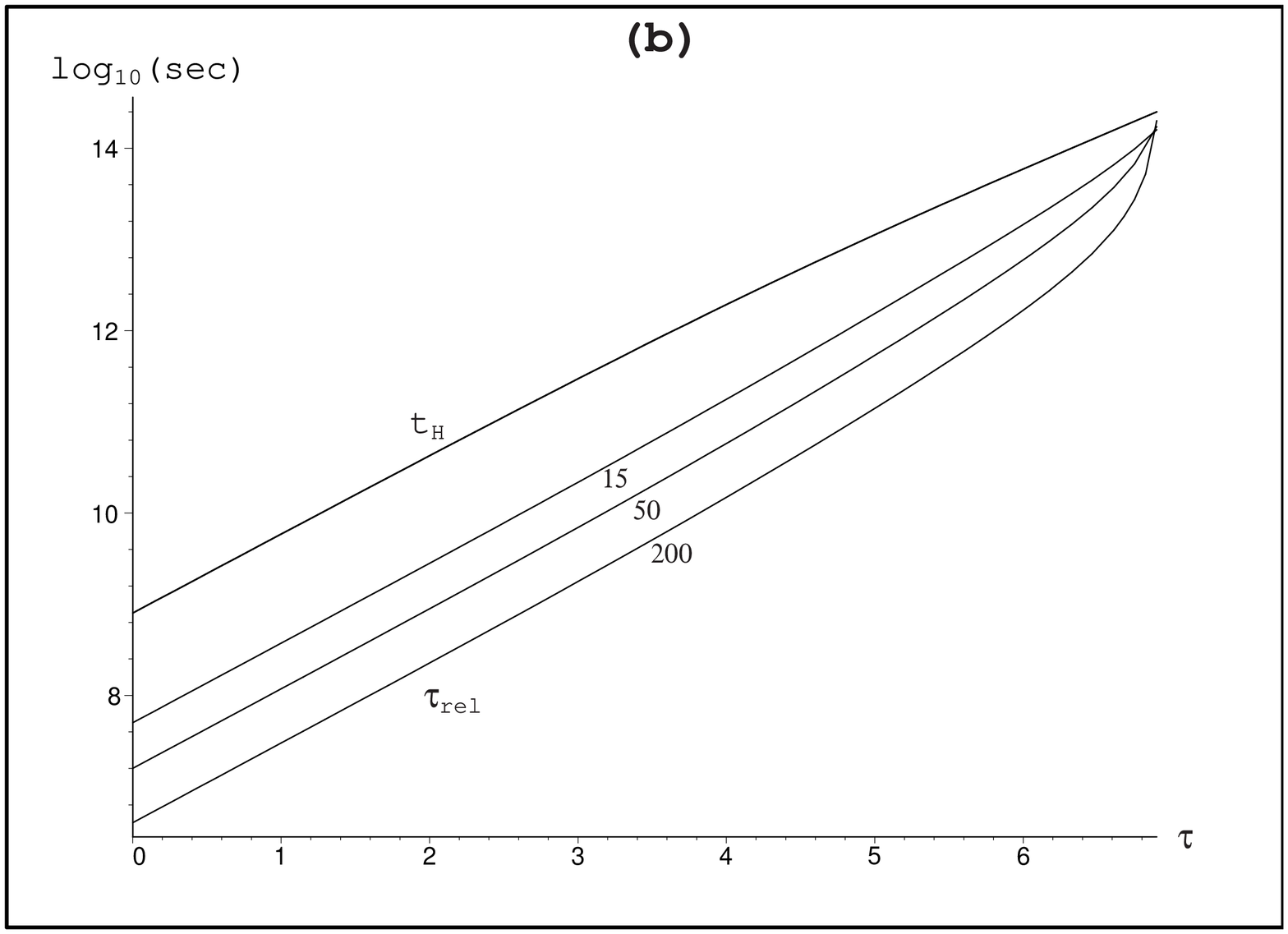}
\vskip0.1cm
\leavevmode
\epsfxsize=2.7in
\epsffile{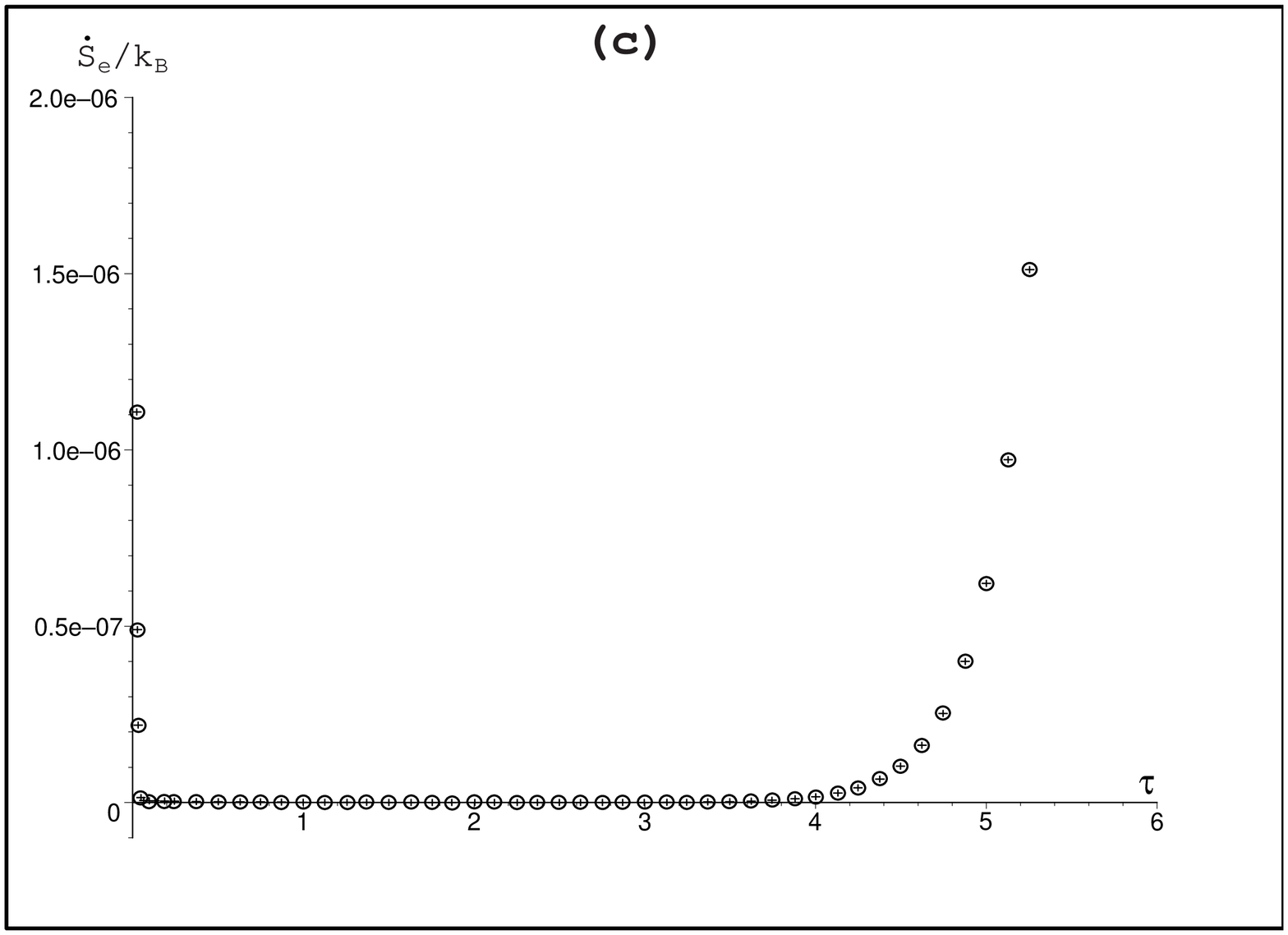}
\end{center}
\caption{${\hbox{Dynamical relaxation time.}} \ $ Panel (a) displays $ \ 
\gamma \ $ obtained from integrating the governing equations for the form 
given by (106), the numbers next to each curve correspond to the chosen 
numerical values of $ \ \mu_0; \ \gamma \ $ diverges for $ \ \mu_0 > 15. \ $ 
Panel (b) displays the corresponding form for $ \ \trel \ $ which overtakes 
$ \ \tH \ $ for $ \ \tau \approx 6.8. \ $ Panel (c) shows $ \ \dSe/\kB \ $ for 
$ \ \mu_0 = 200. \ $ A comparison with figure (1b), shows that requiring $ \ 
\dSe/\kB \ $ to be small is better satisfied with $ \ \gamma \ $ given by eqn. 
(104) than with a constant value $ \ \gamma=0.7 \ $ (as in figure 1b).} 
\label{layers_e}
\end{figure}

\subsection{The truncated equation.}

Considering initial conditions (\ref{trial_ic}) and $ \ \gamma = \gamma_0 
= 0.7, \ $ the integration of the governing equations (\ref{EqOmega}), 
(\ref{EqOmegar}), (\ref{EqSigma}), and (\ref{EqTr2}) yield the curves for 
$ \ Q, \ \Sigma, \ \Omm, \ $ and $ \ \Omr \ $ depicted in figures 7a, 7b 
and 7c. These figures reveal that a physical evolution fails to occur, as 
the growing modes predicted by the qualitative analysis clearly emerge, 
making all these functions undertake an unphysical growth. We tested 
other values of $ \ \gamma_0 \ $ and $ \ \gamma \ $ given by eqn. 
(\ref{gamma_dynrel}), as well as other initial conditions and obtained very 
similar curves to those of figures 7a-c, all failing to comply with the 
criteria for a physically plausible evolution.

\subsection{A baryonic scenario without CDM}

If we had considered only the radiative fluid (baryons, photons and 
electrons) as the matter source of the models, then the bulk of the rest mass 
energy density would have been due to the baryons, so that $ \ \nm \ $ 
would have been identified with $ \ \nbar \ $ instead of $ \ \nw \ $ and 
$ \ \Omm \ $ would correspond to the baryonic $ \ \Ombar. \ $
It is  interesting to examine numerically the consequences of this ``baryonic 
scenario''. Considering the baryon--photon number ratio in eqn. (\ref{nu}), 
the baryonic scenario implies replacing eqns. (\ref{e1_e2}) and (\ref{e1e2}) 
by
\be 
\frac{\epsilon-\delta}{1-\epsilon} = \frac{\mbar c^2}{3 \kB T_i} \ \nubar 
\approx 0.002, \qquad \delta = 1.002 \ \epsilon - 0.002,
\label{e1_e2_b}
\ee
while initial conditions are then given by eqns. (\ref{trial_ic}) and 
(\ref{delta_ic}), but $ \ \delta = 0 \ $ now corresponds to $ \ \epsilon = 
0.00199 \ $ instead of $ \ \epsilon = 0.0128. \ $ Intuitively, we do not 
expect a mayor qualitative change in the resulting graphs, though it is 
reasonable to expect that $ \ \Omm \ $ will be smaller and $ \ \Omr \ $ 
larger, since baryons have less rest mass density (by one or two orders of 
magnitude) than WIMP's and so it should take longer for baryons to dominate 
over radiation. The numerical curves that result are as expected intuitively,
with $ \ \Omm, \ \Omt, \ $ and $ \ S/\Se \ $ having very similar forms as the 
curves of figures 3 and 4, with $ \ \Omr \ $ decreasing slightly slower 
than in the case with WIMP's. Since the obtained curves for $ \ \Omm \ $ in 
the baryonic scenario are so close to those obtained in figures 3 and 4 for 
the case with WIMP's, these curves yield baryon abundances that are clearly 
incompatible with the bounds placed by cosmic nucleosynthesis ($\Ombar \sim
10^{-2}$).        

\begin{figure}
\begin{center}
\leavevmode
\epsfxsize=2.5in
\epsffile{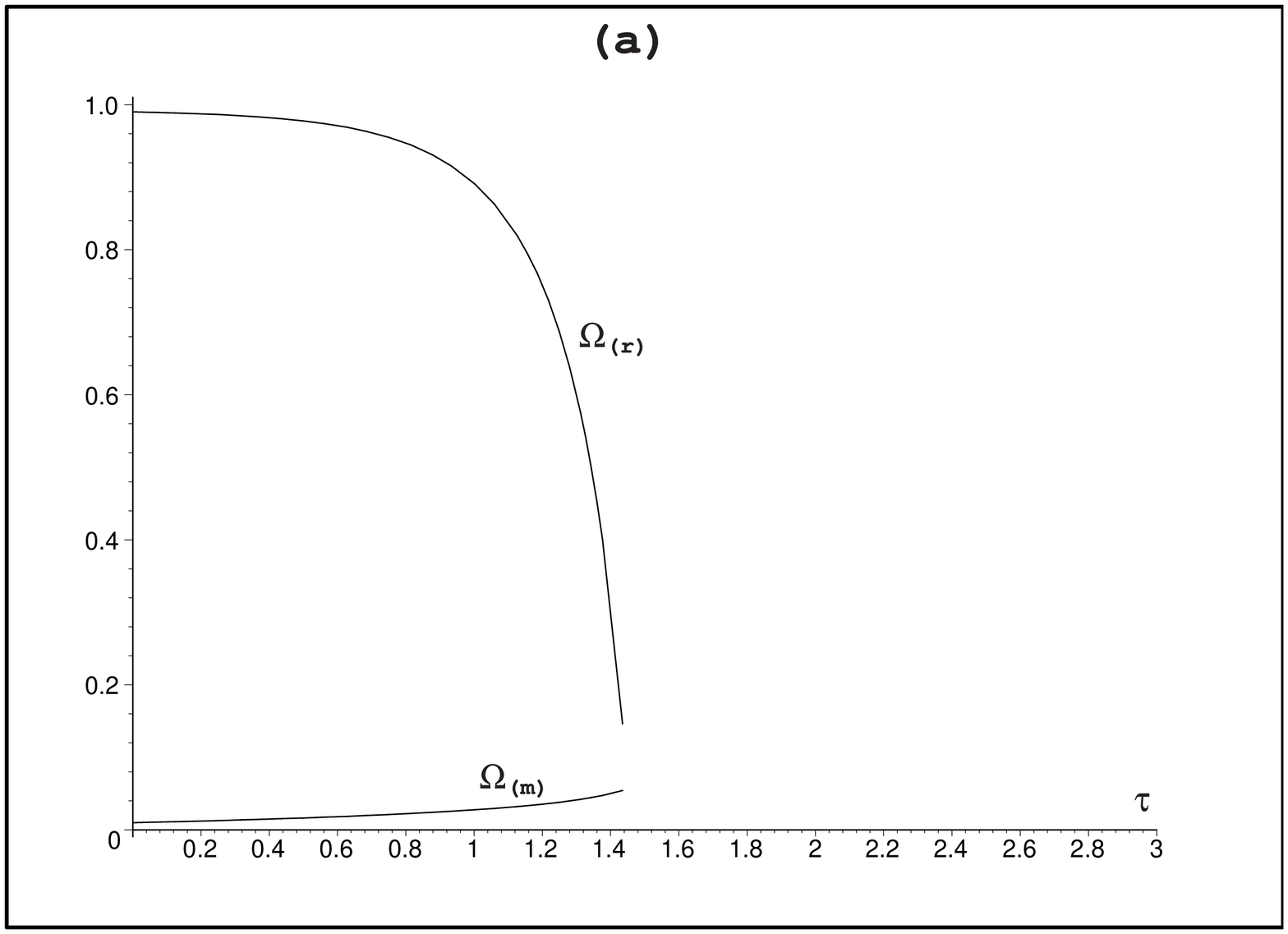}
\leavevmode
\epsfxsize=2.5in
\epsffile{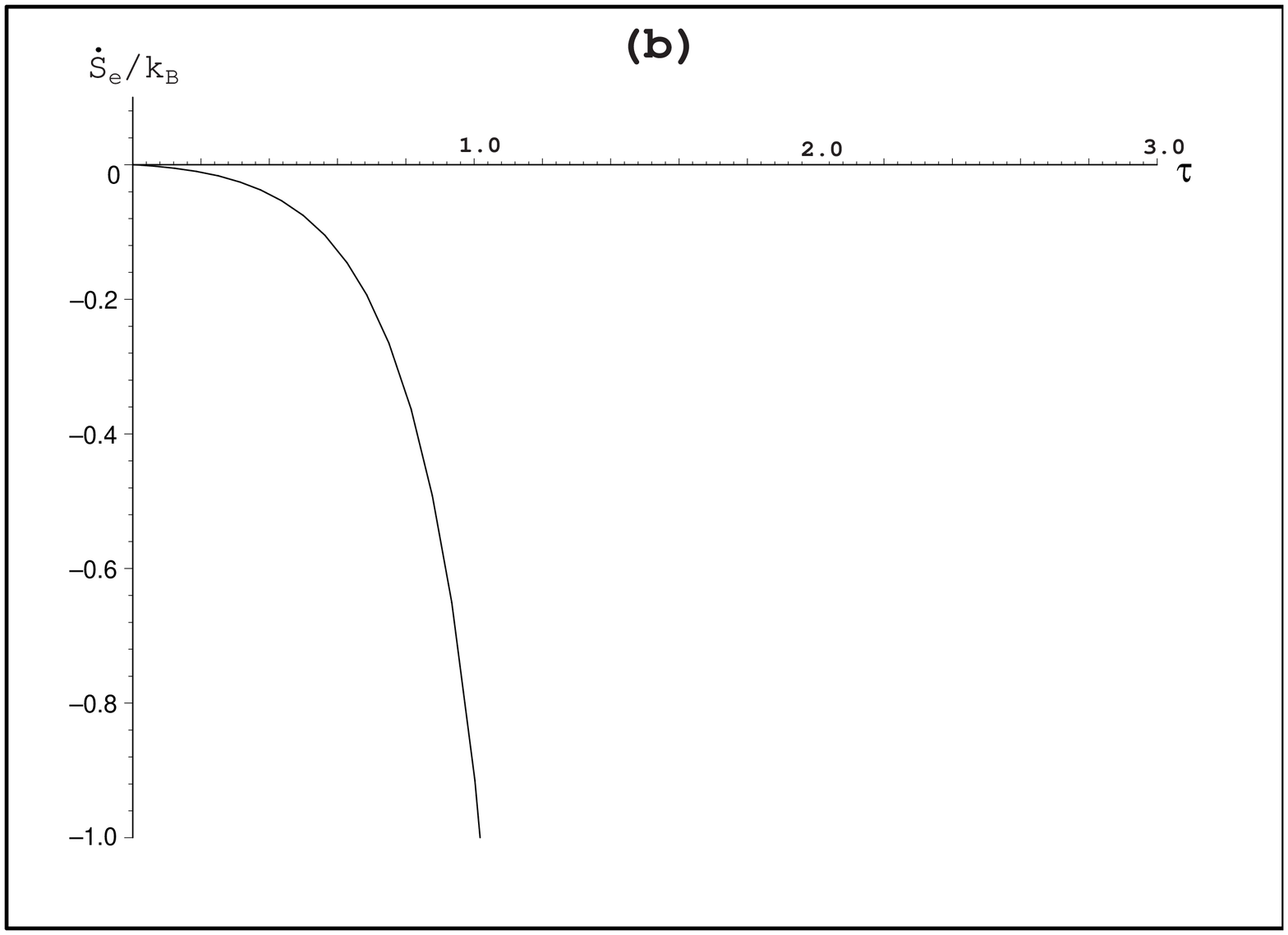}
\vskip0.1cm
\leavevmode
\epsfxsize=2.5in
\epsffile{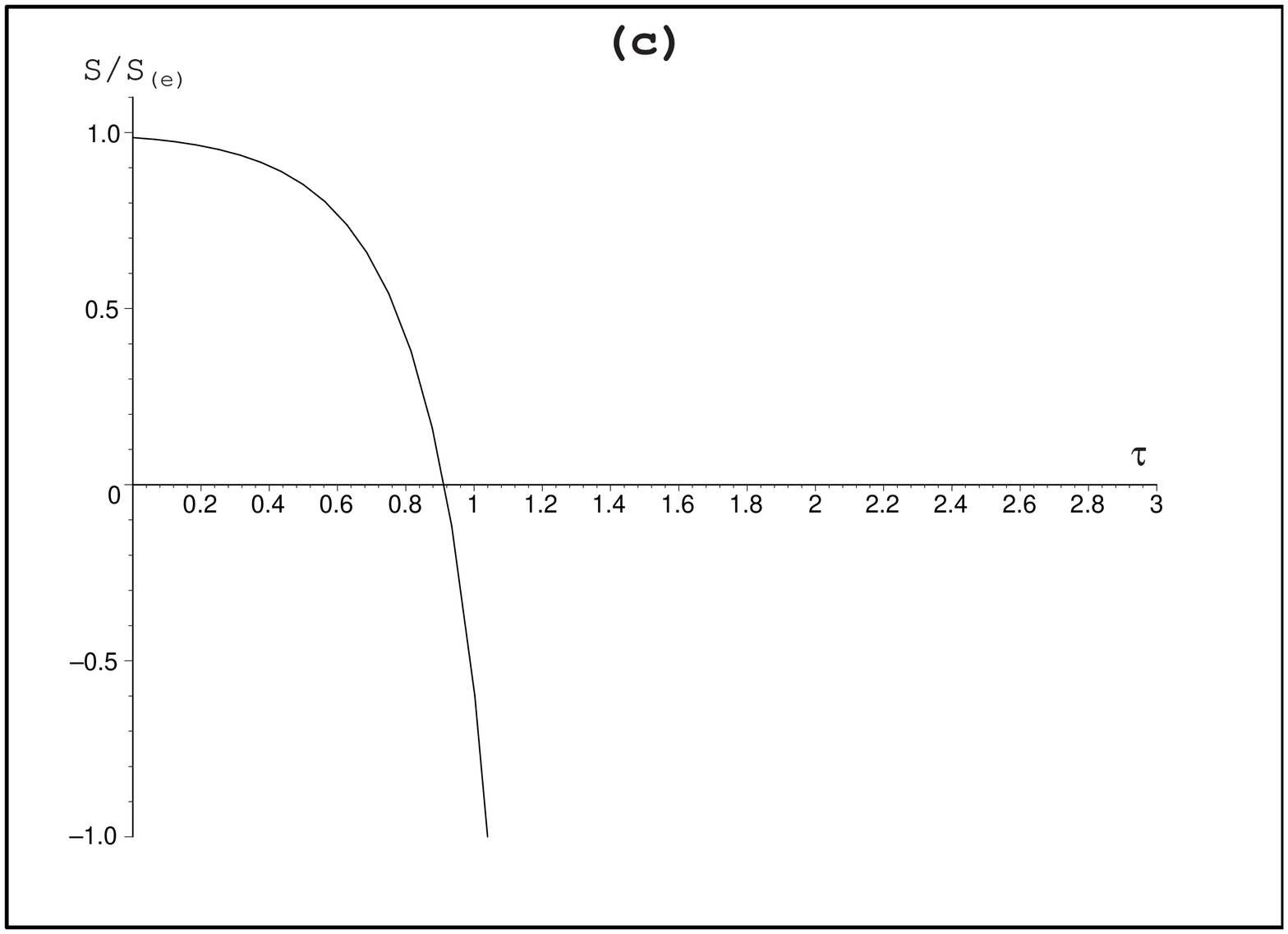}
\end{center}
\caption{${\hbox{The truncated transport equation}}. \ $ Curves obtained by 
integrating the governing equations with initial conditions (104) and $ \ 
\gamma_0 = 0.7, \ $ but using eqn. (83) (``truncated'' transport equation) 
instead of eqn. (64) (full transport equation). The exponential growth terms 
mentioned in section VII lead to an unphysical evolution characterized by 
negative $ \ \dSe \ $ and $ \ \dot S$ during all the evolution range.} 
\label{initfunc_test2}
\end{figure}

\section{Discussion and Conclusion.}\label{d_&_c}

We have studied a class of dissipative Kantowski-Sachs models describing the
cosmological evolution during the radiative era characterized by radiative 
processes involving baryons, electrons and photons, considered as a single 
dissipative ``radiative'' fluid. We also assumed the presence of CDM, in the 
form of a non-relativistic gas of WIMP's (lightest neutralinos). Although 
this gas does not interact with the radiative fluid, it provides the bulk of 
the rest mass energy density and thus it strongly influences the dynamics of 
the models and the resulting values of cosmological timescales, such as $ \ 
\tH. \ $ On the other hand, the radiative fluid provides the bulk of thermal 
and dissipative effects, related to the rate of change and relaxation of the 
radiation entropy to its equilibrium value.  

After defining new normalized variables $ \ \Omm, \ \Omr, \ $ $\Sigma, \ Q, \ $
a set of evo\-lu\-tion equa\-tions has been derived based upon appropriate 
thermodynamical laws and equations of state. The qualitative and numerical 
study of these evolution equations has clarified various aspects of the 
dynamical behavior of the models, their physical viability, as well as a 
peculiar sensitivity to certain initial conditions related to deviation from 
the invariant set $ \ \chi = 0. \ $ We discuss below the main features 
emerging from previous sections.

The definition of the phase space variables $ \ \Omr, \ \Omm, \ $ $\Sigma, \ Q 
\ $ leads in a natural way to ex\-press\-ing the relaxation time $ \ \trel \ $ 
as pro\-por\-tional to $ \ \tH = 3/\Theta \ $ (see equations 
(\ref{trel_prop_H2}) and (\ref{trel_prop_H3})). The understanding of the 
relaxation process can be accomplished by studying the effect of different 
choices of the proportionality factor, $ \ \gamma(\tau) > 0, \ $ on the 
exponential decay of the dissipative stress (related to $ \ Q$) and of the 
photon entropy $ \ S \ $ to its equilibrium value $ \ \Se. \ $ We have 
identified a ``near-Eckart'' regime if this decay is abrupt ($ \ \gamma \ll 
1, \ $ and becoming instantaneous in the limit $ \ \gamma \to 0, \ $ so that 
$ \ \trel \to 0$), while a ``transient regime'' ($\gamma \approx O(1) < 1$) 
can be associated with a slower decay. Both the near-Eckart and the 
transient regimes are compatible with a physically plausible evolution. The 
difference between the two regimes is the timescale of their relaxation 
process: for the transient regime this timescale can be comparable with the 
duration of the radiative era, for the near-Eckart regime it is much shorter 
(about eight orders of magnitude in physical time). This is well illustrated 
by the differences in evolution timescales between figures 5 and figures 1 
to 4. Comparing figures 2b and 5d, it is evident that the relaxation 
timescale of the near-Eckart regime ($\approx 10^{8}$ sec) is much shorter 
than that of the Compton scattering ($\approx 10^{11}$ sec). Therefore the 
near-Eckart regime yields a relaxation that is too swift and so it is 
inadequate to examine the two main radiative processes of the radiative era: 
the Compton scattering and (more so) the Thomson scattering. It is important 
to mention this fact in view of recent claims~\cite{geroch} that a 
transient theory of irreversible thermodynamics is not really necessary 
(see~\cite{pavher} for a comprehensive discussion). It is evident that the 
cosmological study of radiative processes in pre-decoupling times needs to be 
accomplished with a transient regime.

As revealed by figures 1, 2, 3 and 4, a physically plausible evolution is 
possible for all the duration of the radiative era for initial conditions 
given by an initial state very close to $ \ \chi(0) = 0, \ $ hence lying 
very close to the invariant set of zero  curvature $ \ \chi = 0. \ $ All 
models complying with a physical evolution begin  their evolution near the 
equilibrium point $ \ (\Omr, \ \Omm, \ \Sigma, \ Q) = (1,0,0,0), \ $ a 
saddle with positive eigenvalues (i. e., stable) associated with the FLRW 
sub-case of the models, and the proceeding evolution remains very 
close to the invariant set $ \ \chi=0. \ $ This is an extremely interesting 
feature of these models, as it relates a geometric property of KS solutions 
with the constraints imposed by the physics and by observational evidence, 
since recent data from CMB observations  indicates $ \ \Omt \approx 1. \ $ 
By looking at the curves of $ \ \Omm \ $ and $ \ \Omt \ $ with $ \ \delta 
\alt 10^{-5} \ $ in figures 4a and 4c (negative curvature), it is evident 
that for $ \ \tau > 6 \ $ these curves decrease from their values $ \ \Omm 
\ \sim 1 \ $ and $ \ \Omt \sim 1 \ $ around $ \ \tau = 6. \ $ Had we 
plotted these curves for larger values of $ \ \tau, \ $, extending to the 
present era ($\tau \approx 15$), we would have obtained $ \ \Omm \sim \Omt 
\sim 0.3, \ $ in agreement with the currently accepted value of $ \ \Omm, \ 
$ but not of $ \ \Omt. \ $ Of course, the estimated contribution to $ \ 
\Omt, \ $ today, for non-relativistic matter (CDM plus baryons) is only 
$ \ \approx 0.3, \ $  with the remaining two thirds of the critical density 
probably related to a $ \ \Lambda$-type ``dark energy'' interaction whose 
precise nature and properties are still uncertain. However, this discrepancy
with regards to $ \ \Omt \ $ today is not surprising since we did not 
consider any $ \ \Lambda$-type interaction, and so it does not affect our 
results. Also, the models  we are considering are only valid for a specific 
range of cosmological times: $ \ 10^3 \alt z \alt 10^6, \ $ in which this 
``dark energy'' would likely not have been dominant. Still, the close link 
between a physically plausible evolution and $ \ \Omt \ $ near unity is 
remarkable. 

It is interesting to compare our results to those reported
previously~\cite{weber} dealing with the perfect fluid sub-case of the KS 
models examined in this paper (though, these models did not consider a CDM 
component). As reported in~\cite{weber}, there are numerical solutions in 
which both matter and radiation normalized densities, $ \ \Omm, \ \Omr, \ $
decay to zero as the models re-collapse and approach a crunch singularity.  By
looking at the forms of $ \ \Omm \ $ and $ \ \Omr \ $ in figures 4, it is 
evident that such evolution is similar to that depicted by curves associated 
with initial conditions $ \ \delta \agt 10^{-3}. \ $ However, the evolution 
that results from these initial conditions fails to comply with our physical 
criteria, since the entropy $ \ S \ $ is no longer a convex function for all 
of the time range (see figure 4d) and starts decreasing at too early times. In 
the perfect fluid case, these examples satisfy an appropriate equation of 
state and all of the energy conditions and also the photon entropy is simply 
$ \ \Se \ $ and is constant for all times, hence there is no physical reason 
to discard these curves (other than remarking that such behavior of $ \ \Omm, 
\ \Omr \ $ is not observed in the real universe). However, for the dissipative 
source under examination here, $ \ S \ $ and $ \ \Se \ $ are not constant 
and the conditions for a physical evolution are more stringent, hence we can 
apply clear physical criteria to discard these perfect fluid ``strange''
cases.

For initial conditions near the equilibrium point $(\Omr,$ $\Omm, \ \Sigma, \ 
Q) \ = \ (1, 0, 0, 0) \ $ (as in equation (\ref{trial_ic})), the numerical 
curves of the ``equilibrium variables'' $ \ \Omm \ $ and $ \ \Omr \ $ are not 
affected by the choice of $ \ \gamma = \gamma_0, \ $ the constant 
proportionality factor between $ \ \trel \ $ and $ \ \tH. \ $ This is a 
consequence of the fact that for these initial conditions the values of $ \ 
\Sigma, \ Q \ $ and $ \ S - \Se \ $ remain small during all of the evolution. 
Therefore, the  evolution equations for $ \ \Omm \ $ and $ \ \Omr \ $ 
(equations (\ref{EqOmega}) and (\ref{EqOmegar})) are practically unaffected 
by the presence of $ \ \Sigma \ $ and $ \ Q, \ $ and so are insensitive to the 
rate of transiency given by $ \ \gamma \ $ constant. Since $ \ \Sigma \ $ 
and $ \ Q \ $ govern the deviation from the FLRW equilibrium point, this 
decoupling of $ \ \Omm \ $ and $ \ \Omr \ $ from $ \ \Sigma \ $ and $ \ Q \ 
$ is then a consequence of the evolution of the system always remaining  
close to thermal equilibrium. We tested this behavior for a particular form of 
$\ \gamma \ $ (eqn. (\ref{gamma_dynrel}), section \ref{dynrel}): the curves 
for $ \ \Omm, \ \Omr, \ \Sigma, \ $ and $ \ Q \ $ are practically identical 
with those that follow from choosing $ \ \gamma = 0.7 \ $ in figures 1 to 4. 
However, as shown in figures 6a and 6b, for large values of $ \ \mu_0 \ $ 
defined by (\ref{constants}), the obtained relaxation time $ \ \trel \ $ 
behaves similarly to what one would expect of a relaxation parameter for 
the radiative era. Although it has become common practice to simply equate 
$ \ \trel \ $ with a microscopic interaction time, like $ \ t_\gamma \ $ or 
$ \ t_c, \ $ the relaxation time is not a microscopic but a mesoscopic or 
even macroscopic quantity (though it must be qualitatively analogous to 
interaction times~\cite{pavher}). Since it can be extremely cumbersome to 
evaluate $ \ \trel, \ $ it is useful to have a concrete example where this 
relaxation parameter can be adequately approximated by the same dynamical 
equations associated with the models.

Finally, by means of qualitative arguments supported by the numerical 
analysis, we have shown in section VII (figures 7), that the truncated 
equation does not comply with a physically plausible evolution. This is an 
important result, since we have found a concrete example in which a 
truncated transport equation leads to unphysical evolution of dissipative 
fluxes. Although this conclusions strictly applies to the KS models under 
consideration, we must point out that one should be very cautious when 
applying these equations to other models and other equations of state. 

A possible extension of this work would be to consider instead of CDM other 
forms of dark matter, such as ``warm'' dark matter (WDM) or axions. Another 
possibility is to include, together with dark matter, a scalar field 
associated with ``dark energy''. Another route to generalize the present 
work is use a class of metrics associated with a geometry that is less 
restrictive than KS, for example the non-static spherical symmetry (perhaps 
under the assumption of self--similarity). We regard the present 
analysis of the Kantowski-Sachs models as a first step toward an 
understanding of the dynamics  of cosmic matter in more general and 
physically motivated inhomogeneous models.

\section*{Acknowledgements}
This work has been partially supported by the Universidad Nacional Aut\'onoma 
de M\'exico (UNAM) under grant DGAPA-IN-122498.


\begin{thebibliography}{99}

\bibitem{kotu}
E. W. Kolb and M. S. Turner, {\it The Early Universe}, Addison-Wesley 
(1990).

\bibitem{padma}
T. Padmanabhan, {\it Structure formation in the universe}. Cambridge 
University Press, Cambridge, U. K. (1995).

\bibitem{peacock} J. A. Peacock, {\it Cosmological Physics}. Cambridge 
University Press, Cambridge, U. K. (1999).

\bibitem{bern}
J. Bernstein, {\it Kinetic Theory in the Expanding Universe}, Cambridge 
University Press, Cambridge, U. K. (1988).

\bibitem{efsta}
G. Efstathiou, {\it Cosmological Perturbations}. In {\it Physics of the 
Early Universe}, Proceedings of the Thirty Sixth Scottish Universities 
Summer School in Physics. Eds J. A. Peacock, A. F. Heavens and A. T. 
Davies. IOP Publishing Ltd., Bristol, U. K. (1990).

\bibitem{ct}
A. A. Coley and B. O. J. Tupper, {\it J. Math. Phys.}, {\bf 27}, 406 
(1986).

\bibitem{rdm}
M. D. Pollock and N. Caderni, {\it Mon. Not. R. Astr. Soc.}, {\bf 190}, 
509 (1980); J. A. S. Lima and J. Tiomno, {\it Gen. Rel. Grav.}, {\bf 20}, 
1019 (1988); R. A. Sussman, {\it Class. Quantum Grav.}, {\bf 9}, 1891 
(1992).

\bibitem{pm}
V. M\'endez and D. Pav\'on, {\it Mon. Not. R. Astr. Soc.} {\bf 782}, 753 
(1996).

\bibitem{wei}
S. Weinberg, {\it Astrophys J.} {\bf 168}, 175, (1971).

\bibitem{sw}
S. Weinberg, {\it Gravitation and Cosmology}, J. Wiley, N. Y. (1972).

\bibitem{peeb}
P. J. E. Peebles, {\it The Large Scale Structure of the Universe}, 
Princeton University Press, Princeton, U. S. A. (1980).

\bibitem{susstr}
R. A. Sussman and J. Triginer, {\it Class. Quantum Grav.}, {\bf 16}, 
167 (1999).

\bibitem{ksmh}
D. Kramer, H. Stephani, M. A. H. MacCallum, E. Herlt, {\it Exact 
Solutions of Einstein's Field Equations}, Cambridge University Press, 
Cambridge, U. K. (1980).

\bibitem{krs}
A. Krasi\'nski, {\it Inhomogeneous Cosmological Models}, Cambridge 
University Press, Cambridge, U. K. (1997).

\bibitem{Ellis}
John Ellis, Summary of DARK 2002: {\it 4th International Heidelberg 
Conference on Dark Matter in Astro and Particle Physics}, Cape Town, 
South Africa, 4-9 Feb. 2002. e-preprint: {\tt astro-ph/0204059}.

\bibitem{Report} 
G. Jungman, M. Kamionkowski and K. Griest, {\it Phys. Rep.} {\bf 267}, 
195-373 (1996).

\bibitem{Torrente} 
S. Khalil, C. Mu\~noz and E. Torrente-Lujan {\it New Jour. Phys.} {\bf 4}, 
27 (2002). e-preprint: {\tt hep-ph/0202139}.

\bibitem{hl1}
W. A. Hiscock and L. Lindblom, {\it Phys. Rev.} D {\bf 31}, 725 (1985).

\bibitem{hl2}
W. A. Hiscock and L. Lindblom, {\it Phys. Rev.} D {\bf 35}, 3723 (1987).

\bibitem{wi}
W. Israel, {\it Ann. Phys.} (N. Y.), {\bf 100}, 310 (1976).

\bibitem{wjs}
W. Israel and J. Stewart, {\it Ann. Phys.} (N. Y.), {\bf 118}, 341 (1979).

\bibitem{ddj}
D. Pav\'on, D. Jou and J. Casas-V\'azquez, {\it Ann. Inst. H. Poincar\'e. 
Ser. A}, {\bf 36}, 79 (1982).

\bibitem{jcl}
D. Jou, J. Casas-V\'azquez and G. Lebon, {\it Extended Irreversible 
Thermodynamics}, 2nd edition, Springer, Berlin (1996).

\bibitem{hl3}
W. A. Hiscock and L. Lindblom, {\it Contemporary Mathematics}, {\bf 71}, 
181 (1991).

\bibitem{uwi}
N. Udey and W. Israel, {\it Mon. Not. R. Astr. Soc.}, {\bf 199}, 1137 (1982).

\bibitem{dd1}
D. Pav\'on, D. Jou and J. Casas-V\'azquez, {\it J. Phys. A: Math. Gen.} {\bf 
16}, 775 (1983).

\bibitem{dd2}
D. Jou and D. Pav\'on, {\it Astrophys. J.}, {\bf 291}, 447 (1985).

\bibitem{lan}
P. T. Landsberg and D. Evans, {\it Mathematical Cosmology}, Oxford University 
Press, Oxford, U. K. (1979).

\bibitem{matr}
R. Maartens and J. Triginer, {Phys. Rev.} D {\bf 56}, 4640 (1997).

\bibitem{zim}
W. Zimdahl, {\it Mon. Not. R. Astr. Soc.} {\bf 280}, 1239 (1996).

\bibitem{zp}
R. A. Sussman and D. Pav\'{o}n, {\it Phys. Rev.} D, {\bf 60}, 104023, (1999).

\bibitem{CG} A. Coley and M. Goliath, {\it Phys. Rev.} D {\bf 62}, 043526 
(2000).

\bibitem{mes}
R. Maartens, G. F. R. Ellis and W. R. Stoeger, {\it Phys. Rev.} D, {\bf 51}, 
1525 (1995).

\bibitem{ellisvan} 
G. F. R. Ellis and H. van Helst, {\it Cosmological Models}, Carg\`ese 
Lectures 1998. e-preprint: {\tt gr-qc/9812046}.

\bibitem{geroch}
R. Geroch, {\it On hyperbolic ``theories'' of relativistic dissipative 
fluids.} e-preprint: {\tt gr-qc/0103112}.

\bibitem{pavher}
L. Herrera and D. Pav\'on, {\it Hyperbolic theories of dissipation: why and 
when do we need them ?} e-preprint: {\tt gr-qc/0111112}. To appear in {\it 
Physica A.}

\bibitem{weber}
E. Weber, {\it J. Math. Phys.} {\bf 27}, 1578 (1986).

\end{thebibliography}
\end{document}